\documentclass[aps,pra,superscriptaddress,twocolumn,longbibliography]{revtex4-1}

\usepackage[english]{babel}
\usepackage{amssymb}
\usepackage{algorithm}
\usepackage{algpseudocode}
\usepackage{graphicx}
\usepackage{color}
\usepackage[colorlinks=true,
			urlcolor=blue,
			citecolor=blue,
			linkcolor=blue,
			citebordercolor={1 0 0},
			linkbordercolor={0 0 1}]{hyperref}
\usepackage{quantikz}

\newcommand{\ketbra}[2]{\vert #1 \rangle \langle #2 \vert}
\newcommand{\figref}[1]{Fig.~\ref{#1}}

\makeatletter
\renewcommand\frontmatter@abstractwidth{\dimexpr\textwidth -0in \relax}

\newcommand{\utchem}{Department of Chemistry, University of Toronto, Toronto, Ontario M5G 1Z8, Canada}
\newcommand{\utcomp}{Department of Computer Science, University of Toronto, Toronto, Ontario M5S 2E4, Canada}
\newcommand{\vectorinst}{Vector Institute for Artificial Intelligence, Toronto, Ontario M5S 1M1, Canada}
\newcommand{\cifar}{Lebovic Fellow, Canadian Institute for Advanced Research, Toronto, Ontario M5G 1Z8, Canada}
\newcommand{\wisc}{Department of Chemistry, University of Wisconsin-Madison, 1101 University Ave., Madison, WI 53706, USA}
\newcommand{\yqi}{Yale Quantum Institute, Yale University, P.O. Box 208334, New Haven, CT 06520-8263, USA}
\newcommand{\ychem}{Department of Chemistry, Yale University, P.O. Box 208107, New Haven, CT 06520, USA}
\newcommand{\yenergy}{Energy Sciences Institute, Yale University, P.O. Box 27394, West Haven, CT 06516-7394, USA}
\newcommand{\surrey}{Department of Mathematics, University of Surrey, Guildford, United Kingdom}
\newcommand{\material}{Department of Materials Science \& Engineering, University of Toronto, Toronto, Ontario M5S 3E4, Canada}
\newcommand{\chemical}{Department of Chemical Engineering \& Applied Chemistry, University of Toronto, Toronto, Ontario M5S 3E5, Canada}

\begin{document}

\title{Variational quantum iterative power algorithms for global optimization}
\author{Thi Ha Kyaw}
\thanks{These authors contributed equally to this work.\\
\urlstyle{same}
\url{thihakyaw@cs.toronto.edu}\,
\url{micheline.soley@yale.edu}}
\affiliation{\utchem}
\affiliation{\utcomp}
\affiliation{Current address: LG Electronics Toronto AI Lab, Toronto, Ontario M5V 1M3, Canada}
\author{Micheline  B. Soley}
\thanks{These authors contributed equally to this work.\\
\urlstyle{same}
\url{thihakyaw@cs.toronto.edu}\,
\url{micheline.soley@yale.edu}}
\affiliation{\wisc}
\affiliation{\yqi}
\affiliation{\ychem}
\author{Brandon Allen}
\affiliation{\yqi}
\affiliation{\ychem}
\author{Paul Bergold}
\affiliation{\surrey}
\author{Chong Sun}
\affiliation{\utcomp}
\author{Victor S. Batista}
\email{victor.batista@yale.edu}
\affiliation{\yqi}
\affiliation{\ychem}
\affiliation{\yenergy}
\author{Al\'an Aspuru-Guzik}
\email{alan@apsuru.com}
\affiliation{\utchem}
\affiliation{\utcomp}
\affiliation{\vectorinst}
\affiliation{\material}
\affiliation{\chemical}
\affiliation{\cifar}

\date{\today}

\begin{abstract}
We introduce a family of variational quantum algorithms called quantum iterative power algorithms (QIPA) that outperform existing hybrid near-term quantum algorithms of the same kind. We demonstrate the capabilities of QIPA as applied to three different global-optimization numerical experiments: the ground-state optimization of the $H_2$ molecular dissociation, search of the transmon qubit ground-state, and biprime factorization. Since our algorithm is hybrid, quantum/classical technologies such as error mitigation and adaptive variational ansatzes can easily be incorporated into the algorithm.
Due to the shallow quantum circuit requirements, we anticipate  large-scale implementation and adoption of the proposed algorithm across current major quantum hardware.
\end{abstract}

\maketitle

\section{Introduction}
Quantum computers promise exponential speedup over classical counterparts in solving certain tasks \cite{Shor1994algorithms}. When fault-tolerant general-purpose quantum computers become available, adiabatic state preparation and quantum phase estimation may become the standard quantum routines for determining the ground-state energy of sophisticated physical Hamiltonians \cite{Aspuru-Guzik2005simulated,Georgescu2014quantum,Cao2019quantum,McArdle2020quantum}. However, such schemes are very costly in terms of required overhead and hence are not suitable for the current era of noisy intermediate-scale quantum (NISQ) hardware \cite{Preskill2018quantum,Arute2019quantum,Zhong2020quantum,Wu2021strong,Madsen2022quantum}. This limitation of quantum computers today shifts central attention towards low-depth hybrid quantum-classical algorithms, known as NISQ algorithms \cite{Bharti2022noisy,cerezo2021variational,Tilly2021variational,fedorov2022vqe}. The variational quantum eigensolver (VQE) \cite{Peruzzo2014variational,McClean2016theory} serves as a  prototypical example, as an algorithm that computes the expectation value of a Hamiltonian, which is measured on a quantum machine, resulting in a cost function with a set of variational parameters, which are optimized using classical computers. The process is repeated until the cost function reaches its local minimum.

On the other hand, the variational quantum simulator~\cite{li2017efficient} has been proposed for hybrid quantum-classical simulations of quantum dynamics based on the McLachlan's variational principle \cite{mcardle2019variational,yuan2019theory}, including quantum imaginary time evolution (QITE) to prepare ground states~\cite{moll2018quantum,mcardle2019variational,yuan2019theory,gomes2020efficient,Motta2020determining,nishi2021implementation,selvarajan2021prime,huang2022efficient,yeter2022quantum}. Here, we introduce the ``quantum iterative power algorithm'' (QIPA) inspired by the variational quantum simulator to provide an efficient method to the general problem of global optimization with near term quantum computers.

Global optimization is central to many important problems in science and engineering, from back-propagation in machine learning \cite{bottou2018optimization,sun2019survey} and molecular geometry optimization/protein structure prediction \cite{wales1999global,hardin2002ab,dill2008protein,dorn2014three,kuhlman2019advances} to route planning and control of drone/unmanned aerial vehicles \cite{coutinho2018unmanned,otto2018optimization}. However, the brute force approach of considering each possible element of a search space often becomes computationally intractable. For example, identification of the optimal configuration of a protein faces Levinthal’s paradox \cite{levinthal1969fold,karplus1997levinthal} — that the native configuration must be identified out of about $10^{300}$ possibilities. This has inspired a broad array of both classical \cite{hartmann2004new,venter2010review} and quantum computing \cite{Grover.1996.212,Durr.1996.9607014v2,Bulger.2003.517,Farhi.2014.1411.4028,Farhi.2019.1602.07674v2,Zhou.2019.1812.01041v2,Kadowaki.1998.5355,Farhi.2000.0001106,Farhi.2001.472,PerdomoOrtiz.2012.571,Babbush.2014.6603,Albash.2018.031016,Temme.2011.87,Daskin:2019aa,moll2018quantum,mcardle2019variational,gomes2020efficient,Motta2020determining,nishi2021implementation,selvarajan2021prime,huang2022efficient,yeter2022quantum} optimizers. Recently, we have shown that tensor trains \cite{Oseledets.2010.70,Oseledets.2011.2295} (also known as matrix product states \cite{Ostlund.1995.3537}) provide a way to vastly reduce the computational cost of exploring low-rank optimization cost functions, and have employed the approach to introduce an optimization algorithm that deterministically explores the full search space in data-compressed form, the tensor-train ``iterative power algorithm (IPA)'' \cite{soley2021iterative}.

We recognize the strategy of tensor-train IPA can be implemented on quantum computers to enable global optimization of an even broader class of optimization problems. In tensor-train IPA \cite{soley2021iterative}, the optimization cost function of interest is taken to be a potential energy surface. A density is initialized in the potential energy surface, and an oracle is iteratively applied in a sifting approach akin to imaginary time propagation (with infinite mass) to localize the density as a delta function at the global minimum position. The expectation value of position then gives the location of the global minimum. Tensor-train IPA represents the density and potential energy surface as tensor trains to avoid calculation of the cost function everywhere in search space, which is efficient for representation of problems amenable to low-rank representations, such as prime factorization or molecular geometry optimization \cite{soley2021iterative}. However, the tensor-train strategy faces the roadblock that highly-coupled systems cannot be efficiently represented in low-rank tensor-train format. In contrast, quantum computers excel in the simulation of highly-coupled systems, as the coupling or entanglement between qubits is limited only by the choice of ansatz \cite{jozsa2003role,vidal2008class,gross2009most,evenbly2014class,bluvstein2022quantum,haghshenas2022variational,Sim2019expressibility}.

\begin{figure}[t]
	\centering
	\includegraphics[width=1.\columnwidth]{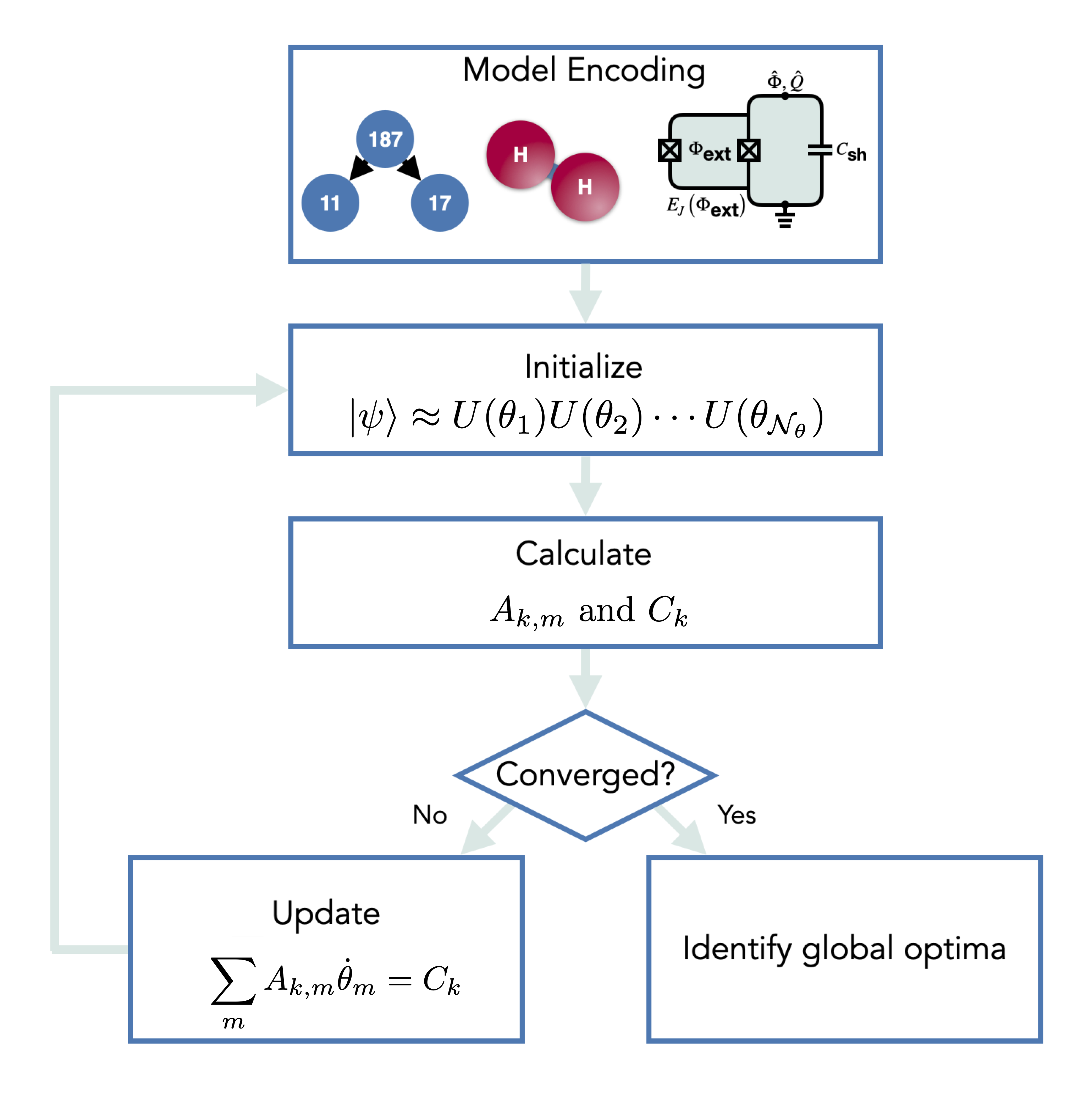}
	\caption{Sketch of the quantum iterative power algorithm. First, the physical problem is mapped into the language of the  quantum computer. Second, initialization of a parameterized wavefunction $\ket{\phi(\theta(\tau))}$ is achieved by using a certain quantum ansatz circuit. Third, based on the ansatz choice, Hadamard test measurements are performed to obtain the $A$ matrix and $C$ vector on the quantum computer. Fourth, the new $\theta$ parameters are obtained from $A$ and $C$ on the classical computer. If the desired convergence is obtained, the program is stopped and the global minima are identified. Otherwise, the step of evaluating $A$ and $C$ is repeated to obtain new angles $\theta$.}\label{fig:intro}
\end{figure}

The quantum iterative power algorithm (QIPA) takes advantage of the high degree of entanglement possible on quantum computers with a hybrid variational scheme. In standard variational approaches such as the variational quantum eigensolver (VQE) \cite{Peruzzo2014variational,yung2014transistor,wecker2015progress,McClean2016theory,shen2017quantum,moll2018quantum,romero2018strategies,cerezo2021variational,fedorov2022vqe}, classical optimizers are used to determine the parameters of a quantum circuit, which are used to prepare trial wavefunctions measured to obtain expectation values. Analogously, the variational quantum simulator~\cite{li2017efficient} evolves the parameters that define the time-evolved wavefunction by using a classical computer that integrates the Euler-Lagrange equation obtained from the Schr\"odinger equation with the McLachlan’s variational principle. Parameters required by the Euler-Lagrange equation are obtained with a quantum circuit with a small number of quantum operations. QIPA generalizes the variational quantum approach to evolve an arbitrary initial density distribution so that it would become localized at the global minimum of a given cost function (see \figref{fig:intro}). As in IPA, the propagator of QIPA is not limited to the imaginary time quantum propagator (with infinite mass) enabling the use of other propagators that are maximal at the minimum of the cost function.

\figref{fig:intro} shows the overall work flow of the QIPA algorithm. First, we select parameters $\theta_1,\theta_2,\dots,\theta_{\mathcal{N}_\theta}$ corresponding to the initial state. Second, we use a quantum co-processor and the Hadamard test to calculate the parameters $A_{k,m}$, and $C_k$ introduced by the Euler-Lagrange equation $\sum_m A_{k,m}\dot{\theta}_m=C_k$. Next, we update the parameters $\theta_j$ by numerical integration of the Euler-Lagrange equation using a classical computer. Having updated the parameters, the process is iterated until convergence to obtain parameters $\theta_j$ corresponding to a distribution function localized at the global minimum. The converged parameters are then used by a quantum circuit to prepare the final state and obtain the expectation value of the global minimum coordinates.

The paper is organized, as follows. First, we introduce the quantum iterative power algorithm from a general perspective. Next, we apply QIPA to a wide-range of applications and we compare the performance of QIPA with that of QITE, including a molecular Hamiltonian example, the $\text{H}_2$ complex, quantum computer-aided design, and biprime factorization. Finally, the Discussion and Conclusions section includes a discussion of results and ideas for further development of the QIPA algorithm. The Tequila package \cite{Kottmann2021tequila} has been used to implement all the quantum circuit operations, using Qulacs \cite{Suzuki2021qulacs} as a quantum backend. 

\section{Oracle Function}\label{sec:qipa}
The main idea behind our proposed algorithm is the realization that an arbitrary cooling function \cite{zeng2021universal,xu2014demon} or strictly positive oracle function that is maximized at the location of the global minimum of a potential energy surface $V$ can be used to amplify the global minimum amplitude of any initial state $\ket{\psi(0)}$ with finite amplitude at the global minimum. Here, we show that oracles  defined by concatenated exponential functions, 
\begin{equation}\label{eq:ExponentialFunctions}
    \alpha_n(-\tau V)
    =e^{a_n\alpha_{n-1}(-V\tau)},
\end{equation}
with $n>1,\,\alpha_0(y)=y$ and real constants $a_1,\dots,a_n\neq 0$, provide effective algorithms based on a generalization of the McLachlan’s variational principle (Appendix~\ref{Appen:qipa}). The calculations reported in this paper are based on the oracle defined by the double exponential $\alpha_2(-\tau V)=e^{e^{-\tau V}}$, which we obtain for the choice $n=2$ and $a_2=a_1=1$. 

A particular case of global optimization involves the search of the ground state of a Hamiltonian $H$, a problem that is typically solved by imaginary time propagation. QIPA can solve that problem analogously by simply replacing $V$ by $H$ in the definition of the oracle, $\alpha_n$. In that case, the normalized oracle function $f(H;\tau)$ acts on the initial wavefunction $\ket{\psi(0)}$ (onwards, $\hbar=1$), as follows:
\begin{align}\label{eq:general_oracle}
	\ket{\psi(\tau)}
	&=f(H;\tau)\ket{\psi(0)}\nonumber\\
	&=\frac{U_n(\tau)\ket{\psi(0)}}{\sqrt{\langle U_n(\tau)\psi(0)\vert U_n(\tau)\psi(0)\rangle}},
\end{align}
where $U_n(\tau)=\alpha_n(-H\tau)$.
\begin{figure}[ht]
	\begin{algorithm}[H]
		\begin{algorithmic}[1] 
			\Require Hamiltonian $H$ and initial state $\ket{\psi(0)}=\ket{0}^{\otimes \mathcal{N}}$
			\State Start with an ansatz $\ket{\phi(\theta_0)}=U(\theta_0)\ket{\Bar{0}}$ at time $\tau=0$;
			\State Evaluate Hadamard tests to form the matrix $A(\tau)$ and the vector $C(\tau)$ (quantum computer subroutine);
			\State Compute approximate solution $\xi_\tau$ of $A(\tau)\dot{\theta}(\tau)=C(\tau)$ via the CG method (classical computer subroutine);
			\State Update the parameter as $\theta(\tau+\delta\tau)\leftarrow\theta(\tau)+\xi_\tau\delta\tau$ and set $\tau\leftarrow\tau+\delta\tau$;
			\State Repeat steps $2-4$ until $\tau=\tau_{\textrm{total}}$ or the convergence criteria is met;
		\end{algorithmic} 
		\caption{Variational quantum iterative power algorithm}\label{alg:VQIP}
	\end{algorithm}
\end{figure}
In particular, $\alpha_1$ corresponds to the standard imaginary time evolution, which is widely used in quantum Monte Carlo algorithms \cite{Foulkes2001quantum,LeBellac2004equilibrium}. Refs.~\cite{mcardle2019variational,yuan2019theory} show that one can perform imaginary time evolution \cite{Motta2020determining} with unitary gates defined by Eq.~(\ref{eq:general_oracle}) with $n=1$ that evolve the initial state according to the Wick-rotated Schr\"odinger equation, obtained by the McLachlan's variational principle:
\begin{align}
	\frac{\mathrm{d}}{\mathrm{d}\tau}\ket{\psi(\tau)}
	=-\left(H-E_1(\tau)\right)\ket{\psi(\tau)},
\end{align}
where $E_1(\tau)=\langle\psi(\tau)\mid H\mid\psi(\tau)\rangle$. Here, we introduce a family of near-term quantum algorithms defined by $\alpha_n$ with $n>1$ that evolve the initial state according to the generalized Wick-like-rotated Schr\"odinger equation (Appendix~\ref{Appen:qipa}):
\begin{align}
		\frac{\mathrm{d}}{\mathrm{d}\tau}\ket{\psi(\tau)}
		&=-\prod_{k=1}^{n}a_k\Big(H\exp(S_{n-1})\label{eq:general}\\
		&\qquad-\operatorname{Re}\langle H\exp(S_{n-1})\psi(\tau)\mid\psi(\tau)\rangle\Big)\ket{\psi(\tau)},\nonumber
	\end{align}
	where $S_{n-1}=\sum_{k=1}^{n-1}a_k\alpha_{k-1}(-H\tau)$. 
\begin{figure*}[th]
	\centering
	\includegraphics[scale=0.9]{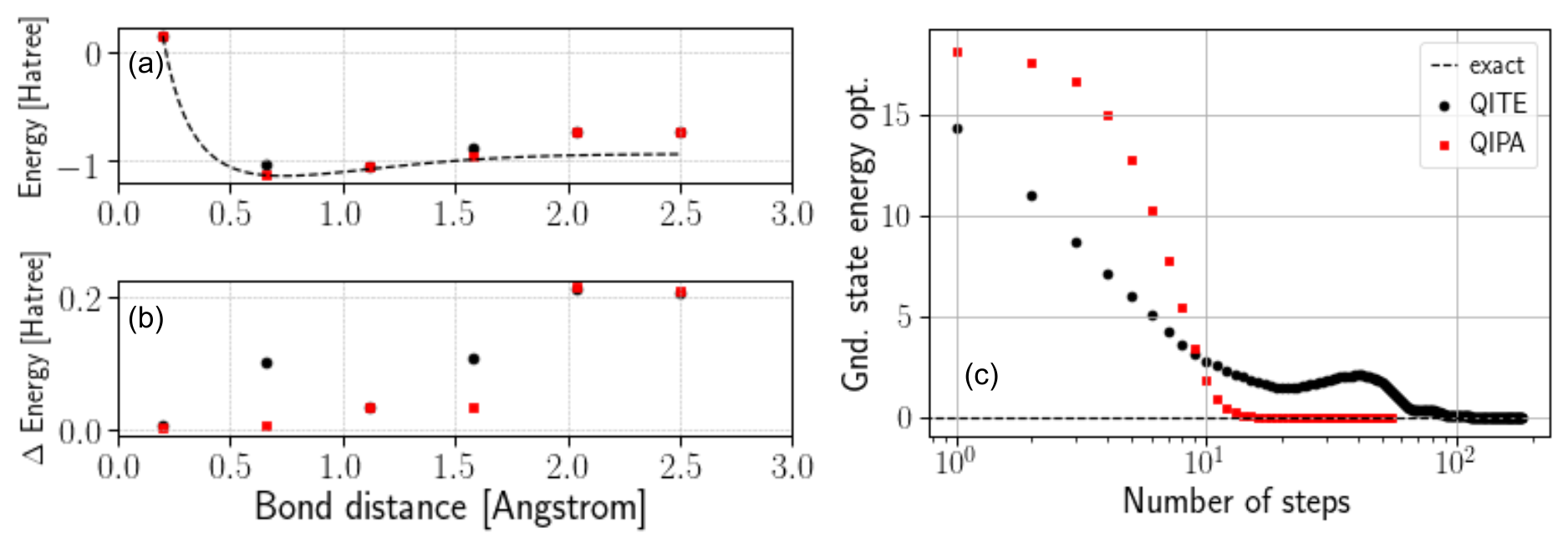}
	\caption{$\text{H}_2$ energy dissociation curve in the minimal basis set (sto-3g). (a) Exact diagonalization result/ full CI (black solid line) is seen with data points from QIPA (red squares) and QITE (black dots) runs for different bond distances. (b) Absolute energy difference between the exact energy and the QIPA and QITE results in Hatrees, corresponding to the data in (a). (c) Ground-state energy optimization plot for a flux tunable transmon at the external flux $f=0.25$ as a function of the number of iteration steps for both QIPA and QITE. In all results, both QIPA and QITE are run with the same time step. Here QIPA runs require significantly fewer steps to reach the convergence criteria.}\label{fig:H2_transmon_qipa}
\end{figure*}

With the choice $n=2$, we arrive at a double exponential function and the following Wick-like-rotated Schr\"odinger equation:
\begin{align}
	\frac{\mathrm{d}}{\mathrm{d}\tau}\ket{\psi(\tau)}
	=-\left(He^{- H\tau}-E_2(\tau)\right)\ket{\psi(\tau)},
\end{align}
with $E_2(\tau)=\langle\psi(\tau)\mid He^{- H\tau}\mid\psi(\tau)\rangle$. 
According to the McLachlan's variational principle, when we constrain the equation of motion as such
\begin{align}
	\delta\left\|\left(\partial/\partial\tau+\left[He^{- H\tau}-E_2(\tau)\right]\right)\ket{\psi(\tau)}\right\|^2
	=0,
\end{align}
the result is equivalent in finding a solution of the linear equation
\begin{align}\label{eq:qipa_EOM}
	\sum_mA_{k,m}\dot{\theta}_m
	=C_k,
\end{align}
where the entries of the symmetric and positive semi-definite matrix $A$ and the right-hand side $C$ can be computed on a quantum computer by deploying the Hadamard test. The value of $\theta$ is updated with $\dot{\theta}$ for a short time step $\delta\tau>0$ according to the Euler method as $\theta(\tau+\delta\tau)\approx\theta(\tau)+\dot{\theta}(\tau)\delta\tau$. The underlying assumption is that we can approximate $\ket{\psi(\tau)}$ by $\ket{\phi(\theta(\tau))}=U(\theta_1(\tau))U(\theta_2(\tau))\cdots U(\theta_{\mathcal{N}_\theta}(\tau))\ket{\bar{0}}$, where $\ket{\bar{0}}=\ket{0}^{\otimes\mathcal{N}}$ and
$U(\theta_1(\tau)),\dots,U(\theta_{\mathcal{N}_\theta}(\tau))$ are parameterized quantum circuits (PQCs), with $\theta=(\theta_1,\dots,\theta_{\mathcal{N}_\theta})$ the corresponding real-valued parameter vector. 

In general, for an $\mathcal{N}$-qubit system with Hamiltonian $H$ with $\mathcal{N}_H\ge 1$ Pauli words and a parameterized wavefunction $\ket{\phi (\theta)}$ (where $\tau$ dependency $\theta(\tau)$ is understood throughout) with $\mathcal{N}_\theta\ge 1$ parameters, the upper bound for the number of distinct measurements $\mathcal{N}_{A}$ required to obtain the matrix $A$ for QIPA via the Hadamard test and the number of gates required are $\mathcal{N}_\theta(\mathcal{N}_\theta-1)/2$ and $G_{\mathcal{N}_A}\ge \mathcal{N}_\theta$, respectively. Such an estimate can be understood as the number of times required to completely evaluate all the $A$ matrix elements since $A$ is symmetric. Moreover, to obtain the vector $C$, the number of measurements and gates required (assuming a second-order Taylor series expansion of the required function of the Hamiltonian) are $\mathcal{N}_\theta$ and $G_{\mathcal{N}_C} \ge \mathcal{N}_H+\mathcal{N}_H^2+\mathcal{N}_H^3+\mathcal{N}_\theta$, respectively. 
`$>$' sign in $G_{\mathcal{N}_{A}}$ and $G_{\mathcal{N}_{C}}$ holds when two-qubit gates are not parameterized, while `$=$' sign holds when they are parameterized.  
Assuming a polynomial scaling: $\mathcal{N}_H=\mathcal{O}(\mathcal{N}^h)$,\,$\mathcal{N}_\theta=\mathcal{O}(\mathcal{N}^d)$, the leading order becomes $\mathcal{N}_{A}=\mathcal{O}(\mathcal{N}^{d})$ and $\mathcal{N}_C=\mathcal{O}(\mathcal{N}^{\textrm{max}(3h,d)})$, respectively. In comparison, in QITE, one needs $\mathcal{N}_A=\mathcal{O}(\mathcal{N}^{d})$ and $\mathcal{N}_C=\mathcal{O}(\mathcal{N}^{\textrm{max}(h,d)})$, with the same number of Hadamard test measurements required. In general, QIPA yields improved convergence in shorter times compared to QITE, requiring the same number of Hadamard test operations and a higher number of unitary gates.

\section{Results}\label{sec:results}
\subsection{Molecular ground-state search}\label{subsec:molecule}
\begin{figure*}[t]
	\centering
	\includegraphics[width=1\textwidth]{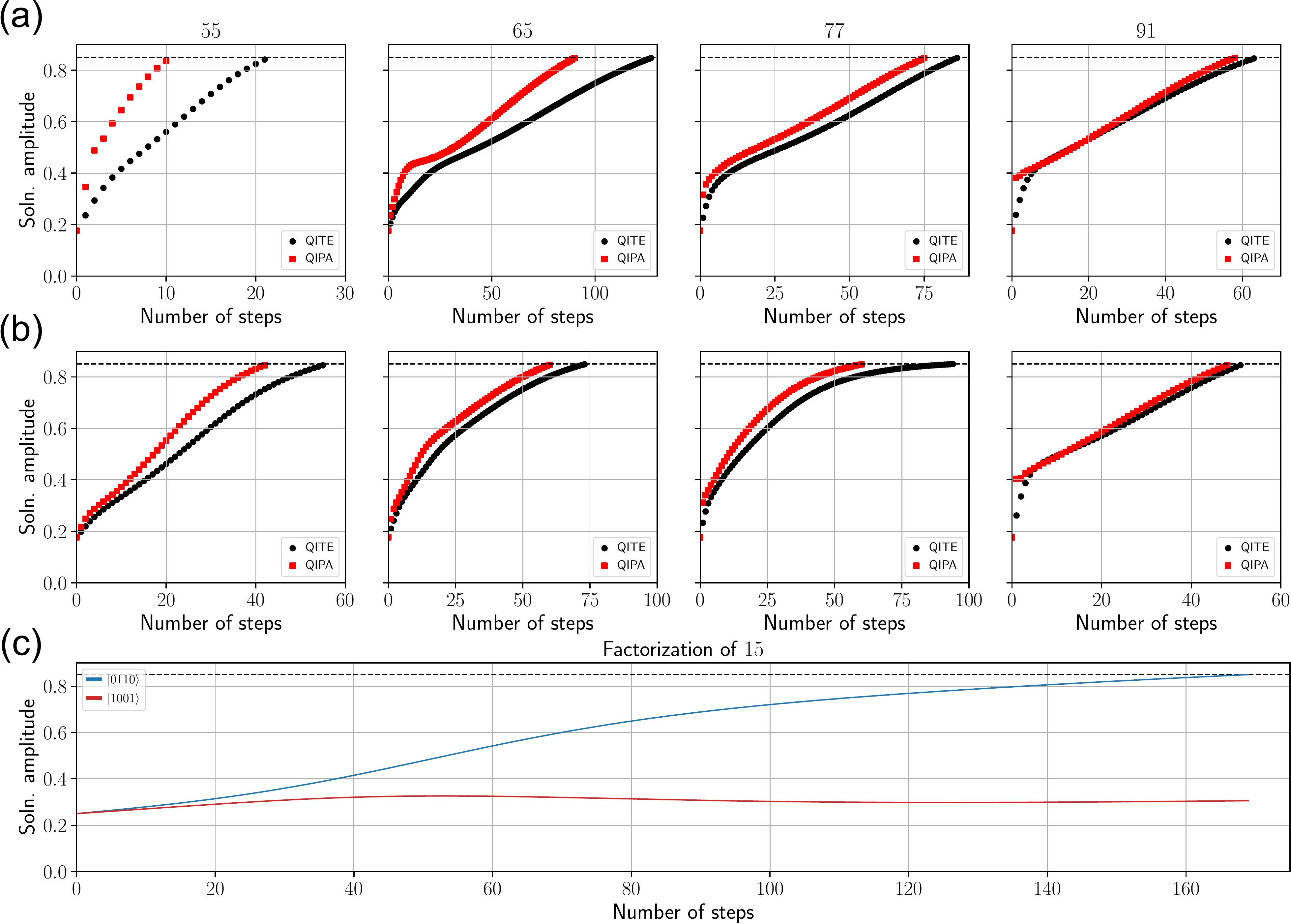}
	\caption{QIPA factorization of biprimes 55, 65, 77, and 91 for the (top) YZ and (center) Y Ansatz as compared to QITE for equal time steps, and (bottom) the amplitude of the wavefunction components corresponding to the prime factors of 15 for varying numbers of time steps in QIPA. The black dashed line represents to the final amplitude of the wavefunction component with the largest magnitude at convergence.
	Here, all the horizontal dashed lines represent pre-defined numerical threshold where our algorithm would stop running.}\label{fig:biprimes1}
\end{figure*}
It would be natural to use a fully fault-tolerant general purpose quantum computer to simulate large-scale quantum chemical molecular systems. However, due to the high overhead requirements of such a quantum simulation, a full-scale quantum chemistry simulation has not yet been seen. Henceforth, it is important to continue to push the frontier of quantum simulation with limited quantum resources.

The Hamiltonian for a chemical system in the second quantization picture has the following general form 
\begin{align}
	H
	=\sum_{ij}h_{ij}e^\dagger_i{a}_j+\sum_{ijkl}V_{ijkl}{a}^\dagger_i{a}^\dagger_k{a}_l{a}_j,
\end{align}
where ${a}^\dagger_i$ is a creation operator that creates an electron on the $i$th orbital, ${a}_i$ is an annihilation operator which removes an electron from the $i$th orbital, and $h_{ij}$ and $V_{ijkl}$ are the one-electron and two-electron interaction coefficients, respectively, which are determined for specific systems. The antisymmetric property of electrons is fulfilled by the anti-commutation relation of the creation and annihilation operators $\{{a}_i,{a}^\dagger_j\}=\delta_{ij},\,\{{a}^\dagger_i,{a}^\dagger_j\}=0$. The above anti-commutation relation precludes direct encoding of a chemical Hamiltonian on a quantum computer, since the operating units of a quantum computer ({\em i.e.}, qubits) obey the commutation relation of spins. The remedy to this discrepancy is to perform a fermion-spin mapping, such as the Jordan--Wigner transformation \cite{jordan1928pauli} presented here as an example. The transformation maps the fermionic creation and annihilation operators to qubit raising and lowering operators ${\sigma}^{\pm}={X}\pm i{Y}$ with a string of ${Z}$ operators to enforce the fermionic anti-commutation properties:
\begin{align}
	{a}_j
	\rightarrow\left(\prod_{l=1}^{j-1}-{Z}_l\right){\sigma}^{-}_j,\quad
	{a}^\dagger_j
	\rightarrow\left(\prod_{l=1}^{j-1}-{Z}_l\right){\sigma}^{+}_j
\end{align}
With the above transformation, the fermionic anti-commutation relation is preserved. For other fermion-spin mapping approaches, the reader is referred to the literature \cite{Tranter2018JCTC, Cao2019ChemRev, McArdle2020RevModPhys}. An alternative method to decompose gates for molecular systems is provided in ref.~\cite{daskin2011decomposition}.

Here, we show that one can efficiently search for the ground-state energy of hydrogen molecule across various bond stretching distances with QIPA with fewer time steps than QITE \cite{moll2018quantum,mcardle2019variational,yuan2019theory,gomes2020efficient,Motta2020determining,nishi2021implementation,selvarajan2021prime,huang2022efficient,yeter2022quantum}. Results are shown in \figref{fig:H2_transmon_qipa}(a) comparing QIPA and QITE. For consistency and fair comparison, we use the same time step for each bond distance for both QIPA and QITE runs. The error difference between the exact energy obtained from full configuration interaction (CI) calculations and the QIPA and QITE results can be seen in \figref{fig:H2_transmon_qipa}(b). QIPA features less error for all bond distances considered. 

\subsection{Quantum computer-aided designs}\label{subsec:transmon}
As the number of high-quality qubits inside a quantum processing unit (QPU) grows over time, it is expected that eventually no classical supercomputer will be able to simulate, verify, and cross-check the inner working mechanism and data obtained from the QPU. This is commonly known as ``quantum advantage.'' Once such an event occurs, from a practical point-of-view, it is beneficial to make use of existing quantum hardware that is already well-calibrated to simulate subsets of new QPU designs. Quantum computer-aided designs of superconducting qubits \cite{kyaw2021quantum} and photonic chips \cite{kottmann2021quantum} have recently been proposed and experimentally realized in a superconducting qubit architecture \cite{liu2021quantum}, but not yet with imaginary-time-like evolution. Here, we show that, with the proposed QIPA, we are able to optimize for the ground-state-energy search of a flux-tunable superconducting transmon system.

Given an arbitrary classical electrical circuit diagram composed of inductors, capacitors, and Josephson junctions, one can quantize such circuit into a quantum Hamiltonian \cite{kyaw2019towards} via the Legendre transformation. Once we obtain the quantum Hamiltonian, the task is to translate it into a language that a  quantum computer can understand, such as Pauli words or strings. Let us consider the case of a flux-tunable transmon system shown in the main text Figure 1 prior to conversion as an example. The system has the following Hamiltonian: 
\begin{align}\label{eq:transmon}
	{H}_{\text{transmon}}
	=2e^2{N}^2/\mathbf{C}-2E_{\text{J}}\left|\cos\left(2\pi f\right)\right|\cos {\varphi}
\end{align}
Here, $e$ is the electron charge. The normalized external flux $f=\Phi_{\text{ext}}/\Phi_0$ is derived from the external magnetic flux $\Phi_{\text{ext}}$ that penetrates the loop formed by the two Josephson junctions of the transmon. The Josephson energy of the two junctions is equal and given by $E_{\text{J}}$ while $\mathbf{C}$ is the capacitance. The magnetic flux quantum $\Phi_0$ is a fundamental constant that describes the smallest amount of flux that a superconducting loop can sustain. Here, $\varphi$ and $N$ are the phase and number operators, respectively and fulfill the commutation relation $[{\varphi},{N}]=i$. And, the following relations follow:
\begin{align}\label{Eq:raising_lowering}
	[{e^{i\varphi}},{N}]
	=-{e^{i\varphi}},\quad
	{e^{\pm i\varphi}}\ket{n}
	=\ket{n\pm 1},
\end{align}
where $\ket{n_j}$ are the eigenstates of ${N}$. We notice that the operators ${e^{\pm i\varphi_j}}$ are similar to the usual bosonic creation and annihilation operators, without the square root prefactor. They are, in fact, the Susskind--Glogower phase operators \cite{susskind1964quantum}. When we write down the transmon Hamiltonian in the charge number basis, we use \eqref{Eq:raising_lowering} and assign the operators as:
\begin{align}\label{Eq:n_operator}
	{N}
	=\sum_{n=0}^{d-1}\left(n-\frac{d}{2}\right)\ketbra{n}{n},
\end{align}
\begin{align}\label{Eq:cosine_operator}
	\cos{\varphi}
	=\frac{1}{2}\sum_{n=0}^{d-2}(\ketbra{n}{n+1}+\ketbra{n+1}{n}),
\end{align}
\begin{align}\label{Eq:sine_operator}
	\sin{\varphi}
	=\frac{i}{2}\sum_{n=0}^{d-2}(\ketbra{n}{n+1}-\ketbra{n+1}{n}).
\end{align}
In general, the number of Cooper pairs can take on infinitely many integer values. However, for practical purposes, we are only interested in low-lying energy states. In that case, we can truncate the Hilbert space as described above by introducing a finite maximum number of excitations $d=2^k$. The number of data qubits used in the quantum simulation of the transmon qubit is $k\in\mathbb{N}$.

Next, we convert the eigenstates of the number operator $\hat{N}$ into the computational basis states of $k$ data qubits by representing the integer charge number in a preferred encoding \cite{veis2016quantum,mcardle2018quantum,nicolas2019}. This implies a truncation of the physical space to the subspace spanned by $2^k$ Cooper pair numbers. There are combinatorially many ways to map such a state space to a set of qubits. For all the numerical quantum simulation experiments presented in this work, we have employed the Gray code \cite{nicolas2019,kottmann2021quantum,kyaw2021quantum} due to its resource-efficient representation of tridiagonal quantum matrix operators (see Table~\ref{tab:enc16}). After integer labeling, each level is encoded into a set of bits, which is then mapped to Pauli operators, as follows:
\begin{align}
	\ketbra{0}{1}
	&=(X+iY)/2,\nonumber\\
	\ketbra{1}{0}
	&=(X-iY)/2,\nonumber\\
	\ketbra{0}{0}
	&=(\mathbb{I}+Z)/2,\nonumber\\
	\ketbra{1}{1}
	&=(\mathbb{I}-Z)/2.\nonumber
\end{align}
Here, $X,Y,Z$ are the usual Pauli matrices and $\mathbb{I}$ is the identity. Encodings for the operators seen in \eqref{eq:transmon} truncated at $d=16$ are given in Table~\ref{tab:enc16}. As shown in Fig.~\ref{fig:H2_transmon_qipa}(c), both QITE and QIPA successfully minimize the energy of a given transmon circuit. Moreover, QIPA requires significantly fewer iterations than QITE for global optimization of the energy. These results constitute the successful implementation of two forms of imaginary-time-like evolution to perform quantum computer-aided design.

\subsection{Biprime factorization}\label{subsec:prime}
Prime factorization of biprimes is essential to modern Rivest--Shamir--Adleman (RSA) encryption algorithms \cite{Rivest.1978.120} and is seen as a classic test of the power of quantum computing to address problems that are computationally intensive on classical computers \cite{MartinLopez.2012.773,Jiang.2018.17667,Dash.2018.1805.10478v2,selvarajan2021prime}. Prime factorization algorithms for NISQ quantum computers are essential to demonstrate the promise of quantum computers for this task and complement current approaches such as Shor's algorithm \cite{Shor1994algorithms,MartinLopez.2012.773}, variational quantum factoring (VQF) \cite{Anschuetz.2018.1808.08927v1,karamlou2021analyzing}, exact search \cite{Grover.1996.212,liu2014exact,Dash.2018.1805.10478v2}, QITE \cite{moll2018quantum,mcardle2019variational,yuan2019theory,gomes2020efficient,Motta2020determining,nishi2021implementation,selvarajan2021prime,huang2022efficient,yeter2022quantum}, and quantum annealing \cite{Kadowaki.1998.5355,Farhi.2000.0001106,Farhi.2001.472,PerdomoOrtiz.2012.571,Babbush.2014.6603,Albash.2018.031016,Jiang.2018.17667}. Recently QITE has been used to identify prime factors via global optimization \cite{selvarajan2021prime}, and here we show such an imaginary-time-like approach can be further accelerated by using QIPA.

To solve the prime factorization problem for a given biprime (product of two prime numbers) $N=q^*\times p^*$, we consider the Hamiltonian
\begin{align}
	H_N(q,p)
	=d(N;q,p)^2,\quad
	d(N;q,p)
	=N-q\times p,
\end{align}
defined on the space of prime numbers $q,p\le\sqrt{N}$. This Hamiltonian is a non-negative function that attains its global minimum $H_N(q^*,p^*)=0$ for the unique pair of solutions $q=q^*$ and $p=p^*$. Binary representation \cite{Jiang.2018.17667,sawaya2020resource,kyaw2021quantum,kottmann2021quantum,Anschuetz.2018.1808.08927v1,selvarajan2021prime} of the prime factors yields
\begin{align}\label{H_binary1}
	d(N;q,p)
	=N-\sum_{j,k=0}^Lq_jp_k2^{j+k},
\end{align}
where $(q_L,\dots,q_0),(p_L,\dots,p_0)\in\{0,1\}^L$ are the binary representations of $q$ and $p$, respectively, and $L=\lfloor\log_2(N/2)\rfloor+1$. Assuming that $p^*,q^*>2$ (otherwise $N$ is an even number), we further restrict the search space by setting $q_0=p_0=1$ such that \eqref{H_binary1} can also be written in terms of the combined parameter $\vec x=(q_1,\dots,q_L,p_1,\dots,p_L)$, as follows:
\begin{align}
	d(N;\vec x)
	=N-\left(1+\sum_{j=1}^Lx_j2^j\right)\times\left(1+\sum_{k=1}^{L}x_{L+k}2^k\right),
\end{align}
where $x_l^2=x_l$ holds for all $1\le l\le 2L$ since $x_l\in\{0,1\}$. Equivalently, in terms of the scaled spin parameters $s_l=2x_l-1$, we write
\begin{align}
	d(N;\vec s)
	=N-\left(2^L+\sum_{j=1}^Ls_j2^{j-1}\right)\times\left(2^L+\sum_{k=1}^{L}s_{L+k}2^{k-1}\right)
\end{align}
where $s_l^2=1$ holds since $s_l\in\{-1,1\}$. For example, for $N=15$ we obtain $L=2$ and $d(15;\vec s)=15-\left(4+s_1+2s_2\right)\times\left(4+s_3+2s_4\right)=-1-4s_1-8s_2-4s_3-s_1s_3-2s_2s_3-8s_4-2s_1s_4-4s_2s_4$, which gives the Hamiltonian $H_{15}(\vec s)=d(15;\vec s)^2=186+48s_1+96s_2+84s_1s_2+48s_3+34s_1s_3+68s_2s_3+32s_1s_2s_3+96s_4+68s_1s_4+136s_2s_4+64s_1s_2s_4+84s_3s_4+32s_1s_3 s_4+64s_2s_3s_4+16s_1s_2s_3s_4$, with twin global minima. Alternatively, with previous information about the prime factors of $15$, the integer can be factorized with fewer qubits via the test Hamiltonian \cite{selvarajan2021prime} $H=196-52Z_2-52Z_0-56Z_2Z_0-96Z_1-48Z_2Z_1+16Z_0Z_1+128Z_0Z_1Z_2$, which features a unique ground state $\ket{011}$ that corresponds to the correct factorization of the number $15$ to $3$ and $5$. The resulting circuit implementation is given in Fig.~\ref{fig:Hadamard_test_circuit_example}.

As shown in Fig.~\ref{fig:biprimes1}(a) and (b), we find QIPA identifies the prime factors of biprimes with fewer iterations than QITE for all integers factored. The number of required iterations varies depending on the biprime factored and the ansatz, with speedups shown here of up to 50\%. In addition, QIPA successfully identifies both factors of biprimes. As expected, QIPA succeeds for stable integration with a small, converged time step, for which the twin solutions are readily identified as the two components of the final wavefunction with the largest and equal amplitude. Moreover, QIPA succeeds for {\em unstable} integration with a larger time step, for which the twin solutions are identified as the two largest components of the final wavefunction with {\em unequal} amplitude, as depicted in Fig.~\ref{fig:biprimes1}(c). To our knowledge this ability to identify multiple prime factors is unique among published quantum computing prime factorization results.

\section{Discussion and Conclusions}
In summary, we have presented a family of generalized imaginary-time-like near-term quantum algorithms which we coin the ``quantum iterative power algorithm,'' inspired by its classical counterpart. We have analyzed its convergence rate.

One caveat is that since the proposed algorithm relies heavily on the ansatz circuit used, its convergence rate is difficult to discern in the generic case. We have also determined QIPA's estimated resource count, and demonstrated it can outperform the quantum imaginary time evolution while it reduces the number of required iterations, at the cost of a moderate increase in the number of gates. Furthermore, we have used the three numerical case studies -- quantum chemical molecular simulation of the hydrogen molecule for various bond dissociation distances, quantum computer-aided design of a superconducting transmon, and finding optimal solutions for double prime factorization -- to highlight how QIPA outperforms QITE.

We would like to point out that an additional consideration for such an algorithm besides the choice of the ansatz circuit is setting the right parameter for the time step $\delta t$ at each evolution step. A number of proposals \cite{Gomes2021adaptive,Zhang2020low} suggest the use of an adaptive time step to overcome such an issue. However, there exists an opportunity to develop a systematic way to adjust the time step $\delta\tau$ according to gradient descent, rather than with a heuristic argument on how to adjust $\delta\tau$. 
A major drawback with quantum imaginary-time-like evolution algorithm is that it involves constructing the matrix elements of $A$ with NISQ hardware. Since we are working with noisy quantum hardware, any large fluctuation in the matrix elements would result in suboptimal $\vec{\theta}$ angles. Future work will explore preconditioning the matrix based on errors associated with performing the Hadamard tests.

The generality of the global-optimization method presented here invites further application to other problems that currently have not been explored with QITE-based quantum computing algorithms. Quantum approaches could facilitate identification of reaction pathways and transition states in chemical physics, as well as optimization in a broad range of machine learning applications. The method also provides a general framework for adaptation of a class of classical optimization algorithms to quantum computers to further broaden the range of algorithms amenable to implementation on current NISQ quantum computers.

The main code used to generate data presented here can be found at the following repository \href{https://github.com/aspuru-guzik-group/QIPA/}{https://github.com/aspuru-guzik-group/QIPA/}.

\section*{Acknowledgments}
THK, CS, and AAG acknowledge funding from Dr. Anders G. Fr{\o}seth. AAG also acknowledges support from the Canada 150 Research Chairs Program, the Canada Industrial Research Chair Program, and from Google, Inc. in the form of a Google Focused Award.  MBS acknowledges financial support from the Yale Quantum Institute Postdoctoral Fellowship.
VSB acknowledges support from the Center for Quantum Dynamics on Modular Quantum Devices (NSF Grant No.~CHE-2124511) and computational resources from NERSC and the Yale Center for Research Computing High Performance Computing. We acknowledge fruitful discussions with Jakob Kottmann, Abhinav Anand, Sabre Kais, Eitan Geva, Raja Selvarajan, Suguru Endo, Xiao Yuan, Ellen Mulvihill, and Zi-jian Zhang. Numerical simulations of quantum circuit evaluations for the $\text{H}_2$ and transmon quantum systems were performed on the Niagara supercomputer at the SciNet HPC Consortium \cite{loken2010scinet,ponce2019deploying}. SciNet is funded by the Canada Foundation for Innovation, the Government of Ontario, Ontario Research Fund-Research Excellence, and the University of Toronto. 

\bibliographystyle{apsrev4-1}
\bibliography{references}

\begin{thebibliography}{105}%
\makeatletter
\providecommand \@ifxundefined [1]{%
 \@ifx{#1\undefined}
}%
\providecommand \@ifnum [1]{%
 \ifnum #1\expandafter \@firstoftwo
 \else \expandafter \@secondoftwo
 \fi
}%
\providecommand \@ifx [1]{%
 \ifx #1\expandafter \@firstoftwo
 \else \expandafter \@secondoftwo
 \fi
}%
\providecommand \natexlab [1]{#1}%
\providecommand \enquote  [1]{``#1''}%
\providecommand \bibnamefont  [1]{#1}%
\providecommand \bibfnamefont [1]{#1}%
\providecommand \citenamefont [1]{#1}%
\providecommand \href@noop [0]{\@secondoftwo}%
\providecommand \href [0]{\begingroup \@sanitize@url \@href}%
\providecommand \@href[1]{\@@startlink{#1}\@@href}%
\providecommand \@@href[1]{\endgroup#1\@@endlink}%
\providecommand \@sanitize@url [0]{\catcode `\\12\catcode `\$12\catcode
  `\&12\catcode `\#12\catcode `\^12\catcode `\_12\catcode `\%12\relax}%
\providecommand \@@startlink[1]{}%
\providecommand \@@endlink[0]{}%
\providecommand \url  [0]{\begingroup\@sanitize@url \@url }%
\providecommand \@url [1]{\endgroup\@href {#1}{\urlprefix }}%
\providecommand \urlprefix  [0]{URL }%
\providecommand \Eprint [0]{\href }%
\providecommand \doibase [0]{http://dx.doi.org/}%
\providecommand \selectlanguage [0]{\@gobble}%
\providecommand \bibinfo  [0]{\@secondoftwo}%
\providecommand \bibfield  [0]{\@secondoftwo}%
\providecommand \translation [1]{[#1]}%
\providecommand \BibitemOpen [0]{}%
\providecommand \bibitemStop [0]{}%
\providecommand \bibitemNoStop [0]{.\EOS\space}%
\providecommand \EOS [0]{\spacefactor3000\relax}%
\providecommand \BibitemShut  [1]{\csname bibitem#1\endcsname}%
\let\auto@bib@innerbib\@empty
\bibitem [{\citenamefont {Shor}(1994)}]{Shor1994algorithms}%
  \BibitemOpen
  \bibfield  {author} {\bibinfo {author} {\bibfnamefont {P.~W.}\ \bibnamefont
  {Shor}},\ }in\ \href {\doibase 10.1109/SFCS.1994.365700} {\emph {\bibinfo
  {booktitle} {{Proceedings 35th Annual Symposium on Foundations of Computer
  Science}}}}\ (\bibinfo  {publisher} {IEEE},\ \bibinfo {year} {1994})\ pp.\
  \bibinfo {pages} {124--134}\BibitemShut {NoStop}%
\bibitem [{\citenamefont {Aspuru-Guzik}\ \emph {et~al.}(2005)\citenamefont
  {Aspuru-Guzik}, \citenamefont {Dutoi}, \citenamefont {Love},\ and\
  \citenamefont {Head-Gordon}}]{Aspuru-Guzik2005simulated}%
  \BibitemOpen
  \bibfield  {author} {\bibinfo {author} {\bibfnamefont {A.}~\bibnamefont
  {Aspuru-Guzik}}, \bibinfo {author} {\bibfnamefont {A.~D.}\ \bibnamefont
  {Dutoi}}, \bibinfo {author} {\bibfnamefont {P.~J.}\ \bibnamefont {Love}}, \
  and\ \bibinfo {author} {\bibfnamefont {M.}~\bibnamefont {Head-Gordon}},\
  }\href {\doibase 10.1126/science.1113479} {\bibfield  {journal} {\bibinfo
  {journal} {Science}\ }\textbf {\bibinfo {volume} {309}},\ \bibinfo {pages}
  {1704} (\bibinfo {year} {2005})}\BibitemShut {NoStop}%
\bibitem [{\citenamefont {Georgescu}\ \emph {et~al.}(2014)\citenamefont
  {Georgescu}, \citenamefont {Ashhab},\ and\ \citenamefont
  {Nori}}]{Georgescu2014quantum}%
  \BibitemOpen
  \bibfield  {author} {\bibinfo {author} {\bibfnamefont {I.~M.}\ \bibnamefont
  {Georgescu}}, \bibinfo {author} {\bibfnamefont {S.}~\bibnamefont {Ashhab}}, \
  and\ \bibinfo {author} {\bibfnamefont {F.}~\bibnamefont {Nori}},\ }\href
  {\doibase 10.1103/RevModPhys.86.153} {\bibfield  {journal} {\bibinfo
  {journal} {Rev. Mod. Phys.}\ }\textbf {\bibinfo {volume} {86}},\ \bibinfo
  {pages} {153} (\bibinfo {year} {2014})}\BibitemShut {NoStop}%
\bibitem [{\citenamefont {Cao}\ \emph {et~al.}(2019{\natexlab{a}})\citenamefont
  {Cao}, \citenamefont {Romero}, \citenamefont {Olson}, \citenamefont
  {Degroote}, \citenamefont {Johnson}, \citenamefont
  {Kieferov{\ifmmode\acute{a}\else\'{a}\fi}}, \citenamefont {Kivlichan},
  \citenamefont {Menke}, \citenamefont {Peropadre}, \citenamefont {Sawaya}
  \emph {et~al.}}]{Cao2019quantum}%
  \BibitemOpen
  \bibfield  {author} {\bibinfo {author} {\bibfnamefont {Y.}~\bibnamefont
  {Cao}}, \bibinfo {author} {\bibfnamefont {J.}~\bibnamefont {Romero}},
  \bibinfo {author} {\bibfnamefont {J.~P.}\ \bibnamefont {Olson}}, \bibinfo
  {author} {\bibfnamefont {M.}~\bibnamefont {Degroote}}, \bibinfo {author}
  {\bibfnamefont {P.~D.}\ \bibnamefont {Johnson}}, \bibinfo {author}
  {\bibfnamefont {M.}~\bibnamefont {Kieferov{\ifmmode\acute{a}\else\'{a}\fi}}},
  \bibinfo {author} {\bibfnamefont {I.~D.}\ \bibnamefont {Kivlichan}}, \bibinfo
  {author} {\bibfnamefont {T.}~\bibnamefont {Menke}}, \bibinfo {author}
  {\bibfnamefont {B.}~\bibnamefont {Peropadre}}, \bibinfo {author}
  {\bibfnamefont {N.~P.~D.}\ \bibnamefont {Sawaya}},  \emph {et~al.},\ }\href
  {\doibase 10.1021/acs.chemrev.8b00803} {\bibfield  {journal} {\bibinfo
  {journal} {Chem. Rev.}\ }\textbf {\bibinfo {volume} {119}},\ \bibinfo {pages}
  {10856} (\bibinfo {year} {2019}{\natexlab{a}})}\BibitemShut {NoStop}%
\bibitem [{\citenamefont {McArdle}\ \emph
  {et~al.}(2020{\natexlab{a}})\citenamefont {McArdle}, \citenamefont {Endo},
  \citenamefont {Aspuru-Guzik}, \citenamefont {Benjamin},\ and\ \citenamefont
  {Yuan}}]{McArdle2020quantum}%
  \BibitemOpen
  \bibfield  {author} {\bibinfo {author} {\bibfnamefont {S.}~\bibnamefont
  {McArdle}}, \bibinfo {author} {\bibfnamefont {S.}~\bibnamefont {Endo}},
  \bibinfo {author} {\bibfnamefont {A.}~\bibnamefont {Aspuru-Guzik}}, \bibinfo
  {author} {\bibfnamefont {S.~C.}\ \bibnamefont {Benjamin}}, \ and\ \bibinfo
  {author} {\bibfnamefont {X.}~\bibnamefont {Yuan}},\ }\href {\doibase
  10.1103/RevModPhys.92.015003} {\bibfield  {journal} {\bibinfo  {journal}
  {Rev. Mod. Phys.}\ }\textbf {\bibinfo {volume} {92}},\ \bibinfo {pages}
  {015003} (\bibinfo {year} {2020}{\natexlab{a}})}\BibitemShut {NoStop}%
\bibitem [{\citenamefont {Preskill}(2018)}]{Preskill2018quantum}%
  \BibitemOpen
  \bibfield  {author} {\bibinfo {author} {\bibfnamefont {J.}~\bibnamefont
  {Preskill}},\ }\href {\doibase 10.22331/q-2018-08-06-79} {\bibfield
  {journal} {\bibinfo  {journal} {Quantum}\ }\textbf {\bibinfo {volume} {2}},\
  \bibinfo {pages} {79} (\bibinfo {year} {2018})},\ \Eprint
  {http://arxiv.org/abs/1801.00862v3} {1801.00862v3} \BibitemShut {NoStop}%
\bibitem [{\citenamefont {Arute}\ \emph {et~al.}(2019)\citenamefont {Arute},
  \citenamefont {Arya}, \citenamefont {Babbush}, \citenamefont {Bacon},
  \citenamefont {Bardin}, \citenamefont {Barends}, \citenamefont {Biswas},
  \citenamefont {Boixo}, \citenamefont {Brandao}, \citenamefont {Buell} \emph
  {et~al.}}]{Arute2019quantum}%
  \BibitemOpen
  \bibfield  {author} {\bibinfo {author} {\bibfnamefont {F.}~\bibnamefont
  {Arute}}, \bibinfo {author} {\bibfnamefont {K.}~\bibnamefont {Arya}},
  \bibinfo {author} {\bibfnamefont {R.}~\bibnamefont {Babbush}}, \bibinfo
  {author} {\bibfnamefont {D.}~\bibnamefont {Bacon}}, \bibinfo {author}
  {\bibfnamefont {J.~C.}\ \bibnamefont {Bardin}}, \bibinfo {author}
  {\bibfnamefont {R.}~\bibnamefont {Barends}}, \bibinfo {author} {\bibfnamefont
  {R.}~\bibnamefont {Biswas}}, \bibinfo {author} {\bibfnamefont
  {S.}~\bibnamefont {Boixo}}, \bibinfo {author} {\bibfnamefont {F.~G. S.~L.}\
  \bibnamefont {Brandao}}, \bibinfo {author} {\bibfnamefont {D.~A.}\
  \bibnamefont {Buell}},  \emph {et~al.},\ }\href {\doibase
  10.1038/s41586-019-1666-5} {\bibfield  {journal} {\bibinfo  {journal}
  {Nature}\ }\textbf {\bibinfo {volume} {574}},\ \bibinfo {pages} {505}
  (\bibinfo {year} {2019})}\BibitemShut {NoStop}%
\bibitem [{\citenamefont {Zhong}\ \emph {et~al.}(2020)\citenamefont {Zhong},
  \citenamefont {Wang}, \citenamefont {Deng}, \citenamefont {Chen},
  \citenamefont {Peng}, \citenamefont {Luo}, \citenamefont {Qin}, \citenamefont
  {Wu}, \citenamefont {Ding}, \citenamefont {Hu} \emph
  {et~al.}}]{Zhong2020quantum}%
  \BibitemOpen
  \bibfield  {author} {\bibinfo {author} {\bibfnamefont {H.-S.}\ \bibnamefont
  {Zhong}}, \bibinfo {author} {\bibfnamefont {H.}~\bibnamefont {Wang}},
  \bibinfo {author} {\bibfnamefont {Y.-H.}\ \bibnamefont {Deng}}, \bibinfo
  {author} {\bibfnamefont {M.-C.}\ \bibnamefont {Chen}}, \bibinfo {author}
  {\bibfnamefont {L.-C.}\ \bibnamefont {Peng}}, \bibinfo {author}
  {\bibfnamefont {Y.-H.}\ \bibnamefont {Luo}}, \bibinfo {author} {\bibfnamefont
  {J.}~\bibnamefont {Qin}}, \bibinfo {author} {\bibfnamefont {D.}~\bibnamefont
  {Wu}}, \bibinfo {author} {\bibfnamefont {X.}~\bibnamefont {Ding}}, \bibinfo
  {author} {\bibfnamefont {Y.}~\bibnamefont {Hu}},  \emph {et~al.},\ }\href
  {\doibase 10.1126/science.abe8770} {\bibfield  {journal} {\bibinfo  {journal}
  {Science}\ }\textbf {\bibinfo {volume} {370}},\ \bibinfo {pages} {1460}
  (\bibinfo {year} {2020})}\BibitemShut {NoStop}%
\bibitem [{\citenamefont {Wu}\ \emph {et~al.}(2021)\citenamefont {Wu},
  \citenamefont {Bao}, \citenamefont {Cao}, \citenamefont {Chen}, \citenamefont
  {Chen}, \citenamefont {Chen}, \citenamefont {Chung}, \citenamefont {Deng},
  \citenamefont {Du}, \citenamefont {Fan} \emph {et~al.}}]{Wu2021strong}%
  \BibitemOpen
  \bibfield  {author} {\bibinfo {author} {\bibfnamefont {Y.}~\bibnamefont
  {Wu}}, \bibinfo {author} {\bibfnamefont {W.-S.}\ \bibnamefont {Bao}},
  \bibinfo {author} {\bibfnamefont {S.}~\bibnamefont {Cao}}, \bibinfo {author}
  {\bibfnamefont {F.}~\bibnamefont {Chen}}, \bibinfo {author} {\bibfnamefont
  {M.-C.}\ \bibnamefont {Chen}}, \bibinfo {author} {\bibfnamefont
  {X.}~\bibnamefont {Chen}}, \bibinfo {author} {\bibfnamefont {T.-H.}\
  \bibnamefont {Chung}}, \bibinfo {author} {\bibfnamefont {H.}~\bibnamefont
  {Deng}}, \bibinfo {author} {\bibfnamefont {Y.}~\bibnamefont {Du}}, \bibinfo
  {author} {\bibfnamefont {D.}~\bibnamefont {Fan}},  \emph {et~al.},\ }\href
  {\doibase 10.1103/PhysRevLett.127.180501} {\bibfield  {journal} {\bibinfo
  {journal} {Phys. Rev. Lett.}\ }\textbf {\bibinfo {volume} {127}},\ \bibinfo
  {pages} {180501} (\bibinfo {year} {2021})}\BibitemShut {NoStop}%
\bibitem [{\citenamefont {Madsen}\ \emph {et~al.}(2022)\citenamefont {Madsen},
  \citenamefont {Laudenbach}, \citenamefont {Askarani}, \citenamefont
  {Rortais}, \citenamefont {Vincent}, \citenamefont {Bulmer}, \citenamefont
  {Miatto}, \citenamefont {Neuhaus}, \citenamefont {Helt}, \citenamefont
  {Collins} \emph {et~al.}}]{Madsen2022quantum}%
  \BibitemOpen
  \bibfield  {author} {\bibinfo {author} {\bibfnamefont {L.~S.}\ \bibnamefont
  {Madsen}}, \bibinfo {author} {\bibfnamefont {F.}~\bibnamefont {Laudenbach}},
  \bibinfo {author} {\bibfnamefont {M.~{\relax Falamarzi}.}\ \bibnamefont
  {Askarani}}, \bibinfo {author} {\bibfnamefont {F.}~\bibnamefont {Rortais}},
  \bibinfo {author} {\bibfnamefont {T.}~\bibnamefont {Vincent}}, \bibinfo
  {author} {\bibfnamefont {J.~F.~F.}\ \bibnamefont {Bulmer}}, \bibinfo {author}
  {\bibfnamefont {F.~M.}\ \bibnamefont {Miatto}}, \bibinfo {author}
  {\bibfnamefont {L.}~\bibnamefont {Neuhaus}}, \bibinfo {author} {\bibfnamefont
  {L.~G.}\ \bibnamefont {Helt}}, \bibinfo {author} {\bibfnamefont {M.~J.}\
  \bibnamefont {Collins}},  \emph {et~al.},\ }\href {\doibase
  10.1038/s41586-022-04725-x} {\bibfield  {journal} {\bibinfo  {journal}
  {Nature}\ }\textbf {\bibinfo {volume} {606}},\ \bibinfo {pages} {75}
  (\bibinfo {year} {2022})}\BibitemShut {NoStop}%
\bibitem [{\citenamefont {Bharti}\ \emph {et~al.}(2022)\citenamefont {Bharti},
  \citenamefont {Cervera-Lierta}, \citenamefont {Kyaw}, \citenamefont {Haug},
  \citenamefont {Alperin-Lea}, \citenamefont {Anand}, \citenamefont {Degroote},
  \citenamefont {Heimonen}, \citenamefont {Kottmann}, \citenamefont {Menke}
  \emph {et~al.}}]{Bharti2022noisy}%
  \BibitemOpen
  \bibfield  {author} {\bibinfo {author} {\bibfnamefont {K.}~\bibnamefont
  {Bharti}}, \bibinfo {author} {\bibfnamefont {A.}~\bibnamefont
  {Cervera-Lierta}}, \bibinfo {author} {\bibfnamefont {T.~H.}\ \bibnamefont
  {Kyaw}}, \bibinfo {author} {\bibfnamefont {T.}~\bibnamefont {Haug}}, \bibinfo
  {author} {\bibfnamefont {S.}~\bibnamefont {Alperin-Lea}}, \bibinfo {author}
  {\bibfnamefont {A.}~\bibnamefont {Anand}}, \bibinfo {author} {\bibfnamefont
  {M.}~\bibnamefont {Degroote}}, \bibinfo {author} {\bibfnamefont
  {H.}~\bibnamefont {Heimonen}}, \bibinfo {author} {\bibfnamefont {J.~S.}\
  \bibnamefont {Kottmann}}, \bibinfo {author} {\bibfnamefont {T.}~\bibnamefont
  {Menke}},  \emph {et~al.},\ }\href {\doibase 10.1103/RevModPhys.94.015004}
  {\bibfield  {journal} {\bibinfo  {journal} {Rev. Mod. Phys.}\ }\textbf
  {\bibinfo {volume} {94}},\ \bibinfo {pages} {015004} (\bibinfo {year}
  {2022})}\BibitemShut {NoStop}%
\bibitem [{\citenamefont {Cerezo}\ \emph {et~al.}(2021)\citenamefont {Cerezo},
  \citenamefont {Arrasmith}, \citenamefont {Babbush}, \citenamefont {Benjamin},
  \citenamefont {Endo}, \citenamefont {Fujii}, \citenamefont {McClean},
  \citenamefont {Mitarai}, \citenamefont {Yuan}, \citenamefont {Cincio} \emph
  {et~al.}}]{cerezo2021variational}%
  \BibitemOpen
  \bibfield  {author} {\bibinfo {author} {\bibfnamefont {M.}~\bibnamefont
  {Cerezo}}, \bibinfo {author} {\bibfnamefont {A.}~\bibnamefont {Arrasmith}},
  \bibinfo {author} {\bibfnamefont {R.}~\bibnamefont {Babbush}}, \bibinfo
  {author} {\bibfnamefont {S.~C.}\ \bibnamefont {Benjamin}}, \bibinfo {author}
  {\bibfnamefont {S.}~\bibnamefont {Endo}}, \bibinfo {author} {\bibfnamefont
  {K.}~\bibnamefont {Fujii}}, \bibinfo {author} {\bibfnamefont {J.~R.}\
  \bibnamefont {McClean}}, \bibinfo {author} {\bibfnamefont {K.}~\bibnamefont
  {Mitarai}}, \bibinfo {author} {\bibfnamefont {X.}~\bibnamefont {Yuan}},
  \bibinfo {author} {\bibfnamefont {L.}~\bibnamefont {Cincio}},  \emph
  {et~al.},\ }\href {\doibase 10.1038/s42254-021-00348-9} {\bibfield  {journal}
  {\bibinfo  {journal} {Nat. Rev. Phys.}\ }\textbf {\bibinfo {volume} {3}},\
  \bibinfo {pages} {625} (\bibinfo {year} {2021})}\BibitemShut {NoStop}%
\bibitem [{\citenamefont {Tilly}\ \emph {et~al.}(2021)\citenamefont {Tilly},
  \citenamefont {Chen}, \citenamefont {Cao}, \citenamefont {Picozzi},
  \citenamefont {Setia}, \citenamefont {Li}, \citenamefont {Grant},
  \citenamefont {Wossnig}, \citenamefont {Rungger}, \citenamefont {Booth} \emph
  {et~al.}}]{Tilly2021variational}%
  \BibitemOpen
  \bibfield  {author} {\bibinfo {author} {\bibfnamefont {J.}~\bibnamefont
  {Tilly}}, \bibinfo {author} {\bibfnamefont {H.}~\bibnamefont {Chen}},
  \bibinfo {author} {\bibfnamefont {S.}~\bibnamefont {Cao}}, \bibinfo {author}
  {\bibfnamefont {D.}~\bibnamefont {Picozzi}}, \bibinfo {author} {\bibfnamefont
  {K.}~\bibnamefont {Setia}}, \bibinfo {author} {\bibfnamefont
  {Y.}~\bibnamefont {Li}}, \bibinfo {author} {\bibfnamefont {E.}~\bibnamefont
  {Grant}}, \bibinfo {author} {\bibfnamefont {L.}~\bibnamefont {Wossnig}},
  \bibinfo {author} {\bibfnamefont {I.}~\bibnamefont {Rungger}}, \bibinfo
  {author} {\bibfnamefont {G.~H.}\ \bibnamefont {Booth}},  \emph {et~al.},\
  }\href {https://10.48550/arXiv.2111.05176} {\bibfield  {journal} {\bibinfo
  {journal} {arXiv:2111.05176}\ } (\bibinfo {year} {2021})}\BibitemShut
  {NoStop}%
\bibitem [{\citenamefont {Fedorov}\ \emph {et~al.}(2022)\citenamefont
  {Fedorov}, \citenamefont {Peng}, \citenamefont {Govind},\ and\ \citenamefont
  {Alexeev}}]{fedorov2022vqe}%
  \BibitemOpen
  \bibfield  {author} {\bibinfo {author} {\bibfnamefont {D.~A.}\ \bibnamefont
  {Fedorov}}, \bibinfo {author} {\bibfnamefont {B.}~\bibnamefont {Peng}},
  \bibinfo {author} {\bibfnamefont {N.}~\bibnamefont {Govind}}, \ and\ \bibinfo
  {author} {\bibfnamefont {Y.}~\bibnamefont {Alexeev}},\ }\href {\doibase
  10.1186/s41313-021-00032-6} {\bibfield  {journal} {\bibinfo  {journal}
  {Mater. Theory}\ }\textbf {\bibinfo {volume} {6}},\ \bibinfo {pages} {1}
  (\bibinfo {year} {2022})}\BibitemShut {NoStop}%
\bibitem [{\citenamefont {Peruzzo}\ \emph {et~al.}(2014)\citenamefont
  {Peruzzo}, \citenamefont {McClean}, \citenamefont {Shadbolt}, \citenamefont
  {Yung}, \citenamefont {Zhou}, \citenamefont {Love}, \citenamefont
  {Aspuru-Guzik},\ and\ \citenamefont {O{'}Brien}}]{Peruzzo2014variational}%
  \BibitemOpen
  \bibfield  {author} {\bibinfo {author} {\bibfnamefont {A.}~\bibnamefont
  {Peruzzo}}, \bibinfo {author} {\bibfnamefont {J.}~\bibnamefont {McClean}},
  \bibinfo {author} {\bibfnamefont {P.}~\bibnamefont {Shadbolt}}, \bibinfo
  {author} {\bibfnamefont {M.-H.}\ \bibnamefont {Yung}}, \bibinfo {author}
  {\bibfnamefont {X.-Q.}\ \bibnamefont {Zhou}}, \bibinfo {author}
  {\bibfnamefont {P.~J.}\ \bibnamefont {Love}}, \bibinfo {author}
  {\bibfnamefont {A.}~\bibnamefont {Aspuru-Guzik}}, \ and\ \bibinfo {author}
  {\bibfnamefont {J.~L.}\ \bibnamefont {O{'}Brien}},\ }\href {\doibase
  10.1038/ncomms5213} {\bibfield  {journal} {\bibinfo  {journal} {Nat.
  Commun.}\ }\textbf {\bibinfo {volume} {5}},\ \bibinfo {pages} {1} (\bibinfo
  {year} {2014})}\BibitemShut {NoStop}%
\bibitem [{\citenamefont {McClean}\ \emph {et~al.}(2016)\citenamefont
  {McClean}, \citenamefont {Romero}, \citenamefont {Babbush},\ and\
  \citenamefont {Aspuru-Guzik}}]{McClean2016theory}%
  \BibitemOpen
  \bibfield  {author} {\bibinfo {author} {\bibfnamefont {J.~R.}\ \bibnamefont
  {McClean}}, \bibinfo {author} {\bibfnamefont {J.}~\bibnamefont {Romero}},
  \bibinfo {author} {\bibfnamefont {R.}~\bibnamefont {Babbush}}, \ and\
  \bibinfo {author} {\bibfnamefont {A.}~\bibnamefont {Aspuru-Guzik}},\ }\href
  {\doibase 10.1088/1367-2630/18/2/023023} {\bibfield  {journal} {\bibinfo
  {journal} {New J. Phys.}\ }\textbf {\bibinfo {volume} {18}},\ \bibinfo
  {pages} {023023} (\bibinfo {year} {2016})}\BibitemShut {NoStop}%
\bibitem [{\citenamefont {Li}\ and\ \citenamefont
  {Benjamin}(2017)}]{li2017efficient}%
  \BibitemOpen
  \bibfield  {author} {\bibinfo {author} {\bibfnamefont {Y.}~\bibnamefont
  {Li}}\ and\ \bibinfo {author} {\bibfnamefont {S.~C.}\ \bibnamefont
  {Benjamin}},\ }\href@noop {} {\bibfield  {journal} {\bibinfo  {journal}
  {Physical Review X}\ }\textbf {\bibinfo {volume} {7}},\ \bibinfo {pages}
  {021050} (\bibinfo {year} {2017})}\BibitemShut {NoStop}%
\bibitem [{\citenamefont {McArdle}\ \emph
  {et~al.}(2019{\natexlab{a}})\citenamefont {McArdle}, \citenamefont {Jones},
  \citenamefont {Endo}, \citenamefont {Li}, \citenamefont {Benjamin},\ and\
  \citenamefont {Yuan}}]{mcardle2019variational}%
  \BibitemOpen
  \bibfield  {author} {\bibinfo {author} {\bibfnamefont {S.}~\bibnamefont
  {McArdle}}, \bibinfo {author} {\bibfnamefont {T.}~\bibnamefont {Jones}},
  \bibinfo {author} {\bibfnamefont {S.}~\bibnamefont {Endo}}, \bibinfo {author}
  {\bibfnamefont {Y.}~\bibnamefont {Li}}, \bibinfo {author} {\bibfnamefont
  {S.~C.}\ \bibnamefont {Benjamin}}, \ and\ \bibinfo {author} {\bibfnamefont
  {X.}~\bibnamefont {Yuan}},\ }\href {\doibase 10.1038/s41534-019-0187-2}
  {\bibfield  {journal} {\bibinfo  {journal} {npj Quantum Inf.}\ }\textbf
  {\bibinfo {volume} {5}},\ \bibinfo {pages} {1} (\bibinfo {year}
  {2019}{\natexlab{a}})}\BibitemShut {NoStop}%
\bibitem [{\citenamefont {Yuan}\ \emph {et~al.}(2019)\citenamefont {Yuan},
  \citenamefont {Endo}, \citenamefont {Zhao}, \citenamefont {Li},\ and\
  \citenamefont {Benjamin}}]{yuan2019theory}%
  \BibitemOpen
  \bibfield  {author} {\bibinfo {author} {\bibfnamefont {X.}~\bibnamefont
  {Yuan}}, \bibinfo {author} {\bibfnamefont {S.}~\bibnamefont {Endo}}, \bibinfo
  {author} {\bibfnamefont {Q.}~\bibnamefont {Zhao}}, \bibinfo {author}
  {\bibfnamefont {Y.}~\bibnamefont {Li}}, \ and\ \bibinfo {author}
  {\bibfnamefont {S.~C.}\ \bibnamefont {Benjamin}},\ }\href {\doibase
  https://doi.org/10.48550/arXiv.1812.08767} {\bibfield  {journal} {\bibinfo
  {journal} {Quantum}\ }\textbf {\bibinfo {volume} {3}},\ \bibinfo {pages}
  {191} (\bibinfo {year} {2019})}\BibitemShut {NoStop}%
\bibitem [{\citenamefont {Moll}\ \emph {et~al.}(2018)\citenamefont {Moll},
  \citenamefont {Barkoutsos}, \citenamefont {Bishop}, \citenamefont {Chow},
  \citenamefont {Cross}, \citenamefont {Egger}, \citenamefont {Filipp},
  \citenamefont {Fuhrer}, \citenamefont {Gambetta}, \citenamefont {Ganzhorn}
  \emph {et~al.}}]{moll2018quantum}%
  \BibitemOpen
  \bibfield  {author} {\bibinfo {author} {\bibfnamefont {N.}~\bibnamefont
  {Moll}}, \bibinfo {author} {\bibfnamefont {P.}~\bibnamefont {Barkoutsos}},
  \bibinfo {author} {\bibfnamefont {L.~S.}\ \bibnamefont {Bishop}}, \bibinfo
  {author} {\bibfnamefont {J.~M.}\ \bibnamefont {Chow}}, \bibinfo {author}
  {\bibfnamefont {A.}~\bibnamefont {Cross}}, \bibinfo {author} {\bibfnamefont
  {D.~J.}\ \bibnamefont {Egger}}, \bibinfo {author} {\bibfnamefont
  {S.}~\bibnamefont {Filipp}}, \bibinfo {author} {\bibfnamefont
  {A.}~\bibnamefont {Fuhrer}}, \bibinfo {author} {\bibfnamefont {J.~M.}\
  \bibnamefont {Gambetta}}, \bibinfo {author} {\bibfnamefont {M.}~\bibnamefont
  {Ganzhorn}},  \emph {et~al.},\ }\href {\doibase 10.1088/2058-9565/aab822}
  {\bibfield  {journal} {\bibinfo  {journal} {Quantum Sci. Technol.}\ }\textbf
  {\bibinfo {volume} {3}},\ \bibinfo {pages} {030503} (\bibinfo {year}
  {2018})}\BibitemShut {NoStop}%
\bibitem [{\citenamefont {Gomes}\ \emph {et~al.}(2020)\citenamefont {Gomes},
  \citenamefont {Zhang}, \citenamefont {Berthusen}, \citenamefont {Wang},
  \citenamefont {Ho}, \citenamefont {Orth},\ and\ \citenamefont
  {Yao}}]{gomes2020efficient}%
  \BibitemOpen
  \bibfield  {author} {\bibinfo {author} {\bibfnamefont {N.}~\bibnamefont
  {Gomes}}, \bibinfo {author} {\bibfnamefont {F.}~\bibnamefont {Zhang}},
  \bibinfo {author} {\bibfnamefont {N.~F.}\ \bibnamefont {Berthusen}}, \bibinfo
  {author} {\bibfnamefont {C.-Z.}\ \bibnamefont {Wang}}, \bibinfo {author}
  {\bibfnamefont {K.-M.}\ \bibnamefont {Ho}}, \bibinfo {author} {\bibfnamefont
  {P.~P.}\ \bibnamefont {Orth}}, \ and\ \bibinfo {author} {\bibfnamefont
  {Y.}~\bibnamefont {Yao}},\ }\href {\doibase 10.1021/acs.jctc.0c00666}
  {\bibfield  {journal} {\bibinfo  {journal} {J. Chem. Theory Comput.}\
  }\textbf {\bibinfo {volume} {16}},\ \bibinfo {pages} {6256} (\bibinfo {year}
  {2020})}\BibitemShut {NoStop}%
\bibitem [{\citenamefont {Motta}\ \emph {et~al.}(2020)\citenamefont {Motta},
  \citenamefont {Sun}, \citenamefont {Tan}, \citenamefont {O{'}Rourke},
  \citenamefont {Ye}, \citenamefont {Minnich}, \citenamefont
  {Brand{\ifmmode\tilde{a}\else\~{a}\fi}o},\ and\ \citenamefont
  {Chan}}]{Motta2020determining}%
  \BibitemOpen
  \bibfield  {author} {\bibinfo {author} {\bibfnamefont {M.}~\bibnamefont
  {Motta}}, \bibinfo {author} {\bibfnamefont {C.}~\bibnamefont {Sun}}, \bibinfo
  {author} {\bibfnamefont {A.~T.~K.}\ \bibnamefont {Tan}}, \bibinfo {author}
  {\bibfnamefont {M.~J.}\ \bibnamefont {O{'}Rourke}}, \bibinfo {author}
  {\bibfnamefont {E.}~\bibnamefont {Ye}}, \bibinfo {author} {\bibfnamefont
  {A.~J.}\ \bibnamefont {Minnich}}, \bibinfo {author} {\bibfnamefont {F.~G.
  S.~L.}\ \bibnamefont {Brand{\ifmmode\tilde{a}\else\~{a}\fi}o}}, \ and\
  \bibinfo {author} {\bibfnamefont {G.~K.-L.}\ \bibnamefont {Chan}},\ }\href
  {\doibase 10.1038/s41567-019-0704-4} {\bibfield  {journal} {\bibinfo
  {journal} {Nat. Phys.}\ }\textbf {\bibinfo {volume} {16}},\ \bibinfo {pages}
  {205} (\bibinfo {year} {2020})}\BibitemShut {NoStop}%
\bibitem [{\citenamefont {Nishi}\ \emph {et~al.}(2021)\citenamefont {Nishi},
  \citenamefont {Kosugi},\ and\ \citenamefont
  {Matsushita}}]{nishi2021implementation}%
  \BibitemOpen
  \bibfield  {author} {\bibinfo {author} {\bibfnamefont {H.}~\bibnamefont
  {Nishi}}, \bibinfo {author} {\bibfnamefont {T.}~\bibnamefont {Kosugi}}, \
  and\ \bibinfo {author} {\bibfnamefont {Y.-i.}\ \bibnamefont {Matsushita}},\
  }\href {\doibase 10.1038/s41534-021-00409-y} {\bibfield  {journal} {\bibinfo
  {journal} {npj Quantum Inf.}\ }\textbf {\bibinfo {volume} {7}},\ \bibinfo
  {pages} {1} (\bibinfo {year} {2021})}\BibitemShut {NoStop}%
\bibitem [{\citenamefont {Selvarajan}\ \emph {et~al.}(2021)\citenamefont
  {Selvarajan}, \citenamefont {Dixit}, \citenamefont {Cui}, \citenamefont
  {Humble},\ and\ \citenamefont {Kais}}]{selvarajan2021prime}%
  \BibitemOpen
  \bibfield  {author} {\bibinfo {author} {\bibfnamefont {R.}~\bibnamefont
  {Selvarajan}}, \bibinfo {author} {\bibfnamefont {V.}~\bibnamefont {Dixit}},
  \bibinfo {author} {\bibfnamefont {X.}~\bibnamefont {Cui}}, \bibinfo {author}
  {\bibfnamefont {T.~S.}\ \bibnamefont {Humble}}, \ and\ \bibinfo {author}
  {\bibfnamefont {S.}~\bibnamefont {Kais}},\ }\href {\doibase
  10.1038/s41598-021-00339-x} {\bibfield  {journal} {\bibinfo  {journal} {Sci.
  Rep.}\ }\textbf {\bibinfo {volume} {11}},\ \bibinfo {pages} {1} (\bibinfo
  {year} {2021})}\BibitemShut {NoStop}%
\bibitem [{\citenamefont {Huang}\ \emph {et~al.}(2022)\citenamefont {Huang},
  \citenamefont {Shao}, \citenamefont {Ren}, \citenamefont {Sun},\ and\
  \citenamefont {Lv}}]{huang2022efficient}%
  \BibitemOpen
  \bibfield  {author} {\bibinfo {author} {\bibfnamefont {Y.}~\bibnamefont
  {Huang}}, \bibinfo {author} {\bibfnamefont {Y.}~\bibnamefont {Shao}},
  \bibinfo {author} {\bibfnamefont {W.}~\bibnamefont {Ren}}, \bibinfo {author}
  {\bibfnamefont {J.}~\bibnamefont {Sun}}, \ and\ \bibinfo {author}
  {\bibfnamefont {D.}~\bibnamefont {Lv}},\ }\href
  {https://10.48550/arXiv.2203.11112} {\bibfield  {journal} {\bibinfo
  {journal} {arXiv:2203.11112}\ } (\bibinfo {year} {2022})}\BibitemShut
  {NoStop}%
\bibitem [{\citenamefont {Yeter-Aydeniz}\ \emph {et~al.}(2022)\citenamefont
  {Yeter-Aydeniz}, \citenamefont {Moschandreou},\ and\ \citenamefont
  {Siopsis}}]{yeter2022quantum}%
  \BibitemOpen
  \bibfield  {author} {\bibinfo {author} {\bibfnamefont {K.}~\bibnamefont
  {Yeter-Aydeniz}}, \bibinfo {author} {\bibfnamefont {E.}~\bibnamefont
  {Moschandreou}}, \ and\ \bibinfo {author} {\bibfnamefont {G.}~\bibnamefont
  {Siopsis}},\ }\href {\doibase 10.1103/PhysRevA.105.012412} {\bibfield
  {journal} {\bibinfo  {journal} {Phys. Rev. A}\ }\textbf {\bibinfo {volume}
  {105}},\ \bibinfo {pages} {012412} (\bibinfo {year} {2022})}\BibitemShut
  {NoStop}%
\bibitem [{\citenamefont {Bottou}\ \emph {et~al.}(2018)\citenamefont {Bottou},
  \citenamefont {Curtis},\ and\ \citenamefont
  {Nocedal}}]{bottou2018optimization}%
  \BibitemOpen
  \bibfield  {author} {\bibinfo {author} {\bibfnamefont {L.}~\bibnamefont
  {Bottou}}, \bibinfo {author} {\bibfnamefont {F.~E.}\ \bibnamefont {Curtis}},
  \ and\ \bibinfo {author} {\bibfnamefont {J.}~\bibnamefont {Nocedal}},\ }\href
  {https://epubs.siam.org/doi/10.1137/16M1080173} {\bibfield  {journal}
  {\bibinfo  {journal} {Siam Review}\ }\textbf {\bibinfo {volume} {60}},\
  \bibinfo {pages} {223} (\bibinfo {year} {2018})}\BibitemShut {NoStop}%
\bibitem [{\citenamefont {Sun}\ \emph {et~al.}(2019)\citenamefont {Sun},
  \citenamefont {Cao}, \citenamefont {Zhu},\ and\ \citenamefont
  {Zhao}}]{sun2019survey}%
  \BibitemOpen
  \bibfield  {author} {\bibinfo {author} {\bibfnamefont {S.}~\bibnamefont
  {Sun}}, \bibinfo {author} {\bibfnamefont {Z.}~\bibnamefont {Cao}}, \bibinfo
  {author} {\bibfnamefont {H.}~\bibnamefont {Zhu}}, \ and\ \bibinfo {author}
  {\bibfnamefont {J.}~\bibnamefont {Zhao}},\ }\href {\doibase
  10.1109/TCYB.2019.2950779} {\bibfield  {journal} {\bibinfo  {journal} {IEEE
  Trans. Cybern.}\ }\textbf {\bibinfo {volume} {50}},\ \bibinfo {pages} {3668}
  (\bibinfo {year} {2019})}\BibitemShut {NoStop}%
\bibitem [{\citenamefont {Wales}\ and\ \citenamefont
  {Scheraga}(1999)}]{wales1999global}%
  \BibitemOpen
  \bibfield  {author} {\bibinfo {author} {\bibfnamefont {D.~J.}\ \bibnamefont
  {Wales}}\ and\ \bibinfo {author} {\bibfnamefont {H.~A.}\ \bibnamefont
  {Scheraga}},\ }\href {\doibase 10.1126/science.285.5432.1368} {\bibfield
  {journal} {\bibinfo  {journal} {Science}\ }\textbf {\bibinfo {volume}
  {285}},\ \bibinfo {pages} {1368} (\bibinfo {year} {1999})}\BibitemShut
  {NoStop}%
\bibitem [{\citenamefont {Hardin}\ \emph {et~al.}(2002)\citenamefont {Hardin},
  \citenamefont {Pogorelov},\ and\ \citenamefont
  {Luthey-Schulten}}]{hardin2002ab}%
  \BibitemOpen
  \bibfield  {author} {\bibinfo {author} {\bibfnamefont {C.}~\bibnamefont
  {Hardin}}, \bibinfo {author} {\bibfnamefont {T.~V.}\ \bibnamefont
  {Pogorelov}}, \ and\ \bibinfo {author} {\bibfnamefont {Z.}~\bibnamefont
  {Luthey-Schulten}},\ }\href {\doibase 10.1016/s0959-440x(02)00306-8}
  {\bibfield  {journal} {\bibinfo  {journal} {Curr. Opin. Struct. Biol.}\
  }\textbf {\bibinfo {volume} {12}},\ \bibinfo {pages} {176} (\bibinfo {year}
  {2002})}\BibitemShut {NoStop}%
\bibitem [{\citenamefont {Dill}\ \emph {et~al.}(2008)\citenamefont {Dill},
  \citenamefont {Ozkan}, \citenamefont {Shell},\ and\ \citenamefont
  {Weikl}}]{dill2008protein}%
  \BibitemOpen
  \bibfield  {author} {\bibinfo {author} {\bibfnamefont {K.~A.}\ \bibnamefont
  {Dill}}, \bibinfo {author} {\bibfnamefont {S.~B.}\ \bibnamefont {Ozkan}},
  \bibinfo {author} {\bibfnamefont {M.~S.}\ \bibnamefont {Shell}}, \ and\
  \bibinfo {author} {\bibfnamefont {T.~R.}\ \bibnamefont {Weikl}},\ }\href
  {\doibase 10.1146/annurev.biophys.37.092707.153558} {\bibfield  {journal}
  {\bibinfo  {journal} {Annu. Rev. Biophys.}\ }\textbf {\bibinfo {volume}
  {37}},\ \bibinfo {pages} {289} (\bibinfo {year} {2008})}\BibitemShut
  {NoStop}%
\bibitem [{\citenamefont {Dorn}\ \emph {et~al.}(2014)\citenamefont {Dorn},
  \citenamefont {e~Silva}, \citenamefont {Buriol},\ and\ \citenamefont
  {Lamb}}]{dorn2014three}%
  \BibitemOpen
  \bibfield  {author} {\bibinfo {author} {\bibfnamefont {M.}~\bibnamefont
  {Dorn}}, \bibinfo {author} {\bibfnamefont {M.~B.}\ \bibnamefont {e~Silva}},
  \bibinfo {author} {\bibfnamefont {L.~S.}\ \bibnamefont {Buriol}}, \ and\
  \bibinfo {author} {\bibfnamefont {L.~C.}\ \bibnamefont {Lamb}},\ }\href
  {\doibase 10.1016/j.compbiolchem.2014.10.001} {\bibfield  {journal} {\bibinfo
   {journal} {Computational biology and chemistry}\ }\textbf {\bibinfo {volume}
  {53}},\ \bibinfo {pages} {251} (\bibinfo {year} {2014})}\BibitemShut
  {NoStop}%
\bibitem [{\citenamefont {Kuhlman}\ and\ \citenamefont
  {Bradley}(2019)}]{kuhlman2019advances}%
  \BibitemOpen
  \bibfield  {author} {\bibinfo {author} {\bibfnamefont {B.}~\bibnamefont
  {Kuhlman}}\ and\ \bibinfo {author} {\bibfnamefont {P.}~\bibnamefont
  {Bradley}},\ }\href {\doibase 10.1038/s41580-019-0163-x} {\bibfield
  {journal} {\bibinfo  {journal} {Nat. Rev. Mol. Cell Biol.}\ }\textbf
  {\bibinfo {volume} {20}},\ \bibinfo {pages} {681} (\bibinfo {year}
  {2019})}\BibitemShut {NoStop}%
\bibitem [{\citenamefont {Coutinho}\ \emph {et~al.}(2018)\citenamefont
  {Coutinho}, \citenamefont {Battarra},\ and\ \citenamefont
  {Fliege}}]{coutinho2018unmanned}%
  \BibitemOpen
  \bibfield  {author} {\bibinfo {author} {\bibfnamefont {W.~P.}\ \bibnamefont
  {Coutinho}}, \bibinfo {author} {\bibfnamefont {M.}~\bibnamefont {Battarra}},
  \ and\ \bibinfo {author} {\bibfnamefont {J.}~\bibnamefont {Fliege}},\ }\href
  {\doibase 10.1016/j.cie.2018.04.037} {\bibfield  {journal} {\bibinfo
  {journal} {Comput. Ind. Eng.}\ }\textbf {\bibinfo {volume} {120}},\ \bibinfo
  {pages} {116} (\bibinfo {year} {2018})}\BibitemShut {NoStop}%
\bibitem [{\citenamefont {Otto}\ \emph {et~al.}(2018)\citenamefont {Otto},
  \citenamefont {Agatz}, \citenamefont {Campbell}, \citenamefont {Golden},\
  and\ \citenamefont {Pesch}}]{otto2018optimization}%
  \BibitemOpen
  \bibfield  {author} {\bibinfo {author} {\bibfnamefont {A.}~\bibnamefont
  {Otto}}, \bibinfo {author} {\bibfnamefont {N.}~\bibnamefont {Agatz}},
  \bibinfo {author} {\bibfnamefont {J.}~\bibnamefont {Campbell}}, \bibinfo
  {author} {\bibfnamefont {B.}~\bibnamefont {Golden}}, \ and\ \bibinfo {author}
  {\bibfnamefont {E.}~\bibnamefont {Pesch}},\ }\href {\doibase
  10.1002/net.21818} {\bibfield  {journal} {\bibinfo  {journal} {Networks}\
  }\textbf {\bibinfo {volume} {72}},\ \bibinfo {pages} {411} (\bibinfo {year}
  {2018})}\BibitemShut {NoStop}%
\bibitem [{\citenamefont {Levinthal}(1969)}]{levinthal1969fold}%
  \BibitemOpen
  \bibfield  {author} {\bibinfo {author} {\bibfnamefont {C.}~\bibnamefont
  {Levinthal}},\ }\href@noop {} {\bibfield  {journal} {\bibinfo  {journal}
  {Mossbauer spectroscopy in biological systems}\ }\textbf {\bibinfo {volume}
  {67}},\ \bibinfo {pages} {22} (\bibinfo {year} {1969})}\BibitemShut {NoStop}%
\bibitem [{\citenamefont {Karplus}(1997)}]{karplus1997levinthal}%
  \BibitemOpen
  \bibfield  {author} {\bibinfo {author} {\bibfnamefont {M.}~\bibnamefont
  {Karplus}},\ }\href {\doibase 10.1016/S1359-0278(97)00067-9} {\bibfield
  {journal} {\bibinfo  {journal} {Fold. Des.}\ }\textbf {\bibinfo {volume}
  {2}},\ \bibinfo {pages} {S69} (\bibinfo {year} {1997})}\BibitemShut {NoStop}%
\bibitem [{\citenamefont {Hartmann}\ and\ \citenamefont
  {Rieger}(2004)}]{hartmann2004new}%
  \BibitemOpen
  \bibfield  {author} {\bibinfo {author} {\bibfnamefont {A.~K.}\ \bibnamefont
  {Hartmann}}\ and\ \bibinfo {author} {\bibfnamefont {H.}~\bibnamefont
  {Rieger}},\ }\href
  {https://www.wiley.com/en-us/New+Optimization+Algorithms+in+Physics-p-9783527404063}
  {\emph {\bibinfo {title} {{New Optimization Algorithms in Physics}}}}\
  (\bibinfo  {publisher} {Wiley},\ \bibinfo {address} {Weinheim, Germany},\
  \bibinfo {year} {2004})\BibitemShut {NoStop}%
\bibitem [{\citenamefont {Venter}(2010)}]{venter2010review}%
  \BibitemOpen
  \bibfield  {author} {\bibinfo {author} {\bibfnamefont {G.}~\bibnamefont
  {Venter}},\ }\href {10.1002/9780470686652.eae495} {\emph {\bibinfo {title}
  {{Review of Optimization Techniques}}}}\ (\bibinfo  {publisher} {John Wiley
  {\&} Sons, Ltd},\ \bibinfo {address} {Chichester, England, UK},\ \bibinfo
  {year} {2010})\BibitemShut {NoStop}%
\bibitem [{\citenamefont {Grover}(1996)}]{Grover.1996.212}%
  \BibitemOpen
  \bibfield  {author} {\bibinfo {author} {\bibfnamefont {L.~K.}\ \bibnamefont
  {Grover}},\ }in\ \href {\doibase 10.1145/237814.237866} {\emph {\bibinfo
  {booktitle} {{STOC '96: Proceedings of the twenty-eighth annual ACM symposium
  on Theory of Computing}}}}\ (\bibinfo  {publisher} {Association for Computing
  Machinery},\ \bibinfo {address} {New York, NY, USA},\ \bibinfo {year}
  {1996})\ pp.\ \bibinfo {pages} {212--219}\BibitemShut {NoStop}%
\bibitem [{\citenamefont {D{\"u}rr}\ and\ \citenamefont
  {H{\o}yer}(1996)}]{Durr.1996.9607014v2}%
  \BibitemOpen
  \bibfield  {author} {\bibinfo {author} {\bibfnamefont {C.}~\bibnamefont
  {D{\"u}rr}}\ and\ \bibinfo {author} {\bibfnamefont {P.}~\bibnamefont
  {H{\o}yer}},\ }\href {https://10.48550/arXiv.quant-ph/9607014} {\bibfield
  {journal} {\bibinfo  {journal} {arXiv.quant-ph/9607014}\ } (\bibinfo {year}
  {1996})}\BibitemShut {NoStop}%
\bibitem [{\citenamefont {Bulger}\ \emph {et~al.}(2003)\citenamefont {Bulger},
  \citenamefont {Baritompa},\ and\ \citenamefont {Wood}}]{Bulger.2003.517}%
  \BibitemOpen
  \bibfield  {author} {\bibinfo {author} {\bibfnamefont {D.}~\bibnamefont
  {Bulger}}, \bibinfo {author} {\bibfnamefont {W.~P.}\ \bibnamefont
  {Baritompa}}, \ and\ \bibinfo {author} {\bibfnamefont {G.~R.}\ \bibnamefont
  {Wood}},\ }\href {\doibase 10.1023/A:1023061218864} {\bibfield  {journal}
  {\bibinfo  {journal} {J. Optim. Theory Appl.}\ }\textbf {\bibinfo {volume}
  {116}},\ \bibinfo {pages} {517} (\bibinfo {year} {2003})}\BibitemShut
  {NoStop}%
\bibitem [{\citenamefont {Farhi}\ \emph {et~al.}(2014)\citenamefont {Farhi},
  \citenamefont {Goldstone},\ and\ \citenamefont
  {Gutmann}}]{Farhi.2014.1411.4028}%
  \BibitemOpen
  \bibfield  {author} {\bibinfo {author} {\bibfnamefont {E.}~\bibnamefont
  {Farhi}}, \bibinfo {author} {\bibfnamefont {J.}~\bibnamefont {Goldstone}}, \
  and\ \bibinfo {author} {\bibfnamefont {S.}~\bibnamefont {Gutmann}},\ }\href
  {10.48550/arXiv.1411.4028} {\bibfield  {journal} {\bibinfo  {journal}
  {arXiv1411.4028}\ } (\bibinfo {year} {2014})}\BibitemShut {NoStop}%
\bibitem [{\citenamefont {Farhi}\ and\ \citenamefont
  {Harrow}(2016)}]{Farhi.2019.1602.07674v2}%
  \BibitemOpen
  \bibfield  {author} {\bibinfo {author} {\bibfnamefont {E.}~\bibnamefont
  {Farhi}}\ and\ \bibinfo {author} {\bibfnamefont {A.~W.}\ \bibnamefont
  {Harrow}},\ }\href {10.48550/arXiv.1602.07674} {\bibfield  {journal}
  {\bibinfo  {journal} {arXiv1602.07674}\ } (\bibinfo {year}
  {2016})}\BibitemShut {NoStop}%
\bibitem [{\citenamefont {Zhou}\ \emph {et~al.}(2020)\citenamefont {Zhou},
  \citenamefont {Wang}, \citenamefont {Choi}, \citenamefont {Pichler},\ and\
  \citenamefont {Lukin}}]{Zhou.2019.1812.01041v2}%
  \BibitemOpen
  \bibfield  {author} {\bibinfo {author} {\bibfnamefont {L.}~\bibnamefont
  {Zhou}}, \bibinfo {author} {\bibfnamefont {S.-T.}\ \bibnamefont {Wang}},
  \bibinfo {author} {\bibfnamefont {S.}~\bibnamefont {Choi}}, \bibinfo {author}
  {\bibfnamefont {H.}~\bibnamefont {Pichler}}, \ and\ \bibinfo {author}
  {\bibfnamefont {M.~D.}\ \bibnamefont {Lukin}},\ }\href {\doibase
  10.1103/PhysRevX.10.021067} {\bibfield  {journal} {\bibinfo  {journal} {Phys.
  Rev. X}\ }\textbf {\bibinfo {volume} {10}},\ \bibinfo {pages} {021067}
  (\bibinfo {year} {2020})}\BibitemShut {NoStop}%
\bibitem [{\citenamefont {Kadowaki}\ and\ \citenamefont
  {Nishimori}(1998)}]{Kadowaki.1998.5355}%
  \BibitemOpen
  \bibfield  {author} {\bibinfo {author} {\bibfnamefont {T.}~\bibnamefont
  {Kadowaki}}\ and\ \bibinfo {author} {\bibfnamefont {H.}~\bibnamefont
  {Nishimori}},\ }\href {\doibase 10.1103/PhysRevE.58.5355} {\bibfield
  {journal} {\bibinfo  {journal} {Phys. Rev. E}\ }\textbf {\bibinfo {volume}
  {58}},\ \bibinfo {pages} {5355} (\bibinfo {year} {1998})}\BibitemShut
  {NoStop}%
\bibitem [{\citenamefont {Farhi}\ \emph {et~al.}(2000)\citenamefont {Farhi},
  \citenamefont {Goldstone}, \citenamefont {Gutmann},\ and\ \citenamefont
  {Sipser}}]{Farhi.2000.0001106}%
  \BibitemOpen
  \bibfield  {author} {\bibinfo {author} {\bibfnamefont {E.}~\bibnamefont
  {Farhi}}, \bibinfo {author} {\bibfnamefont {J.}~\bibnamefont {Goldstone}},
  \bibinfo {author} {\bibfnamefont {S.}~\bibnamefont {Gutmann}}, \ and\
  \bibinfo {author} {\bibfnamefont {M.}~\bibnamefont {Sipser}},\ }\href
  {https://10.48550/arXiv.quant-ph/0001106} {\bibfield  {journal} {\bibinfo
  {journal} {arXiv.quant-ph/0001106}\ } (\bibinfo {year} {2000})}\BibitemShut
  {NoStop}%
\bibitem [{\citenamefont {Farhi}\ \emph {et~al.}(2001)\citenamefont {Farhi},
  \citenamefont {Goldstone}, \citenamefont {Gutmann}, \citenamefont {Lapan},
  \citenamefont {Lundgren},\ and\ \citenamefont {Preda}}]{Farhi.2001.472}%
  \BibitemOpen
  \bibfield  {author} {\bibinfo {author} {\bibfnamefont {E.}~\bibnamefont
  {Farhi}}, \bibinfo {author} {\bibfnamefont {J.}~\bibnamefont {Goldstone}},
  \bibinfo {author} {\bibfnamefont {S.}~\bibnamefont {Gutmann}}, \bibinfo
  {author} {\bibfnamefont {J.}~\bibnamefont {Lapan}}, \bibinfo {author}
  {\bibfnamefont {A.}~\bibnamefont {Lundgren}}, \ and\ \bibinfo {author}
  {\bibfnamefont {D.}~\bibnamefont {Preda}},\ }\href {\doibase
  10.1126/science.1057726} {\bibfield  {journal} {\bibinfo  {journal}
  {Science}\ }\textbf {\bibinfo {volume} {292}},\ \bibinfo {pages} {472}
  (\bibinfo {year} {2001})}\BibitemShut {NoStop}%
\bibitem [{\citenamefont {Perdomo-Ortiz}\ \emph {et~al.}(2012)\citenamefont
  {Perdomo-Ortiz}, \citenamefont {Dickson}, \citenamefont {Drew-Brook},
  \citenamefont {Rose},\ and\ \citenamefont
  {Aspuru-Guzik}}]{PerdomoOrtiz.2012.571}%
  \BibitemOpen
  \bibfield  {author} {\bibinfo {author} {\bibfnamefont {A.}~\bibnamefont
  {Perdomo-Ortiz}}, \bibinfo {author} {\bibfnamefont {N.}~\bibnamefont
  {Dickson}}, \bibinfo {author} {\bibfnamefont {M.}~\bibnamefont {Drew-Brook}},
  \bibinfo {author} {\bibfnamefont {G.}~\bibnamefont {Rose}}, \ and\ \bibinfo
  {author} {\bibfnamefont {A.}~\bibnamefont {Aspuru-Guzik}},\ }\href {\doibase
  10.1038/srep00571} {\bibfield  {journal} {\bibinfo  {journal} {Sci. Rep.}\
  }\textbf {\bibinfo {volume} {2}},\ \bibinfo {pages} {1} (\bibinfo {year}
  {2012})}\BibitemShut {NoStop}%
\bibitem [{\citenamefont {Babbush}\ \emph {et~al.}(2014)\citenamefont
  {Babbush}, \citenamefont {Love},\ and\ \citenamefont
  {Aspuru-Guzik}}]{Babbush.2014.6603}%
  \BibitemOpen
  \bibfield  {author} {\bibinfo {author} {\bibfnamefont {R.}~\bibnamefont
  {Babbush}}, \bibinfo {author} {\bibfnamefont {P.~J.}\ \bibnamefont {Love}}, \
  and\ \bibinfo {author} {\bibfnamefont {A.}~\bibnamefont {Aspuru-Guzik}},\
  }\href {\doibase 10.1038/srep06603} {\bibfield  {journal} {\bibinfo
  {journal} {Sci. Rep.}\ }\textbf {\bibinfo {volume} {4}},\ \bibinfo {pages}
  {1} (\bibinfo {year} {2014})}\BibitemShut {NoStop}%
\bibitem [{\citenamefont {Albash}\ and\ \citenamefont
  {Lidar}(2018)}]{Albash.2018.031016}%
  \BibitemOpen
  \bibfield  {author} {\bibinfo {author} {\bibfnamefont {T.}~\bibnamefont
  {Albash}}\ and\ \bibinfo {author} {\bibfnamefont {D.~A.}\ \bibnamefont
  {Lidar}},\ }\href {\doibase 10.1103/PhysRevX.8.031016} {\bibfield  {journal}
  {\bibinfo  {journal} {Phys. Rev. X}\ }\textbf {\bibinfo {volume} {8}},\
  \bibinfo {pages} {031016} (\bibinfo {year} {2018})}\BibitemShut {NoStop}%
\bibitem [{\citenamefont {Temme}\ \emph {et~al.}(2011)\citenamefont {Temme},
  \citenamefont {Osborne}, \citenamefont {Vollbrecht}, \citenamefont {Poulin},\
  and\ \citenamefont {Verstraete}}]{Temme.2011.87}%
  \BibitemOpen
  \bibfield  {author} {\bibinfo {author} {\bibfnamefont {K.}~\bibnamefont
  {Temme}}, \bibinfo {author} {\bibfnamefont {T.~J.}\ \bibnamefont {Osborne}},
  \bibinfo {author} {\bibfnamefont {K.~G.}\ \bibnamefont {Vollbrecht}},
  \bibinfo {author} {\bibfnamefont {D.}~\bibnamefont {Poulin}}, \ and\ \bibinfo
  {author} {\bibfnamefont {F.}~\bibnamefont {Verstraete}},\ }\href {\doibase
  10.1038/nature09770} {\bibfield  {journal} {\bibinfo  {journal} {Nature}\
  }\textbf {\bibinfo {volume} {471}},\ \bibinfo {pages} {87} (\bibinfo {year}
  {2011})}\BibitemShut {NoStop}%
\bibitem [{\citenamefont {Daskin}(2018)}]{Daskin:2019aa}%
  \BibitemOpen
  \bibfield  {author} {\bibinfo {author} {\bibfnamefont {A.}~\bibnamefont
  {Daskin}},\ }\href {https://10.48550/arXiv.1809.01378} {\bibfield  {journal}
  {\bibinfo  {journal} {arXiv:1809.01378}\ } (\bibinfo {year}
  {2018})}\BibitemShut {NoStop}%
\bibitem [{\citenamefont {Oseledets}\ and\ \citenamefont
  {Tyrtyshnikov}(2010)}]{Oseledets.2010.70}%
  \BibitemOpen
  \bibfield  {author} {\bibinfo {author} {\bibfnamefont {I.}~\bibnamefont
  {Oseledets}}\ and\ \bibinfo {author} {\bibfnamefont {E.}~\bibnamefont
  {Tyrtyshnikov}},\ }\href {\doibase 10.1016/j.laa.2009.07.024} {\bibfield
  {journal} {\bibinfo  {journal} {Linear Algebra Appl.}\ }\textbf {\bibinfo
  {volume} {432}},\ \bibinfo {pages} {70} (\bibinfo {year} {2010})}\BibitemShut
  {NoStop}%
\bibitem [{\citenamefont {Oseledets}(2011)}]{Oseledets.2011.2295}%
  \BibitemOpen
  \bibfield  {author} {\bibinfo {author} {\bibfnamefont {I.~V.}\ \bibnamefont
  {Oseledets}},\ }\href {https://epubs.siam.org/doi/10.1137/090752286}
  {\bibfield  {journal} {\bibinfo  {journal} {SIAM J. Sci. Comput.}\ }
  (\bibinfo {year} {2011})}\BibitemShut {NoStop}%
\bibitem [{\citenamefont {{\ifmmode\ddot{O}\else\"{O}\fi}stlund}\ and\
  \citenamefont {Rommer}(1995)}]{Ostlund.1995.3537}%
  \BibitemOpen
  \bibfield  {author} {\bibinfo {author} {\bibfnamefont {S.}~\bibnamefont
  {{\ifmmode\ddot{O}\else\"{O}\fi}stlund}}\ and\ \bibinfo {author}
  {\bibfnamefont {S.}~\bibnamefont {Rommer}},\ }\href {\doibase
  10.1103/PhysRevLett.75.3537} {\bibfield  {journal} {\bibinfo  {journal}
  {Phys. Rev. Lett.}\ }\textbf {\bibinfo {volume} {75}},\ \bibinfo {pages}
  {3537} (\bibinfo {year} {1995})}\BibitemShut {NoStop}%
\bibitem [{\citenamefont {Soley}\ \emph {et~al.}(2021)\citenamefont {Soley},
  \citenamefont {Bergold},\ and\ \citenamefont {Batista}}]{soley2021iterative}%
  \BibitemOpen
  \bibfield  {author} {\bibinfo {author} {\bibfnamefont {M.~B.}\ \bibnamefont
  {Soley}}, \bibinfo {author} {\bibfnamefont {P.}~\bibnamefont {Bergold}}, \
  and\ \bibinfo {author} {\bibfnamefont {V.~S.}\ \bibnamefont {Batista}},\
  }\href {\doibase https://doi.org/10.1021/acs.jctc.1c00292} {\bibfield
  {journal} {\bibinfo  {journal} {J. Chem. Theory Comput.}\ }\textbf {\bibinfo
  {volume} {17}},\ \bibinfo {pages} {3280} (\bibinfo {year}
  {2021})}\BibitemShut {NoStop}%
\bibitem [{\citenamefont {Jozsa}\ and\ \citenamefont
  {Linden}(2003)}]{jozsa2003role}%
  \BibitemOpen
  \bibfield  {author} {\bibinfo {author} {\bibfnamefont {R.}~\bibnamefont
  {Jozsa}}\ and\ \bibinfo {author} {\bibfnamefont {N.}~\bibnamefont {Linden}},\
  }\href {https://www.jstor.org/stable/3560059} {\bibfield  {journal} {\bibinfo
   {journal} {Proc. R. Soc. A: Math. Phys. Eng. Sci.}\ }\textbf {\bibinfo
  {volume} {459}},\ \bibinfo {pages} {2011} (\bibinfo {year}
  {2003})}\BibitemShut {NoStop}%
\bibitem [{\citenamefont {Vidal}(2008)}]{vidal2008class}%
  \BibitemOpen
  \bibfield  {author} {\bibinfo {author} {\bibfnamefont {G.}~\bibnamefont
  {Vidal}},\ }\href {\doibase 10.1103/PhysRevLett.101.110501} {\bibfield
  {journal} {\bibinfo  {journal} {Phys. Rev. Lett.}\ }\textbf {\bibinfo
  {volume} {101}},\ \bibinfo {pages} {110501} (\bibinfo {year}
  {2008})}\BibitemShut {NoStop}%
\bibitem [{\citenamefont {Gross}\ \emph {et~al.}(2009)\citenamefont {Gross},
  \citenamefont {Flammia},\ and\ \citenamefont {Eisert}}]{gross2009most}%
  \BibitemOpen
  \bibfield  {author} {\bibinfo {author} {\bibfnamefont {D.}~\bibnamefont
  {Gross}}, \bibinfo {author} {\bibfnamefont {S.~T.}\ \bibnamefont {Flammia}},
  \ and\ \bibinfo {author} {\bibfnamefont {J.}~\bibnamefont {Eisert}},\ }\href
  {\doibase 10.1103/PhysRevLett.102.190501} {\bibfield  {journal} {\bibinfo
  {journal} {Phys. Rev. Lett.}\ }\textbf {\bibinfo {volume} {102}},\ \bibinfo
  {pages} {190501} (\bibinfo {year} {2009})}\BibitemShut {NoStop}%
\bibitem [{\citenamefont {Evenbly}\ and\ \citenamefont
  {Vidal}(2014)}]{evenbly2014class}%
  \BibitemOpen
  \bibfield  {author} {\bibinfo {author} {\bibfnamefont {G.}~\bibnamefont
  {Evenbly}}\ and\ \bibinfo {author} {\bibfnamefont {G.}~\bibnamefont
  {Vidal}},\ }\href {\doibase 10.1103/PhysRevLett.112.240502} {\bibfield
  {journal} {\bibinfo  {journal} {Phys. Rev. Lett.}\ }\textbf {\bibinfo
  {volume} {112}},\ \bibinfo {pages} {240502} (\bibinfo {year}
  {2014})}\BibitemShut {NoStop}%
\bibitem [{\citenamefont {Bluvstein}\ \emph {et~al.}(2022)\citenamefont
  {Bluvstein}, \citenamefont {Levine}, \citenamefont {Semeghini}, \citenamefont
  {Wang}, \citenamefont {Ebadi}, \citenamefont {Kalinowski}, \citenamefont
  {Keesling}, \citenamefont {Maskara}, \citenamefont {Pichler}, \citenamefont
  {Greiner} \emph {et~al.}}]{bluvstein2022quantum}%
  \BibitemOpen
  \bibfield  {author} {\bibinfo {author} {\bibfnamefont {D.}~\bibnamefont
  {Bluvstein}}, \bibinfo {author} {\bibfnamefont {H.}~\bibnamefont {Levine}},
  \bibinfo {author} {\bibfnamefont {G.}~\bibnamefont {Semeghini}}, \bibinfo
  {author} {\bibfnamefont {T.~T.}\ \bibnamefont {Wang}}, \bibinfo {author}
  {\bibfnamefont {S.}~\bibnamefont {Ebadi}}, \bibinfo {author} {\bibfnamefont
  {M.}~\bibnamefont {Kalinowski}}, \bibinfo {author} {\bibfnamefont
  {A.}~\bibnamefont {Keesling}}, \bibinfo {author} {\bibfnamefont
  {N.}~\bibnamefont {Maskara}}, \bibinfo {author} {\bibfnamefont
  {H.}~\bibnamefont {Pichler}}, \bibinfo {author} {\bibfnamefont
  {M.}~\bibnamefont {Greiner}},  \emph {et~al.},\ }\href {\doibase
  10.1038/s41586-022-04592-6} {\bibfield  {journal} {\bibinfo  {journal}
  {Nature}\ }\textbf {\bibinfo {volume} {604}},\ \bibinfo {pages} {451}
  (\bibinfo {year} {2022})}\BibitemShut {NoStop}%
\bibitem [{\citenamefont {Haghshenas}\ \emph {et~al.}(2022)\citenamefont
  {Haghshenas}, \citenamefont {Gray}, \citenamefont {Potter},\ and\
  \citenamefont {Chan}}]{haghshenas2022variational}%
  \BibitemOpen
  \bibfield  {author} {\bibinfo {author} {\bibfnamefont {R.}~\bibnamefont
  {Haghshenas}}, \bibinfo {author} {\bibfnamefont {J.}~\bibnamefont {Gray}},
  \bibinfo {author} {\bibfnamefont {A.~C.}\ \bibnamefont {Potter}}, \ and\
  \bibinfo {author} {\bibfnamefont {G.~K.-L.}\ \bibnamefont {Chan}},\ }\href
  {\doibase 10.1103/PhysRevX.12.011047} {\bibfield  {journal} {\bibinfo
  {journal} {Phys. Rev. X}\ }\textbf {\bibinfo {volume} {12}},\ \bibinfo
  {pages} {011047} (\bibinfo {year} {2022})}\BibitemShut {NoStop}%
\bibitem [{\citenamefont {Sim}\ \emph {et~al.}(2019)\citenamefont {Sim},
  \citenamefont {Johnson},\ and\ \citenamefont
  {Aspuru-Guzik}}]{Sim2019expressibility}%
  \BibitemOpen
  \bibfield  {author} {\bibinfo {author} {\bibfnamefont {S.}~\bibnamefont
  {Sim}}, \bibinfo {author} {\bibfnamefont {P.~D.}\ \bibnamefont {Johnson}}, \
  and\ \bibinfo {author} {\bibfnamefont {A.}~\bibnamefont {Aspuru-Guzik}},\
  }\href {\doibase 10.1002/qute.201900070} {\bibfield  {journal} {\bibinfo
  {journal} {Adv. Quantum Technol.}\ }\textbf {\bibinfo {volume} {2}},\
  \bibinfo {pages} {1900070} (\bibinfo {year} {2019})}\BibitemShut {NoStop}%
\bibitem [{\citenamefont {Yung}\ \emph {et~al.}(2014)\citenamefont {Yung},
  \citenamefont {Casanova}, \citenamefont {Mezzacapo}, \citenamefont {Mcclean},
  \citenamefont {Lamata}, \citenamefont {Aspuru-Guzik},\ and\ \citenamefont
  {Solano}}]{yung2014transistor}%
  \BibitemOpen
  \bibfield  {author} {\bibinfo {author} {\bibfnamefont {M.-H.}\ \bibnamefont
  {Yung}}, \bibinfo {author} {\bibfnamefont {J.}~\bibnamefont {Casanova}},
  \bibinfo {author} {\bibfnamefont {A.}~\bibnamefont {Mezzacapo}}, \bibinfo
  {author} {\bibfnamefont {J.}~\bibnamefont {Mcclean}}, \bibinfo {author}
  {\bibfnamefont {L.}~\bibnamefont {Lamata}}, \bibinfo {author} {\bibfnamefont
  {A.}~\bibnamefont {Aspuru-Guzik}}, \ and\ \bibinfo {author} {\bibfnamefont
  {E.}~\bibnamefont {Solano}},\ }\href {\doibase 10.1038/srep03589} {\bibfield
  {journal} {\bibinfo  {journal} {Sci. Rep.}\ }\textbf {\bibinfo {volume}
  {4}},\ \bibinfo {pages} {1} (\bibinfo {year} {2014})}\BibitemShut {NoStop}%
\bibitem [{\citenamefont {Wecker}\ \emph {et~al.}(2015)\citenamefont {Wecker},
  \citenamefont {Hastings},\ and\ \citenamefont {Troyer}}]{wecker2015progress}%
  \BibitemOpen
  \bibfield  {author} {\bibinfo {author} {\bibfnamefont {D.}~\bibnamefont
  {Wecker}}, \bibinfo {author} {\bibfnamefont {M.~B.}\ \bibnamefont
  {Hastings}}, \ and\ \bibinfo {author} {\bibfnamefont {M.}~\bibnamefont
  {Troyer}},\ }\href {\doibase 10.1103/PhysRevA.92.042303} {\bibfield
  {journal} {\bibinfo  {journal} {Phys. Rev. A}\ }\textbf {\bibinfo {volume}
  {92}},\ \bibinfo {pages} {042303} (\bibinfo {year} {2015})}\BibitemShut
  {NoStop}%
\bibitem [{\citenamefont {Shen}\ \emph {et~al.}(2017)\citenamefont {Shen},
  \citenamefont {Zhang}, \citenamefont {Zhang}, \citenamefont {Zhang},
  \citenamefont {Yung},\ and\ \citenamefont {Kim}}]{shen2017quantum}%
  \BibitemOpen
  \bibfield  {author} {\bibinfo {author} {\bibfnamefont {Y.}~\bibnamefont
  {Shen}}, \bibinfo {author} {\bibfnamefont {X.}~\bibnamefont {Zhang}},
  \bibinfo {author} {\bibfnamefont {S.}~\bibnamefont {Zhang}}, \bibinfo
  {author} {\bibfnamefont {J.-N.}\ \bibnamefont {Zhang}}, \bibinfo {author}
  {\bibfnamefont {M.-H.}\ \bibnamefont {Yung}}, \ and\ \bibinfo {author}
  {\bibfnamefont {K.}~\bibnamefont {Kim}},\ }\href {\doibase
  10.1103/PhysRevA.95.020501} {\bibfield  {journal} {\bibinfo  {journal} {Phys.
  Rev. A}\ }\textbf {\bibinfo {volume} {95}},\ \bibinfo {pages} {020501}
  (\bibinfo {year} {2017})}\BibitemShut {NoStop}%
\bibitem [{\citenamefont {Romero}\ \emph {et~al.}(2018)\citenamefont {Romero},
  \citenamefont {Babbush}, \citenamefont {McClean}, \citenamefont {Hempel},
  \citenamefont {Love},\ and\ \citenamefont
  {Aspuru-Guzik}}]{romero2018strategies}%
  \BibitemOpen
  \bibfield  {author} {\bibinfo {author} {\bibfnamefont {J.}~\bibnamefont
  {Romero}}, \bibinfo {author} {\bibfnamefont {R.}~\bibnamefont {Babbush}},
  \bibinfo {author} {\bibfnamefont {J.~R.}\ \bibnamefont {McClean}}, \bibinfo
  {author} {\bibfnamefont {C.}~\bibnamefont {Hempel}}, \bibinfo {author}
  {\bibfnamefont {P.~J.}\ \bibnamefont {Love}}, \ and\ \bibinfo {author}
  {\bibfnamefont {A.}~\bibnamefont {Aspuru-Guzik}},\ }\href {\doibase
  https://doi.org/10.1088/2058-9565/aad3e4} {\bibfield  {journal} {\bibinfo
  {journal} {Quantum Sci. Technol.}\ }\textbf {\bibinfo {volume} {4}},\
  \bibinfo {pages} {014008} (\bibinfo {year} {2018})}\BibitemShut {NoStop}%
\bibitem [{\citenamefont {Kottmann}\ \emph
  {et~al.}(2021{\natexlab{a}})\citenamefont {Kottmann}, \citenamefont
  {Alperin-Lea}, \citenamefont {Tamayo-Mendoza}, \citenamefont
  {Cervera-Lierta}, \citenamefont {Lavigne}, \citenamefont {Yen}, \citenamefont
  {Verteletskyi}, \citenamefont {Schleich}, \citenamefont {Anand},
  \citenamefont {Degroote} \emph {et~al.}}]{Kottmann2021tequila}%
  \BibitemOpen
  \bibfield  {author} {\bibinfo {author} {\bibfnamefont {J.~S.}\ \bibnamefont
  {Kottmann}}, \bibinfo {author} {\bibfnamefont {S.}~\bibnamefont
  {Alperin-Lea}}, \bibinfo {author} {\bibfnamefont {T.}~\bibnamefont
  {Tamayo-Mendoza}}, \bibinfo {author} {\bibfnamefont {A.}~\bibnamefont
  {Cervera-Lierta}}, \bibinfo {author} {\bibfnamefont {C.}~\bibnamefont
  {Lavigne}}, \bibinfo {author} {\bibfnamefont {T.-C.}\ \bibnamefont {Yen}},
  \bibinfo {author} {\bibfnamefont {V.}~\bibnamefont {Verteletskyi}}, \bibinfo
  {author} {\bibfnamefont {P.}~\bibnamefont {Schleich}}, \bibinfo {author}
  {\bibfnamefont {A.}~\bibnamefont {Anand}}, \bibinfo {author} {\bibfnamefont
  {M.}~\bibnamefont {Degroote}},  \emph {et~al.},\ }\href {\doibase
  10.1088/2058-9565/abe567} {\bibfield  {journal} {\bibinfo  {journal} {Quantum
  Sci. Technol.}\ }\textbf {\bibinfo {volume} {6}},\ \bibinfo {pages} {024009}
  (\bibinfo {year} {2021}{\natexlab{a}})}\BibitemShut {NoStop}%
\bibitem [{\citenamefont {Suzuki}\ \emph {et~al.}(2021)\citenamefont {Suzuki},
  \citenamefont {Kawase}, \citenamefont {Masumura}, \citenamefont {Hiraga},
  \citenamefont {Nakadai}, \citenamefont {Chen}, \citenamefont {Nakanishi},
  \citenamefont {Mitarai}, \citenamefont {Imai}, \citenamefont {Tamiya} \emph
  {et~al.}}]{Suzuki2021qulacs}%
  \BibitemOpen
  \bibfield  {author} {\bibinfo {author} {\bibfnamefont {Y.}~\bibnamefont
  {Suzuki}}, \bibinfo {author} {\bibfnamefont {Y.}~\bibnamefont {Kawase}},
  \bibinfo {author} {\bibfnamefont {Y.}~\bibnamefont {Masumura}}, \bibinfo
  {author} {\bibfnamefont {Y.}~\bibnamefont {Hiraga}}, \bibinfo {author}
  {\bibfnamefont {M.}~\bibnamefont {Nakadai}}, \bibinfo {author} {\bibfnamefont
  {J.}~\bibnamefont {Chen}}, \bibinfo {author} {\bibfnamefont {K.~M.}\
  \bibnamefont {Nakanishi}}, \bibinfo {author} {\bibfnamefont {K.}~\bibnamefont
  {Mitarai}}, \bibinfo {author} {\bibfnamefont {R.}~\bibnamefont {Imai}},
  \bibinfo {author} {\bibfnamefont {S.}~\bibnamefont {Tamiya}},  \emph
  {et~al.},\ }\href {\doibase 10.22331/q-2021-10-06-559} {\bibfield  {journal}
  {\bibinfo  {journal} {Quantum}\ }\textbf {\bibinfo {volume} {5}},\ \bibinfo
  {pages} {559} (\bibinfo {year} {2021})},\ \Eprint
  {http://arxiv.org/abs/2011.13524v4} {2011.13524v4} \BibitemShut {NoStop}%
\bibitem [{\citenamefont {Zeng}\ \emph {et~al.}(2021)\citenamefont {Zeng},
  \citenamefont {Sun},\ and\ \citenamefont {Yuan}}]{zeng2021universal}%
  \BibitemOpen
  \bibfield  {author} {\bibinfo {author} {\bibfnamefont {P.}~\bibnamefont
  {Zeng}}, \bibinfo {author} {\bibfnamefont {J.}~\bibnamefont {Sun}}, \ and\
  \bibinfo {author} {\bibfnamefont {X.}~\bibnamefont {Yuan}},\ }\href
  {https://arxiv.org/abs/2109.15304} {\bibfield  {journal} {\bibinfo  {journal}
  {arXiv:2109.15304}\ } (\bibinfo {year} {2021})}\BibitemShut {NoStop}%
\bibitem [{\citenamefont {Xu}\ \emph {et~al.}(2014)\citenamefont {Xu},
  \citenamefont {Yung}, \citenamefont {Xu}, \citenamefont {Boixo},
  \citenamefont {Zhou}, \citenamefont {Li}, \citenamefont {Aspuru-Guzik},\ and\
  \citenamefont {Guo}}]{xu2014demon}%
  \BibitemOpen
  \bibfield  {author} {\bibinfo {author} {\bibfnamefont {J.-S.}\ \bibnamefont
  {Xu}}, \bibinfo {author} {\bibfnamefont {M.-H.}\ \bibnamefont {Yung}},
  \bibinfo {author} {\bibfnamefont {X.-Y.}\ \bibnamefont {Xu}}, \bibinfo
  {author} {\bibfnamefont {S.}~\bibnamefont {Boixo}}, \bibinfo {author}
  {\bibfnamefont {Z.-W.}\ \bibnamefont {Zhou}}, \bibinfo {author}
  {\bibfnamefont {C.-F.}\ \bibnamefont {Li}}, \bibinfo {author} {\bibfnamefont
  {A.}~\bibnamefont {Aspuru-Guzik}}, \ and\ \bibinfo {author} {\bibfnamefont
  {G.-C.}\ \bibnamefont {Guo}},\ }\href {\doibase
  https://doi.org/10.1038/nphoton.2013.354} {\bibfield  {journal} {\bibinfo
  {journal} {Nat. Photonics}\ }\textbf {\bibinfo {volume} {8}},\ \bibinfo
  {pages} {113} (\bibinfo {year} {2014})}\BibitemShut {NoStop}%
\bibitem [{\citenamefont {Foulkes}\ \emph {et~al.}(2001)\citenamefont
  {Foulkes}, \citenamefont {Mitas}, \citenamefont {Needs},\ and\ \citenamefont
  {Rajagopal}}]{Foulkes2001quantum}%
  \BibitemOpen
  \bibfield  {author} {\bibinfo {author} {\bibfnamefont {W.~M.~C.}\
  \bibnamefont {Foulkes}}, \bibinfo {author} {\bibfnamefont {L.}~\bibnamefont
  {Mitas}}, \bibinfo {author} {\bibfnamefont {R.~J.}\ \bibnamefont {Needs}}, \
  and\ \bibinfo {author} {\bibfnamefont {G.}~\bibnamefont {Rajagopal}},\ }\href
  {\doibase 10.1103/RevModPhys.73.33} {\bibfield  {journal} {\bibinfo
  {journal} {Rev. Mod. Phys.}\ }\textbf {\bibinfo {volume} {73}},\ \bibinfo
  {pages} {33} (\bibinfo {year} {2001})}\BibitemShut {NoStop}%
\bibitem [{\citenamefont {Le~Bellac}\ \emph {et~al.}(2004)\citenamefont
  {Le~Bellac}, \citenamefont {Mortessagne},\ and\ \citenamefont
  {Batrouni}}]{LeBellac2004equilibrium}%
  \BibitemOpen
  \bibfield  {author} {\bibinfo {author} {\bibfnamefont {M.}~\bibnamefont
  {Le~Bellac}}, \bibinfo {author} {\bibfnamefont {F.}~\bibnamefont
  {Mortessagne}}, \ and\ \bibinfo {author} {\bibfnamefont {G.~G.}\ \bibnamefont
  {Batrouni}},\ }\href {\doibase 10.1017/CBO9780511606571} {\emph {\bibinfo
  {title} {{Equilibrium and Non-Equilibrium Statistical Thermodynamics}}}}\
  (\bibinfo  {publisher} {Cambridge University Press},\ \bibinfo {address}
  {Cambridge, England, UK},\ \bibinfo {year} {2004})\BibitemShut {NoStop}%
\bibitem [{\citenamefont {Jordan}\ and\ \citenamefont
  {Wigner}(1928)}]{jordan1928pauli}%
  \BibitemOpen
  \bibfield  {author} {\bibinfo {author} {\bibfnamefont {P.}~\bibnamefont
  {Jordan}}\ and\ \bibinfo {author} {\bibfnamefont {E.}~\bibnamefont
  {Wigner}},\ }\href@noop {} {\bibfield  {journal} {\bibinfo  {journal} {Z.
  Physik}\ }\textbf {\bibinfo {volume} {47}},\ \bibinfo {pages} {631} (\bibinfo
  {year} {1928})}\BibitemShut {NoStop}%
\bibitem [{\citenamefont {Tranter}\ \emph {et~al.}(2018)\citenamefont
  {Tranter}, \citenamefont {Love}, \citenamefont {Mintert},\ and\ \citenamefont
  {Coveney}}]{Tranter2018JCTC}%
  \BibitemOpen
  \bibfield  {author} {\bibinfo {author} {\bibfnamefont {A.}~\bibnamefont
  {Tranter}}, \bibinfo {author} {\bibfnamefont {P.~J.}\ \bibnamefont {Love}},
  \bibinfo {author} {\bibfnamefont {F.}~\bibnamefont {Mintert}}, \ and\
  \bibinfo {author} {\bibfnamefont {P.~V.}\ \bibnamefont {Coveney}},\ }\href
  {\doibase 10.1021/acs.jctc.8b00450} {\bibfield  {journal} {\bibinfo
  {journal} {J. Chem. Theory Comput.}\ }\textbf {\bibinfo {volume} {14}},\
  \bibinfo {pages} {5617} (\bibinfo {year} {2018})}\BibitemShut {NoStop}%
\bibitem [{\citenamefont {Cao}\ \emph {et~al.}(2019{\natexlab{b}})\citenamefont
  {Cao}, \citenamefont {Romero}, \citenamefont {Olson}, \citenamefont
  {Degroote}, \citenamefont {Johnson}, \citenamefont {Kieferová},
  \citenamefont {Kivlichan}, \citenamefont {Menke}, \citenamefont {Peropadre},
  \citenamefont {Sawaya}, \citenamefont {Sim}, \citenamefont {Veis},\ and\
  \citenamefont {Aspuru-Guzik}}]{Cao2019ChemRev}%
  \BibitemOpen
  \bibfield  {author} {\bibinfo {author} {\bibfnamefont {Y.}~\bibnamefont
  {Cao}}, \bibinfo {author} {\bibfnamefont {J.}~\bibnamefont {Romero}},
  \bibinfo {author} {\bibfnamefont {J.~P.}\ \bibnamefont {Olson}}, \bibinfo
  {author} {\bibfnamefont {M.}~\bibnamefont {Degroote}}, \bibinfo {author}
  {\bibfnamefont {P.~D.}\ \bibnamefont {Johnson}}, \bibinfo {author}
  {\bibfnamefont {M.}~\bibnamefont {Kieferová}}, \bibinfo {author}
  {\bibfnamefont {I.~D.}\ \bibnamefont {Kivlichan}}, \bibinfo {author}
  {\bibfnamefont {T.}~\bibnamefont {Menke}}, \bibinfo {author} {\bibfnamefont
  {B.}~\bibnamefont {Peropadre}}, \bibinfo {author} {\bibfnamefont {N.~P.~D.}\
  \bibnamefont {Sawaya}}, \bibinfo {author} {\bibfnamefont {S.}~\bibnamefont
  {Sim}}, \bibinfo {author} {\bibfnamefont {L.}~\bibnamefont {Veis}}, \ and\
  \bibinfo {author} {\bibfnamefont {A.}~\bibnamefont {Aspuru-Guzik}},\ }\href
  {\doibase 10.1021/acs.chemrev.8b00803} {\bibfield  {journal} {\bibinfo
  {journal} {Chem. Rev.}\ }\textbf {\bibinfo {volume} {119}},\ \bibinfo {pages}
  {10856} (\bibinfo {year} {2019}{\natexlab{b}})}\BibitemShut {NoStop}%
\bibitem [{\citenamefont {McArdle}\ \emph
  {et~al.}(2020{\natexlab{b}})\citenamefont {McArdle}, \citenamefont {Endo},
  \citenamefont {Aspuru-Guzik}, \citenamefont {Benjamin},\ and\ \citenamefont
  {Yuan}}]{McArdle2020RevModPhys}%
  \BibitemOpen
  \bibfield  {author} {\bibinfo {author} {\bibfnamefont {S.}~\bibnamefont
  {McArdle}}, \bibinfo {author} {\bibfnamefont {S.}~\bibnamefont {Endo}},
  \bibinfo {author} {\bibfnamefont {A.}~\bibnamefont {Aspuru-Guzik}}, \bibinfo
  {author} {\bibfnamefont {S.~C.}\ \bibnamefont {Benjamin}}, \ and\ \bibinfo
  {author} {\bibfnamefont {X.}~\bibnamefont {Yuan}},\ }\href {\doibase
  10.1103/RevModPhys.92.015003} {\bibfield  {journal} {\bibinfo  {journal}
  {Rev. Mod. Phys.}\ }\textbf {\bibinfo {volume} {92}},\ \bibinfo {pages}
  {015003} (\bibinfo {year} {2020}{\natexlab{b}})}\BibitemShut {NoStop}%
\bibitem [{\citenamefont {Daskin}\ and\ \citenamefont
  {Kais}(2011)}]{daskin2011decomposition}%
  \BibitemOpen
  \bibfield  {author} {\bibinfo {author} {\bibfnamefont {A.}~\bibnamefont
  {Daskin}}\ and\ \bibinfo {author} {\bibfnamefont {S.}~\bibnamefont {Kais}},\
  }\href {\doibase 10.1063/1.3575402} {\bibfield  {journal} {\bibinfo
  {journal} {J. Chem. Phys.}\ }\textbf {\bibinfo {volume} {134}},\ \bibinfo
  {pages} {144112} (\bibinfo {year} {2011})}\BibitemShut {NoStop}%
\bibitem [{\citenamefont {Kyaw}\ \emph {et~al.}(2021)\citenamefont {Kyaw},
  \citenamefont {Menke}, \citenamefont {Sim}, \citenamefont {Anand},
  \citenamefont {Sawaya}, \citenamefont {Oliver}, \citenamefont {Guerreschi},\
  and\ \citenamefont {Aspuru-Guzik}}]{kyaw2021quantum}%
  \BibitemOpen
  \bibfield  {author} {\bibinfo {author} {\bibfnamefont {T.~H.}\ \bibnamefont
  {Kyaw}}, \bibinfo {author} {\bibfnamefont {T.}~\bibnamefont {Menke}},
  \bibinfo {author} {\bibfnamefont {S.}~\bibnamefont {Sim}}, \bibinfo {author}
  {\bibfnamefont {A.}~\bibnamefont {Anand}}, \bibinfo {author} {\bibfnamefont
  {N.~P.}\ \bibnamefont {Sawaya}}, \bibinfo {author} {\bibfnamefont {W.~D.}\
  \bibnamefont {Oliver}}, \bibinfo {author} {\bibfnamefont {G.~G.}\
  \bibnamefont {Guerreschi}}, \ and\ \bibinfo {author} {\bibfnamefont
  {A.}~\bibnamefont {Aspuru-Guzik}},\ }\href {\doibase
  10.1103/PhysRevApplied.16.044042} {\bibfield  {journal} {\bibinfo  {journal}
  {Phys. Rev. Appl.}\ }\textbf {\bibinfo {volume} {16}},\ \bibinfo {pages}
  {044042} (\bibinfo {year} {2021})}\BibitemShut {NoStop}%
\bibitem [{\citenamefont {Kottmann}\ \emph
  {et~al.}(2021{\natexlab{b}})\citenamefont {Kottmann}, \citenamefont {Krenn},
  \citenamefont {Kyaw}, \citenamefont {Alperin-Lea},\ and\ \citenamefont
  {Aspuru-Guzik}}]{kottmann2021quantum}%
  \BibitemOpen
  \bibfield  {author} {\bibinfo {author} {\bibfnamefont {J.}~\bibnamefont
  {Kottmann}}, \bibinfo {author} {\bibfnamefont {M.}~\bibnamefont {Krenn}},
  \bibinfo {author} {\bibfnamefont {T.~H.}\ \bibnamefont {Kyaw}}, \bibinfo
  {author} {\bibfnamefont {S.}~\bibnamefont {Alperin-Lea}}, \ and\ \bibinfo
  {author} {\bibfnamefont {A.}~\bibnamefont {Aspuru-Guzik}},\ }\href {\doibase
  https://doi.org/10.1088/2058-9565/abfc94} {\bibfield  {journal} {\bibinfo
  {journal} {Quantum Sci. and Technol.}\ }\textbf {\bibinfo {volume} {6}},\
  \bibinfo {pages} {035010} (\bibinfo {year} {2021}{\natexlab{b}})}\BibitemShut
  {NoStop}%
\bibitem [{\citenamefont {Liu}\ \emph {et~al.}(2021)\citenamefont {Liu},
  \citenamefont {Chen}, \citenamefont {Wang}, \citenamefont {Li}, \citenamefont
  {Shang}, \citenamefont {Ying}, \citenamefont {Wang}, \citenamefont {Peng},
  \citenamefont {Zhu}, \citenamefont {Lu} \emph {et~al.}}]{liu2021quantum}%
  \BibitemOpen
  \bibfield  {author} {\bibinfo {author} {\bibfnamefont {F.-M.}\ \bibnamefont
  {Liu}}, \bibinfo {author} {\bibfnamefont {M.-C.}\ \bibnamefont {Chen}},
  \bibinfo {author} {\bibfnamefont {C.}~\bibnamefont {Wang}}, \bibinfo {author}
  {\bibfnamefont {S.-W.}\ \bibnamefont {Li}}, \bibinfo {author} {\bibfnamefont
  {Z.-X.}\ \bibnamefont {Shang}}, \bibinfo {author} {\bibfnamefont
  {C.}~\bibnamefont {Ying}}, \bibinfo {author} {\bibfnamefont {J.-W.}\
  \bibnamefont {Wang}}, \bibinfo {author} {\bibfnamefont {C.-Z.}\ \bibnamefont
  {Peng}}, \bibinfo {author} {\bibfnamefont {X.}~\bibnamefont {Zhu}}, \bibinfo
  {author} {\bibfnamefont {C.-Y.}\ \bibnamefont {Lu}},  \emph {et~al.},\ }\href
  {https://arxiv.org/abs/2109.00994} {\bibfield  {journal} {\bibinfo  {journal}
  {arXiv:2109.00994}\ } (\bibinfo {year} {2021})}\BibitemShut {NoStop}%
\bibitem [{\citenamefont {Kyaw}(2019)}]{kyaw2019towards}%
  \BibitemOpen
  \bibfield  {author} {\bibinfo {author} {\bibfnamefont {T.~H.}\ \bibnamefont
  {Kyaw}},\ }\href {https://link.springer.com/book/10.1007/978-3-030-19658-5}
  {\emph {\bibinfo {title} {{Towards a Scalable Quantum Computing Platform in
  the Ultrastrong Coupling Regime}}}}\ (\bibinfo  {publisher} {Springer},\
  \bibinfo {year} {2019})\BibitemShut {NoStop}%
\bibitem [{\citenamefont {Susskind}\ and\ \citenamefont
  {Glogower}(1964)}]{susskind1964quantum}%
  \BibitemOpen
  \bibfield  {author} {\bibinfo {author} {\bibfnamefont {L.}~\bibnamefont
  {Susskind}}\ and\ \bibinfo {author} {\bibfnamefont {J.}~\bibnamefont
  {Glogower}},\ }\href {\doibase
  https://doi.org/10.1103/PhysicsPhysiqueFizika.1.49} {\bibfield  {journal}
  {\bibinfo  {journal} {Phys. Phys. Fiz.}\ }\textbf {\bibinfo {volume} {1}},\
  \bibinfo {pages} {49} (\bibinfo {year} {1964})}\BibitemShut {NoStop}%
\bibitem [{\citenamefont {Veis}\ \emph {et~al.}(2016)\citenamefont {Veis},
  \citenamefont {Vi{\v{s}}{\v{n}}{\'a}k}, \citenamefont {Nishizawa},
  \citenamefont {Nakai},\ and\ \citenamefont {Pittner}}]{veis2016quantum}%
  \BibitemOpen
  \bibfield  {author} {\bibinfo {author} {\bibfnamefont {L.}~\bibnamefont
  {Veis}}, \bibinfo {author} {\bibfnamefont {J.}~\bibnamefont
  {Vi{\v{s}}{\v{n}}{\'a}k}}, \bibinfo {author} {\bibfnamefont {H.}~\bibnamefont
  {Nishizawa}}, \bibinfo {author} {\bibfnamefont {H.}~\bibnamefont {Nakai}}, \
  and\ \bibinfo {author} {\bibfnamefont {J.}~\bibnamefont {Pittner}},\ }\href
  {\doibase https://doi.org/10.1002/qua.25176} {\bibfield  {journal} {\bibinfo
  {journal} {Int. J. Quantum Chem.}\ }\textbf {\bibinfo {volume} {116}},\
  \bibinfo {pages} {1328} (\bibinfo {year} {2016})}\BibitemShut {NoStop}%
\bibitem [{\citenamefont {McArdle}\ \emph
  {et~al.}(2019{\natexlab{b}})\citenamefont {McArdle}, \citenamefont {Mayorov},
  \citenamefont {Shan}, \citenamefont {Benjamin},\ and\ \citenamefont
  {Yuan}}]{mcardle2018quantum}%
  \BibitemOpen
  \bibfield  {author} {\bibinfo {author} {\bibfnamefont {S.}~\bibnamefont
  {McArdle}}, \bibinfo {author} {\bibfnamefont {A.}~\bibnamefont {Mayorov}},
  \bibinfo {author} {\bibfnamefont {X.}~\bibnamefont {Shan}}, \bibinfo {author}
  {\bibfnamefont {S.}~\bibnamefont {Benjamin}}, \ and\ \bibinfo {author}
  {\bibfnamefont {X.}~\bibnamefont {Yuan}},\ }\href {\doibase
  10.1039/c9sc01313j} {\bibfield  {journal} {\bibinfo  {journal} {Chem. Sci.}\
  }\textbf {\bibinfo {volume} {10}},\ \bibinfo {pages} {5725} (\bibinfo {year}
  {2019}{\natexlab{b}})}\BibitemShut {NoStop}%
\bibitem [{\citenamefont {Sawaya}\ \emph
  {et~al.}(2020{\natexlab{a}})\citenamefont {Sawaya}, \citenamefont {Menke},
  \citenamefont {Kyaw}, \citenamefont {Johri}, \citenamefont {Aspuru-Guzik},\
  and\ \citenamefont {Guerreschi}}]{nicolas2019}%
  \BibitemOpen
  \bibfield  {author} {\bibinfo {author} {\bibfnamefont {N.~P.~D.}\
  \bibnamefont {Sawaya}}, \bibinfo {author} {\bibfnamefont {T.}~\bibnamefont
  {Menke}}, \bibinfo {author} {\bibfnamefont {T.~H.}\ \bibnamefont {Kyaw}},
  \bibinfo {author} {\bibfnamefont {S.}~\bibnamefont {Johri}}, \bibinfo
  {author} {\bibfnamefont {A.}~\bibnamefont {Aspuru-Guzik}}, \ and\ \bibinfo
  {author} {\bibfnamefont {G.~G.}\ \bibnamefont {Guerreschi}},\ }\href
  {\doibase 10.1038/s41534-020-0278-0} {\bibfield  {journal} {\bibinfo
  {journal} {npj Quantum Inf.}\ }\textbf {\bibinfo {volume} {6}},\ \bibinfo
  {pages} {49} (\bibinfo {year} {2020}{\natexlab{a}})}\BibitemShut {NoStop}%
\bibitem [{\citenamefont {Rivest}\ \emph {et~al.}(1978)\citenamefont {Rivest},
  \citenamefont {Shamir},\ and\ \citenamefont {Adleman}}]{Rivest.1978.120}%
  \BibitemOpen
  \bibfield  {author} {\bibinfo {author} {\bibfnamefont {R.~L.}\ \bibnamefont
  {Rivest}}, \bibinfo {author} {\bibfnamefont {A.}~\bibnamefont {Shamir}}, \
  and\ \bibinfo {author} {\bibfnamefont {L.}~\bibnamefont {Adleman}},\ }\href
  {\doibase 10.1145/359340.359342} {\bibfield  {journal} {\bibinfo  {journal}
  {Commun. ACM}\ }\textbf {\bibinfo {volume} {21}},\ \bibinfo {pages} {120}
  (\bibinfo {year} {1978})}\BibitemShut {NoStop}%
\bibitem [{\citenamefont {Mart{\'\i}n-L{\'o}pez}\ \emph
  {et~al.}(2012)\citenamefont {Mart{\'\i}n-L{\'o}pez}, \citenamefont {Laing},
  \citenamefont {Lawson}, \citenamefont {Alvarez}, \citenamefont {Zhou},\ and\
  \citenamefont {O'Brien}}]{MartinLopez.2012.773}%
  \BibitemOpen
  \bibfield  {author} {\bibinfo {author} {\bibfnamefont {E.}~\bibnamefont
  {Mart{\'\i}n-L{\'o}pez}}, \bibinfo {author} {\bibfnamefont {A.}~\bibnamefont
  {Laing}}, \bibinfo {author} {\bibfnamefont {T.}~\bibnamefont {Lawson}},
  \bibinfo {author} {\bibfnamefont {R.}~\bibnamefont {Alvarez}}, \bibinfo
  {author} {\bibfnamefont {X.-Q.}\ \bibnamefont {Zhou}}, \ and\ \bibinfo
  {author} {\bibfnamefont {J.~L.}\ \bibnamefont {O'Brien}},\ }\href {\doibase
  10.1038/nphoton.2012.259} {\bibfield  {journal} {\bibinfo  {journal} {Nat.
  Photon.}\ }\textbf {\bibinfo {volume} {6}},\ \bibinfo {pages} {773} (\bibinfo
  {year} {2012})}\BibitemShut {NoStop}%
\bibitem [{\citenamefont {Jiang}\ \emph {et~al.}(2018)\citenamefont {Jiang},
  \citenamefont {Britt}, \citenamefont {McCaskey}, \citenamefont {Humble},\
  and\ \citenamefont {Kais}}]{Jiang.2018.17667}%
  \BibitemOpen
  \bibfield  {author} {\bibinfo {author} {\bibfnamefont {S.}~\bibnamefont
  {Jiang}}, \bibinfo {author} {\bibfnamefont {K.~A.}\ \bibnamefont {Britt}},
  \bibinfo {author} {\bibfnamefont {A.~J.}\ \bibnamefont {McCaskey}}, \bibinfo
  {author} {\bibfnamefont {T.~S.}\ \bibnamefont {Humble}}, \ and\ \bibinfo
  {author} {\bibfnamefont {S.}~\bibnamefont {Kais}},\ }\href {\doibase
  10.1038/s41598-018-36058-z} {\bibfield  {journal} {\bibinfo  {journal} {Sci.
  Rep.}\ }\textbf {\bibinfo {volume} {8}},\ \bibinfo {pages} {17667} (\bibinfo
  {year} {2018})}\BibitemShut {NoStop}%
\bibitem [{\citenamefont {Dash}\ \emph {et~al.}(2018)\citenamefont {Dash},
  \citenamefont {Sarmah}, \citenamefont {Behera},\ and\ \citenamefont
  {Panigrahi}}]{Dash.2018.1805.10478v2}%
  \BibitemOpen
  \bibfield  {author} {\bibinfo {author} {\bibfnamefont {A.}~\bibnamefont
  {Dash}}, \bibinfo {author} {\bibfnamefont {D.}~\bibnamefont {Sarmah}},
  \bibinfo {author} {\bibfnamefont {B.~K.}\ \bibnamefont {Behera}}, \ and\
  \bibinfo {author} {\bibfnamefont {P.~K.}\ \bibnamefont {Panigrahi}},\ }\href
  {https://arxiv.org/abs/1805.10478} {\bibfield  {journal} {\bibinfo  {journal}
  {arXiv:1805.10478v2}\ } (\bibinfo {year} {2018})}\BibitemShut {NoStop}%
\bibitem [{\citenamefont {Anschuetz}\ \emph {et~al.}(2018)\citenamefont
  {Anschuetz}, \citenamefont {Olson}, \citenamefont {Aspuru-Guzik},\ and\
  \citenamefont {Cao}}]{Anschuetz.2018.1808.08927v1}%
  \BibitemOpen
  \bibfield  {author} {\bibinfo {author} {\bibfnamefont {E.~R.}\ \bibnamefont
  {Anschuetz}}, \bibinfo {author} {\bibfnamefont {J.~P.}\ \bibnamefont
  {Olson}}, \bibinfo {author} {\bibfnamefont {A.}~\bibnamefont {Aspuru-Guzik}},
  \ and\ \bibinfo {author} {\bibfnamefont {Y.}~\bibnamefont {Cao}},\ }\href
  {https://arxiv.org/abs/1808.08927} {\bibfield  {journal} {\bibinfo  {journal}
  {arXiv:1808.08927v1}\ } (\bibinfo {year} {2018})}\BibitemShut {NoStop}%
\bibitem [{\citenamefont {Karamlou}\ \emph {et~al.}(2021)\citenamefont
  {Karamlou}, \citenamefont {Simon}, \citenamefont {Katabarwa}, \citenamefont
  {Scholten}, \citenamefont {Peropadre},\ and\ \citenamefont
  {Cao}}]{karamlou2021analyzing}%
  \BibitemOpen
  \bibfield  {author} {\bibinfo {author} {\bibfnamefont {A.~H.}\ \bibnamefont
  {Karamlou}}, \bibinfo {author} {\bibfnamefont {W.~A.}\ \bibnamefont {Simon}},
  \bibinfo {author} {\bibfnamefont {A.}~\bibnamefont {Katabarwa}}, \bibinfo
  {author} {\bibfnamefont {T.~L.}\ \bibnamefont {Scholten}}, \bibinfo {author}
  {\bibfnamefont {B.}~\bibnamefont {Peropadre}}, \ and\ \bibinfo {author}
  {\bibfnamefont {Y.}~\bibnamefont {Cao}},\ }\href {\doibase
  10.1038/s41534-021-00478-z} {\bibfield  {journal} {\bibinfo  {journal} {npj
  Quantum Inf.}\ }\textbf {\bibinfo {volume} {7}},\ \bibinfo {pages} {1}
  (\bibinfo {year} {2021})}\BibitemShut {NoStop}%
\bibitem [{\citenamefont {Liu}(2014)}]{liu2014exact}%
  \BibitemOpen
  \bibfield  {author} {\bibinfo {author} {\bibfnamefont {Y.}~\bibnamefont
  {Liu}},\ }\href {\doibase 10.1007/s10773-014-2055-3} {\bibfield  {journal}
  {\bibinfo  {journal} {Int. J. Theor. Phys.}\ }\textbf {\bibinfo {volume}
  {53}},\ \bibinfo {pages} {2571} (\bibinfo {year} {2014})}\BibitemShut
  {NoStop}%
\bibitem [{\citenamefont {Sawaya}\ \emph
  {et~al.}(2020{\natexlab{b}})\citenamefont {Sawaya}, \citenamefont {Menke},
  \citenamefont {Kyaw}, \citenamefont {Johri}, \citenamefont {Aspuru-Guzik},\
  and\ \citenamefont {Guerreschi}}]{sawaya2020resource}%
  \BibitemOpen
  \bibfield  {author} {\bibinfo {author} {\bibfnamefont {N.~P.}\ \bibnamefont
  {Sawaya}}, \bibinfo {author} {\bibfnamefont {T.}~\bibnamefont {Menke}},
  \bibinfo {author} {\bibfnamefont {T.~H.}\ \bibnamefont {Kyaw}}, \bibinfo
  {author} {\bibfnamefont {S.}~\bibnamefont {Johri}}, \bibinfo {author}
  {\bibfnamefont {A.}~\bibnamefont {Aspuru-Guzik}}, \ and\ \bibinfo {author}
  {\bibfnamefont {G.~G.}\ \bibnamefont {Guerreschi}},\ }\href {\doibase
  https://doi.org/10.1038/s41534-020-0278-0} {\bibfield  {journal} {\bibinfo
  {journal} {npj Quantum Inf.}\ }\textbf {\bibinfo {volume} {6}},\ \bibinfo
  {pages} {1} (\bibinfo {year} {2020}{\natexlab{b}})}\BibitemShut {NoStop}%
\bibitem [{\citenamefont {Gomes}\ \emph {et~al.}(2021)\citenamefont {Gomes},
  \citenamefont {Mukherjee}, \citenamefont {Zhang}, \citenamefont {Iadecola},
  \citenamefont {Wang}, \citenamefont {Ho}, \citenamefont {Orth},\ and\
  \citenamefont {Yao}}]{Gomes2021adaptive}%
  \BibitemOpen
  \bibfield  {author} {\bibinfo {author} {\bibfnamefont {N.}~\bibnamefont
  {Gomes}}, \bibinfo {author} {\bibfnamefont {A.}~\bibnamefont {Mukherjee}},
  \bibinfo {author} {\bibfnamefont {F.}~\bibnamefont {Zhang}}, \bibinfo
  {author} {\bibfnamefont {T.}~\bibnamefont {Iadecola}}, \bibinfo {author}
  {\bibfnamefont {C.-Z.}\ \bibnamefont {Wang}}, \bibinfo {author}
  {\bibfnamefont {K.-M.}\ \bibnamefont {Ho}}, \bibinfo {author} {\bibfnamefont
  {P.~P.}\ \bibnamefont {Orth}}, \ and\ \bibinfo {author} {\bibfnamefont
  {Y.-X.}\ \bibnamefont {Yao}},\ }\href {\doibase 10.1002/qute.202100114}
  {\bibfield  {journal} {\bibinfo  {journal} {Adv. Quantum Technol.}\ }\textbf
  {\bibinfo {volume} {4}},\ \bibinfo {pages} {2100114} (\bibinfo {year}
  {2021})}\BibitemShut {NoStop}%
\bibitem [{\citenamefont {Zhang}\ \emph {et~al.}(2020)\citenamefont {Zhang},
  \citenamefont {Sun}, \citenamefont {Yuan},\ and\ \citenamefont
  {Yung}}]{Zhang2020low}%
  \BibitemOpen
  \bibfield  {author} {\bibinfo {author} {\bibfnamefont {Z.-J.}\ \bibnamefont
  {Zhang}}, \bibinfo {author} {\bibfnamefont {J.}~\bibnamefont {Sun}}, \bibinfo
  {author} {\bibfnamefont {X.}~\bibnamefont {Yuan}}, \ and\ \bibinfo {author}
  {\bibfnamefont {M.-H.}\ \bibnamefont {Yung}},\ }\href
  {10.48550/arXiv.2011.05283} {\bibfield  {journal} {\bibinfo  {journal}
  {arXiv2011.05283}\ } (\bibinfo {year} {2020})}\BibitemShut {NoStop}%
\bibitem [{\citenamefont {Loken}\ \emph {et~al.}(2010)\citenamefont {Loken},
  \citenamefont {Gruner}, \citenamefont {Groer}, \citenamefont {Peltier},
  \citenamefont {Bunn}, \citenamefont {Craig}, \citenamefont {Henriques},
  \citenamefont {Dempsey}, \citenamefont {Yu}, \citenamefont {Chen} \emph
  {et~al.}}]{loken2010scinet}%
  \BibitemOpen
  \bibfield  {author} {\bibinfo {author} {\bibfnamefont {C.}~\bibnamefont
  {Loken}}, \bibinfo {author} {\bibfnamefont {D.}~\bibnamefont {Gruner}},
  \bibinfo {author} {\bibfnamefont {L.}~\bibnamefont {Groer}}, \bibinfo
  {author} {\bibfnamefont {R.}~\bibnamefont {Peltier}}, \bibinfo {author}
  {\bibfnamefont {N.}~\bibnamefont {Bunn}}, \bibinfo {author} {\bibfnamefont
  {M.}~\bibnamefont {Craig}}, \bibinfo {author} {\bibfnamefont
  {T.}~\bibnamefont {Henriques}}, \bibinfo {author} {\bibfnamefont
  {J.}~\bibnamefont {Dempsey}}, \bibinfo {author} {\bibfnamefont {C.-H.}\
  \bibnamefont {Yu}}, \bibinfo {author} {\bibfnamefont {J.}~\bibnamefont
  {Chen}},  \emph {et~al.},\ }\href {\doibase
  https://doi.org/10.1088/1742-6596/256/1/012026} {\bibfield  {journal}
  {\bibinfo  {journal} {J. Phys.: Conf. Ser.}\ }\textbf {\bibinfo {volume}
  {256}},\ \bibinfo {pages} {012026} (\bibinfo {year} {2010})}\BibitemShut
  {NoStop}%
\bibitem [{\citenamefont {Ponce}\ \emph {et~al.}(2019)\citenamefont {Ponce},
  \citenamefont {van Zon}, \citenamefont {Northrup}, \citenamefont {Gruner},
  \citenamefont {Chen}, \citenamefont {Ertinaz}, \citenamefont {Fedoseev},
  \citenamefont {Groer}, \citenamefont {Mao}, \citenamefont {Mundim} \emph
  {et~al.}}]{ponce2019deploying}%
  \BibitemOpen
  \bibfield  {author} {\bibinfo {author} {\bibfnamefont {M.}~\bibnamefont
  {Ponce}}, \bibinfo {author} {\bibfnamefont {R.}~\bibnamefont {van Zon}},
  \bibinfo {author} {\bibfnamefont {S.}~\bibnamefont {Northrup}}, \bibinfo
  {author} {\bibfnamefont {D.}~\bibnamefont {Gruner}}, \bibinfo {author}
  {\bibfnamefont {J.}~\bibnamefont {Chen}}, \bibinfo {author} {\bibfnamefont
  {F.}~\bibnamefont {Ertinaz}}, \bibinfo {author} {\bibfnamefont
  {A.}~\bibnamefont {Fedoseev}}, \bibinfo {author} {\bibfnamefont
  {L.}~\bibnamefont {Groer}}, \bibinfo {author} {\bibfnamefont
  {F.}~\bibnamefont {Mao}}, \bibinfo {author} {\bibfnamefont {B.~C.}\
  \bibnamefont {Mundim}},  \emph {et~al.},\ }\href {\doibase
  https://doi.org/10.1145/3332186.3332195} {\bibfield  {journal} {\bibinfo
  {journal} {Proceedings of the Practice and Experience in Advanced Research
  Computing on Rise of the Machines (learning)}\ }\textbf {\bibinfo {volume}
  {34}},\ \bibinfo {pages} {1} (\bibinfo {year} {2019})}\BibitemShut {NoStop}%
\bibitem [{Note1()}]{Note1}%
  \BibitemOpen
  \bibinfo {note} {In practice, the Hamiltonian may be expressed as the
  weighted sum $H=\DOTSB \sum@ \slimits@ _i\lambda _i h_i$ with real
  coefficients $\lambda _i$ and tensor products of Pauli matrices $h_i$, since
  Pauli matrices form a complete basis.}\BibitemShut {Stop}%
\bibitem [{\citenamefont {McLachlan}(1964)}]{mclachlan1964variational}%
  \BibitemOpen
  \bibfield  {author} {\bibinfo {author} {\bibfnamefont {A.}~\bibnamefont
  {McLachlan}},\ }\href {\doibase https://doi.org/10.1080/00268976400100041}
  {\bibfield  {journal} {\bibinfo  {journal} {Mol. Phys.}\ }\textbf {\bibinfo
  {volume} {8}},\ \bibinfo {pages} {39} (\bibinfo {year} {1964})}\BibitemShut
  {NoStop}%
\bibitem [{\citenamefont {Broeckhove}\ \emph {et~al.}(1988)\citenamefont
  {Broeckhove}, \citenamefont {Lathouwers}, \citenamefont {Kesteloot},\ and\
  \citenamefont {Van~Leuven}}]{broeckhove1988equivalence}%
  \BibitemOpen
  \bibfield  {author} {\bibinfo {author} {\bibfnamefont {J.}~\bibnamefont
  {Broeckhove}}, \bibinfo {author} {\bibfnamefont {L.}~\bibnamefont
  {Lathouwers}}, \bibinfo {author} {\bibfnamefont {E.}~\bibnamefont
  {Kesteloot}}, \ and\ \bibinfo {author} {\bibfnamefont {P.}~\bibnamefont
  {Van~Leuven}},\ }\href {\doibase
  https://doi.org/10.1016/0009-2614(88)80380-4} {\bibfield  {journal} {\bibinfo
   {journal} {Chem. Phys. Lett.}\ }\textbf {\bibinfo {volume} {149}},\ \bibinfo
  {pages} {547} (\bibinfo {year} {1988})}\BibitemShut {NoStop}%
\bibitem [{\citenamefont {Poulin}\ \emph {et~al.}(2011)\citenamefont {Poulin},
  \citenamefont {Qarry}, \citenamefont {Somma},\ and\ \citenamefont
  {Verstraete}}]{poulin2011quantum}%
  \BibitemOpen
  \bibfield  {author} {\bibinfo {author} {\bibfnamefont {D.}~\bibnamefont
  {Poulin}}, \bibinfo {author} {\bibfnamefont {A.}~\bibnamefont {Qarry}},
  \bibinfo {author} {\bibfnamefont {R.}~\bibnamefont {Somma}}, \ and\ \bibinfo
  {author} {\bibfnamefont {F.}~\bibnamefont {Verstraete}},\ }\href {\doibase
  10.1103/PhysRevLett.106.170501} {\bibfield  {journal} {\bibinfo  {journal}
  {Phys. Rev. Lett.}\ }\textbf {\bibinfo {volume} {106}},\ \bibinfo {pages}
  {170501} (\bibinfo {year} {2011})}\BibitemShut {NoStop}%
\bibitem [{\citenamefont {Hestenes}\ and\ \citenamefont
  {Stiefel}(1952)}]{hestenes1952methods}%
  \BibitemOpen
  \bibfield  {author} {\bibinfo {author} {\bibfnamefont {M.~R.}\ \bibnamefont
  {Hestenes}}\ and\ \bibinfo {author} {\bibfnamefont {E.}~\bibnamefont
  {Stiefel}},\ }\href@noop {} {\bibfield  {journal} {\bibinfo  {journal} {J.
  Res. Natl. Bur. Stand. (U. S.)}\ }\textbf {\bibinfo {volume} {49}},\ \bibinfo
  {pages} {409} (\bibinfo {year} {1952})}\BibitemShut {NoStop}%
\bibitem [{\citenamefont {Ekert}\ \emph {et~al.}(2002)\citenamefont {Ekert},
  \citenamefont {Alves}, \citenamefont {Oi}, \citenamefont {Horodecki},
  \citenamefont {Horodecki},\ and\ \citenamefont {Kwek}}]{ekert2002direct}%
  \BibitemOpen
  \bibfield  {author} {\bibinfo {author} {\bibfnamefont {A.~K.}\ \bibnamefont
  {Ekert}}, \bibinfo {author} {\bibfnamefont {C.~M.}\ \bibnamefont {Alves}},
  \bibinfo {author} {\bibfnamefont {D.~K.}\ \bibnamefont {Oi}}, \bibinfo
  {author} {\bibfnamefont {M.}~\bibnamefont {Horodecki}}, \bibinfo {author}
  {\bibfnamefont {P.}~\bibnamefont {Horodecki}}, \ and\ \bibinfo {author}
  {\bibfnamefont {L.~C.}\ \bibnamefont {Kwek}},\ }\href {\doibase
  https://doi.org/10.1103/PhysRevLett.88.217901} {\bibfield  {journal}
  {\bibinfo  {journal} {Phys. Rev. Lett.}\ }\textbf {\bibinfo {volume} {88}},\
  \bibinfo {pages} {217901} (\bibinfo {year} {2002})}\BibitemShut {NoStop}%
\end{thebibliography}%

\onecolumngrid
\appendix
\setcounter{equation}{0}
\setcounter{table}{0}
\makeatletter

\renewcommand{\theequation}{A\arabic{equation}}
\renewcommand{\thefigure}{\arabic{figure}}

\section{Quantum Iterative Power Algorithm}\label{Appen:qipa}
This section outlines a quantum version of a generalized IPA method~\cite{soley2021iterative}, the so-called quantum iterative power algorithm (QIPA) inspired by the imaginary time propagation (Appendix~\ref{Appen:qite}) and the variational quantum simulator~\cite{li2017efficient}. 
We introduce a series of oracles defined by concatenated exponential functions which evolve the initial state according to a Wick-like-rotated Schr\"odinger equation obtained from a generalized McLachlan's variational principle. Then, we focus on the particular case of a double-exponential function that provides an efficient implementation of QIPA. For completeness, we include a section introducing the quantum imaginary time evolution (QITE) method obtained from the McLachlan's variational principle.

\subsection{Quantum Imaginary Time Evolution}\label{Appen:qite}
\subsubsection{McLachlan Variational Principle Approach}
Let us consider a many-body system subject to the Hermitian Hamiltonian $H$ \footnote{In practice, the Hamiltonian may be expressed as the weighted sum $H=\sum_i\lambda_i h_i$ with real coefficients $\lambda_i$ and tensor products of Pauli matrices $h_i$, since Pauli matrices form a complete basis.}. Given an initial state $\ket{\psi(0)}$, non-unitary quantum imaginary time evolution is defined by
\begin{align}
	\ket{\psi(\tau)}
	=\frac{e^{- H\tau}\ket{\psi(0)}}{\sqrt{\bra{\psi(0)}e^{-2 H\tau}\ket{\psi(0)}}}.
\end{align}
Note that the denominator is a normalization factor. Since a key to our proposal is the realization that there is nothing except the requirement that there be a continuous, integrable, strictly positive oracle $U(x)$ that is maximized at the global minima of $H(x)=V(x)$ to prevent us from assuming a particular form of the oracle; we present quantum imaginary time evolution more generally in terms of non-unitary time evolution according to an oracle, as follows:
\begin{align}
	\ket{\psi(\tau)}
	=\frac{\alpha(- H\tau)\ket{\psi(0)}}{\sqrt{\bra{\psi(0)}\alpha(-2 H\tau)\ket{\psi(0)}}},
\end{align}
where $\alpha$ could be any strictly increasing positive function. When $\alpha(y)=e^y$, we recover imaginary time evolution, which corresponds directly to the Wick rotation $(\tau=-it)$. With that choice of $\alpha(y)$, we obtain that the above quantum state satisfies the Wick-like rotated Schr\"odinger equation:
\begin{align}\label{eq:qite_EOM}
	\frac{\partial\ket{\psi(\tau)}}{\partial\tau}
	=-\left(H-E_1(\tau)\right)\ket{\psi(\tau)},
\end{align}
where $E_1(\tau)=\bra{\psi(\tau)}H\ket{\psi(\tau)}$. Even though $\ket{\psi(\tau)}$ is a valid wavefunction that can be represented in a  quantum computer, the non-unitary time evolution cannot be straightforwardly mapped to a quantum circuit based solely on unitary gates. Here, the McLachlan's variational principle \cite{mclachlan1964variational,broeckhove1988equivalence,mcardle2019variational,yuan2019theory} comes to the rescue and demands that 
\begin{align}
	\delta\left\|(\partial/\partial\tau+ [H-E_1(\tau)])\ket{\psi(\tau)}\right\|
	=0,
\end{align}
with $\|\cdot\|$ representing the $L^2$-norm and $\delta$ the functional derivative. In the following, we intend to simulate the action of non-unitary dynamics, \eqref{eq:qite_EOM}, on a  quantum computer via McLachlan's variational principle.

In variational quantum simulations, instead of directly encoding the quantum state $\ket{\psi(\tau)}$ at time $\tau$, we approximate it with a parameterized quantum circuit $\ket{\psi(\tau)}\approx\ket{\phi(\theta(\tau))}$ with a real-valued parameter vector $\theta(\tau)=(\theta_1 (\tau),\theta_2 (\tau),\dots,\theta_{\mathcal{N}_\theta}(\tau))$. We assume that physically relevant quantum states span a restricted region of the full Hilbert space \cite{poulin2011quantum} for a given time interval, such that the trial state parameterized by $\theta$ is sufficient to prepare a desired quantum state by applying a sequence of parameterized unitary gates $U(\theta)=U_{\mathcal{N}_\theta}(\theta_{\mathcal{N}_\theta})\cdots U_k(\theta_k)\cdots U_1(\theta_1)$ to the initial state $\ket{\Bar{0}}=\ket{0\cdots 0}$. Thus, we have $\ket{\phi(\theta)}=U(\theta)\ket{\Bar{0}}$, where $U(\theta)$ is referred to as the ansatz, and $U_k (\theta_k)$ is the $k$th unitary gate controlled by classical parameter $\theta_k$. Here, we are only concerned with single- or two-qubit gates, which is sufficient for universal quantum computing.\\

According to McLachlan's variational principle, we require $\partial\|(\partial/\partial\tau+(H-E_\tau))\ket{\phi(\theta(\tau))}\|/\partial\Dot{\theta}_k=0$. We have
\begin{align}
	\delta\|(\partial/\partial\tau+[H-E_1(\tau)])\ket{\phi(\theta(\tau))}\|^2
	&=\sum_{m,n}\frac{\partial\bra{\phi(\theta(\tau))}}{\partial\theta_m}\frac{\partial\ket{\phi(\theta(\tau))}}{\partial\theta_n}\Dot{\theta}_m\Dot{\theta}_n+\bra{\phi(\theta(\tau))}^2(H-E_1(\tau))^2\ket{\phi(\theta(\tau))}\dots\nonumber\\
	&\qquad+\sum_m\frac{\partial\bra{\phi(\theta(\tau))}}{\partial\theta_m}(H-E_1(\tau))\ket{\phi(\theta(\tau))}\Dot{\theta}_m\dots\\
	&\qquad+\sum_m\bra{\phi(\theta(\tau))}(H-E_1(\tau))\frac{\partial\ket{\phi(\theta(\tau))}}{\partial\theta_m}\Dot{\theta}_m.\nonumber
\end{align}
By differentiating with respect to $\Dot{\theta}_k$, we obtain
\begin{align}
	\partial\|(\partial/\partial\tau+[H-E_1(\tau)])\ket{\phi(\theta(\tau))}\|^2/\partial\Dot{\theta}_k
	&=\sum_m\left(\frac{\partial\bra{\phi(\theta(\tau))}}{\partial\theta_k}\frac{\partial\ket{\phi(\theta(\tau))}}{\partial\theta_m}+\frac{\partial\bra{\phi(\theta(\tau))}}{\partial\theta_m}\frac{\partial\ket{\phi(\theta(\tau))}}{\partial\theta_k}\right)\Dot{\theta}_m\dots\\
	&\qquad+\frac{\partial\bra{\phi(\theta(\tau)))}}{\partial\theta_k} H\ket{\phi(\theta(\tau))}+\bra{\phi(\theta(\tau))} H\frac{\partial\ket{\phi(\theta(\tau))}}{\partial\theta_k},\nonumber
\end{align}
where we use $\braket{\phi(\theta(\tau))}{\phi(\theta(\tau))}=1$. Finally, we conclude that 
\begin{align}
	\partial\|(\partial/\partial\tau+[H-E_1(\tau)])\ket{\phi(\theta(\tau))}\|/\partial\Dot{\theta}_k
	=0
\end{align}
is equivalent to the following linear equation
\begin{align}\label{eq:QITEExpression}
	\sum_{m}A_{k,m}\Dot{\theta}_m
	=C_k
\end{align}
with 
\begin{align}\label{eq:AandC}
	A_{k,m}
	=\textrm{Re}\left(\frac{\partial\bra{\phi(\theta(\tau))}}{\partial\theta_k}\frac{\partial\ket{\phi(\theta(\tau))}}{\partial\theta_m}\right)
	\quad\text{and}\quad
	C_k
	=-\textrm{\textcolor{black}{Re}}\left(\frac{\partial\bra{\phi(\theta(\tau))}}{\partial\theta_k}H\ket{\phi(\theta(\tau))}\right).
\end{align}
We note that
\begin{align}
	\frac{\partial\bra{\phi(\theta(\tau))}}{\partial\theta_k}H\ket{\phi(\theta(\tau))}
	=\sum_\alpha\lambda_\alpha\frac{\partial\bra{\phi(\theta(\tau))}}{\partial\theta_k}h_\alpha\ket{\phi(\theta(\tau))},
\end{align}
where the $h_\alpha$ are Pauli matrices and $\lambda_\alpha$ are corresponding coefficients. Hence, quantum imaginary time evolution reduces to solving the linear equation $A \Dot{\theta}= C$ for $\Dot{\theta}$ which could be accomplished by inversion of the matrix $A$, as follows $\Dot{\theta}= A^{-1}C$. In our numerical simulations, however, we solve the linear equation via the conjugate gradient (CG) method \cite{hestenes1952methods} using a subroutine from the SciPy library with a tolerance for convergence of $10^{-6}$. That approach by-passes the need of inverting the matrix $A$.

\subsubsection{Quantum circuit evaluations of \texorpdfstring{$A$}{A} and \texorpdfstring{$C$}{C}}\label{sec:QuantumCircuitEvaluation}
\begin{figure}[t]
	\centering
	\includegraphics[scale=1.0]{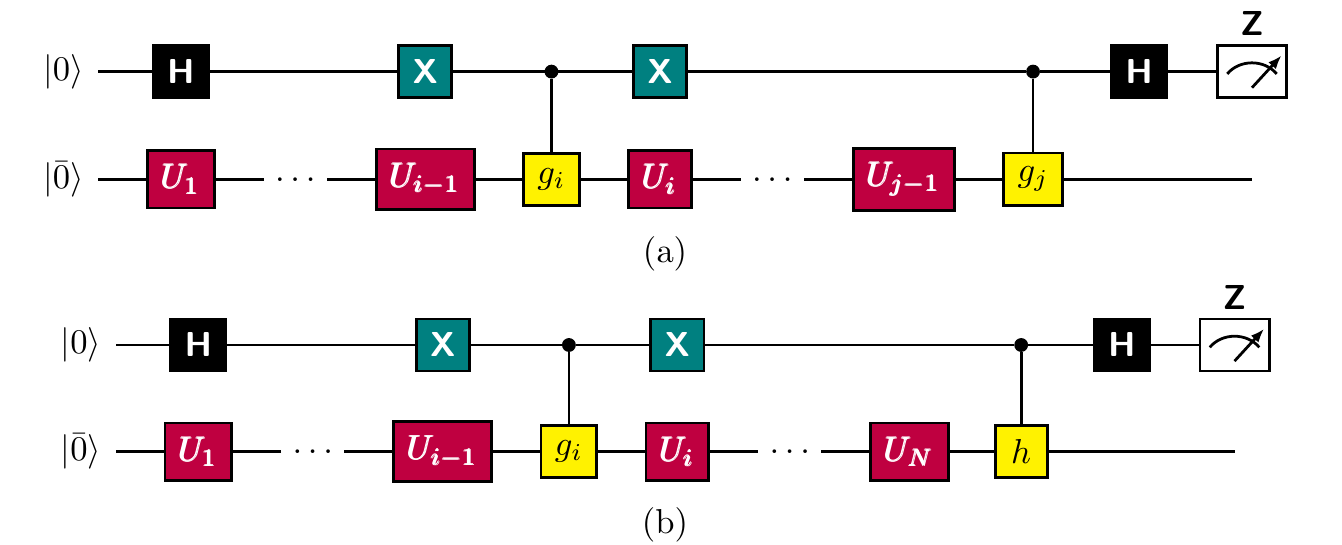}
	\caption{(a) Quantum circuit to evaluate $\textrm{Re}(\bra{\bar{0}}\bar{U}^\dagger_{i}\bar{U}_{j}\ket{\bar{0}})$ as a probability of finding the ancillary qubit in $0$. (b) Quantum circuit to evaluate $\textrm{Re}(\bra{\bar{0}}\bar{U}_{i}^\dagger h U\ket{\bar{0}} )$ as a probability of finding the ancillary qubit in $0$.}\label{fig:Hadamard_test_circuit}
\end{figure}
We efficiently evaluate the components of $A$ and $C$ following Refs.~\cite{romero2018strategies,ekert2002direct,mcardle2019variational} by implementing the Hadamard test with an additional ancilla qubit. Recall that $\ket{\phi(\theta)}=U_{\mathcal{N}_\theta}(\theta_{\mathcal{N}_\theta})\cdots U_1 (\theta_1)\ket{\bar{0}}$ and that in the variational ansatz circuit we are only concerned with single- and two-qubit unitary gates $U_n(\theta_n)$, namely rotational or controlled rotational gates. The required derivatives of the ansatz wavefunction are then determined as follows.

Suppose $U_n(\theta_n)$ is a single-qubit rotational gate $R_{\theta_n}^{Z}=e^{-i\theta_nZ/2}$, with derivative $\partial U_n(\theta_n)/\partial\theta_n=-(i/2)\times Ze^{-i\theta_n Z/2}$. If $U_n(\theta_n)$ is a two-qubit controlled rotational gate $\ketbra{0}{0}\otimes I+\ketbra{1}{1}\otimes R^{Z} _{\theta_n}$, the derivative is given by $\partial U_n(\theta_n)/\partial\theta_n =\ketbra{1}{1}\otimes\partial R_{\theta_n}^{Z}/\partial\theta_n=(-i/2)\times\ketbra{1}{1}\otimes Z e^{-i\theta_n Z/2}$. In our numerical simulations, we do not consider parameterized two-qubit gates for simplicity. 
Using the notation 
\begin{align}
	\bar{U}_{n}
	&=U_{\mathcal{N}_\theta}(\theta_{\mathcal{N}_\theta})\cdots U_{n+1}(\theta_{n+1})U_n (\theta_n)g_{n}U_{n-1}(\theta_{n-1})\cdots U_2(\theta_2)U_1(\theta_1),
\end{align}
we conclude that
\begin{align}
	\bar{U}_{n}
	&=U_{\mathcal{N}_\theta}(\theta_{\mathcal{N}_\theta})\cdots U_{n+1}(\theta_{n+1})\left(2i\times\partial U_n(\theta_n)/\partial\theta_n\right)U_{n-1}(\theta_{n-1})\cdots U_2(\theta_2)U_1(\theta_1).
\end{align}
This implies
\begin{align}
	\frac{\partial\ket{\phi(\theta(\tau))}}{\partial\theta_n}
	=(-i/2) \bar{U}_{n}\ket{\bar{0}}.\label{eq:psi_derivative}
\end{align}
Consequently, we have
\begin{align}
	A_{k,m}
	=\textrm{Re}\left(\frac{\partial\bra{\phi(\theta(\tau))}}{\partial\theta_k}\frac{\partial\ket{\phi(\theta(\tau))}}{\partial\theta_m}\right)
	=\frac{1}{4}\textrm{Re}\left(\bra{\bar{0}}\bar{U}^\dagger_{k}\bar{U}_{m}\ket{\bar{0}}\right),
\end{align}
and
\begin{align}
	C_k
	=\textrm{\textcolor{black}{Re}}\left(-\frac{\partial\bra{\phi(\theta(\tau))}}{\partial\theta_k} H\ket{\phi(\theta(\tau))}\right)
	&=-\textrm{\textcolor{black}{Re}}\left(\sum_\alpha\lambda_\alpha\frac{\partial\bra{\phi(\theta(\tau))}}{\partial\theta_k}h_\alpha\ket{\phi(\theta(\tau))}\right)\\
	&=-\frac{1}{2}\textrm{\textcolor{black}{Re}}\left(i\sum_{\alpha}\lambda_\alpha\bra{\bar{0}}\bar{U}_{k}^\dagger h_\alpha U\ket{\bar{0}}\right).
\end{align}
Since we are evaluating $\textrm{Re}(\bra{\bar{0}}\bar{U}^\dagger_{k}\bar{U}_{m}\ket{\bar{0}})$ and $\textrm{Re}(\bra{\bar{0}}\bar{U}^\dagger_{k}h_\alpha {U}\ket{\bar{0}})$, one can implement them on a  quantum computer by carrying out the Hadamard tests shown in \figref{fig:Hadamard_test_circuit}.

\subsubsection{Parameter update}
We make use of the Euler method (First order Taylor series expansion) to update the variational parameters as 
\begin{align}
	\theta(\tau+\delta\tau)
	\simeq\theta(\tau)+\dot{\theta}(\tau)\delta\tau
	\approx\theta(\tau)+\xi_\tau\delta\tau,
\end{align}
where $\xi_\tau$ is the numerical solution to $A(\tau)\dot{\theta}(\tau)=C(\tau)$. One needs to repeat this procedure $\mathcal{N}_T=\tau_{\textrm{total}}/\delta\tau$ times to simulate imaginary-time-like evolution. The difference between the above parameter update and the gradient descent method is that the latter uses
\begin{align}
	\theta(\tau+\delta\tau)
	\simeq\theta(\tau)+C(\tau)\delta\tau,
\end{align}
which only considers information about the average energy at each time step without taking into account of information about the ansatz circuit itself.

\subsection{General formulation of a cooling function}
As mentioned in the main text, the main ingredient of QIPA is the choice of a suitable cooling function or oracle to quickly reach the optimal solution. Here, we show that oracles defined by the concatenated exponential functions introduced by Eq.~\eqref{eq:ExponentialFunctions} evolve the initial state according to the generalized Wick-like rotated Schr\"odinger equation introduced by Eq.~\eqref{eq:general}.

We apply the chain rule for derivatives to \eqref{eq:ExponentialFunctions} to obtain
\begin{align}
	\frac{\mathrm{d}}{\mathrm{d}y}\alpha_n(y)
	=\alpha_n(y)a_n\alpha'_{n-1}(y)
	&=\alpha_n(y)a_n\alpha_{n-1}(y)a_{n-1}\alpha'_{n-2}(y)\nonumber\\
	&=\alpha_n(y)a_n\alpha_{n-1}(y)a_{n-1}\alpha_{n-2}(y)\cdots a_1\alpha'_0(y)\nonumber\\[2mm]
	&=a_n\cdots a_{1}\exp\left(a_n\alpha_{n-1}(y)+\dots+a_1\alpha_0(y)\right)\nonumber\\
	&=\prod_{k=1}^na_k\exp\left(\sum_{k=1}^na_k\alpha_{k-1}(y)\right).
\end{align}
For a given initial wave function $\ket{\psi(0)}$, let us introduce the auxiliary functions
\begin{align}
	g_{1,n}(\tau)
	&=U_n(\tau)\ket{\psi(0)},\nonumber\\[2mm]
	g_{2,n}(\tau)
	&=\sqrt{\langle U_n(\tau)\psi(0)\mid U_n(\tau)\psi(0)\rangle}.
\end{align}
Note that the normalized time-evolved wavefunction $\ket{\psi(\tau)}$ can be written as
\begin{align}
	\ket{\psi(\tau)}
	&=\frac{g_{1,n}(\tau)}{g_{2,n}(\tau)}
	=\frac{U_n(\tau)\ket{\psi_0}}{\sqrt{\langle U_n(\tau)\psi(0)\mid U_n(\tau)\psi(0)\rangle}}.
\end{align}
For the derivatives of the above functions we obtain
\begin{align}
	\frac{\mathrm{d}}{\mathrm{d}\tau}g_{1,n}(\tau)
	&=\frac{\mathrm{d}}{\mathrm{d}\tau}U_n(\tau)\ket{\psi(0)}
	=- HU_n(\tau)a_n\alpha'_{n-1}(- H\tau)\ket{\psi(0)}
	=- Ha_n\alpha'_{n-1}(- H\tau)g_{1,n}(\tau),\nonumber\\[2mm]
	\frac{\mathrm{d}}{\mathrm{d}\tau}g_{2,n}(\tau)
	&=\frac{1}{2}\left(\left\langle U_n(\tau)\psi(0)\mid U_n(\tau)\psi(0)\right\rangle\right)^{-1/2}\left(\left\langle\frac{\mathrm{d}}{\mathrm{d}\tau}U_n(\tau)\psi(0)\mid U_n(\tau)\psi(0)\right\rangle+\left\langle U_n(\tau)\psi(0)\mid\frac{\mathrm{d}}{\mathrm{d}\tau}U_n(\tau)\psi(0)\right\rangle\right)\nonumber\\
	&=-\frac{1}{2}g_{2,n}(\tau)^{-1}a_n\left(\left\langle H\alpha'_{n-1}(- H\tau)U_n(\tau)\psi(0)\mid U_n(\tau)\psi(0)\right\rangle+\left\langle U_n(\tau)\psi(0)\mid H\alpha'_{n-1}(- H\tau)U_n(\tau)\psi(0)\right\rangle\right),\nonumber\\[2mm]
	\frac{\mathrm{d}}{\mathrm{d}\tau}\ket{\psi(\tau)}
	&=\frac{\mathrm{d}g_{1,n}(\tau)/\mathrm{d}\tau}{g_{2,n}(\tau)}-\frac{g_{1,n}(\tau)\mathrm{d}g_{2,n}(\tau)/\mathrm{d}\tau}{g_{2,n}(\tau)^2}\nonumber\\[1mm]
	&=- Ha_n\alpha'_{n-1}(- H\tau)\ket{\psi(\tau)}+\frac{a_n}{2}\left(\left\langle H\alpha'_{n-1}(- H\tau)\psi(\tau)\mid\psi(\tau)\right\rangle+\left\langle\psi(\tau)\mid H\alpha'_{n-1}(- H\tau)\psi(\tau)\right\rangle\ket{\psi(\tau)}\right)\nonumber\\[1mm]
	&=-a_n\left( H\alpha'_{n-1}(- H\tau)-\operatorname{Re}\left\langle H\alpha'_{n-1}(- H\tau)\psi(\tau)\mid\psi(\tau)\right\rangle\right)\ket{\psi(\tau)}\nonumber\\[1mm]
	&=-\prod_{k=1}^{n}a_k\left( H\exp\left(\sum_{k=1}^{n-1}a_k\alpha_{k-1}(- H\tau)\right)-\operatorname{Re}\left\langle H\exp\left(\sum_{k=1}^{n-1}a_k\alpha_{k-1}(- H\tau)\right)\psi(\tau)\mid\psi(\tau)\right\rangle\right)\ket{\psi(\tau)}.
\end{align}
which is identical to Eq.~\eqref{eq:general}.
As previously mentioned, the standard quantum imaginary time evolution is recovered for the choice $n=1,\,a_1=1$ (standard exponential function, cf. \eqref{eq:qite_EOM}), as follows:
\begin{align}
	\frac{\mathrm{d}}{\mathrm{d}\tau}\ket{\psi(\tau)}
	&=-\left(H-E_1(\tau)\right)\ket{\psi(\tau)},\quad
	E_1(\tau)
	=\langle\psi(\tau)\mid H\mid\psi(\tau)\rangle.
\end{align}
Moreover, for the choice $n=2,\,a_1=1,\,a_2=1$ (double exponential function) we obtain
\begin{align}
	\frac{\mathrm{d}}{\mathrm{d}\tau}\ket{\psi(\tau)}
	&=-\left(He^{- H\tau}-E_2(\tau)\right)\ket{\psi(\tau)},\quad
	E_2(\tau)
	=\langle\psi(\tau)\mid He^{- H\tau}\mid\psi(\tau)\rangle.
\end{align}
We see that the resulting equation for double-exponential QIPA takes a similar form to the original Wick-rotated equation of motion, but with more rapid convergence to the ground state of $H$.

\subsection{McLachlan's variational principle for the double-exponential function}
The McLachlan's variational principle is equivalent to the following minimization problem: 
\begin{align}
	\delta\left\|\left(\partial/\partial\tau+\left[He^{- H\tau}-E_2(\tau)\right]\right)|\ket{\psi(\tau)}\right\|^2
	=0
\end{align}
Since $\theta$ is real, we have
\begin{equation}\label{eq:square_of_mcl2}
	\begin{split}
		&\left\|\left(\partial/\partial\tau+\left[He^{- H\tau}-E_2(\tau)\right]\right)|\phi(\theta(\tau))\rangle\right\|^2\\
		&\qquad=\sum_{m,n}\frac{\partial\langle\phi(\theta(\tau))|}{\partial\theta_m}\frac{\partial|\phi(\theta(\tau))\rangle}{\partial\theta_n}\dot{\theta}_m\dot{\theta}_n+\langle\phi(\theta(\tau))|^2\left[He^{- H\tau}-E_2(\tau)\right]^2|\phi(\theta(\tau))\rangle\dots\\
		&\qquad\quad+\sum_m\frac{\partial\langle\phi(\theta(\tau))|}{\partial\theta_m}\left[He^{- H\tau}-E_2(\tau)\right]|\phi(\theta(\tau))\rangle\dot{\theta}_m+\sum_m\langle\phi(\theta(\tau))|\left[He^{- H\tau}-E_2(\tau)\right]\frac{\partial|\phi(\theta(\tau))\rangle}{\partial\theta_m}\dot{\theta}_m.
	\end{split}
\end{equation}
Differentiating \eqref{eq:square_of_mcl2} with respect to $\dot{\theta}_k$, we obtain
\begin{equation}\label{eq:deriv_mcl2}
	\begin{split}
		&\partial\left\|\left(\partial/\partial\tau+\left[He^{- H\tau}-E_2(\tau)\right]\right)|\phi(\theta(\tau))\rangle\right\|^2/\partial\dot{\theta}_k\\[2mm]
		&\qquad=\sum_m\left(\frac{\partial\langle\phi(\theta(\tau))|}{\partial\theta_k}\frac{\partial|\phi(\theta(\tau))\rangle}{\partial\theta_m}+\frac{\partial\langle\phi(\theta(\tau))|}{\partial\theta_m}\frac{\partial|\phi(\theta(\tau))\rangle}{\partial\theta_k}\right)\dot{\theta}_m\dots\\
		&\qquad\quad+\frac{\partial\langle\phi(\theta(\tau))|}{\partial\theta_k}\left[He^{- H\tau}-E_2(\tau)\right]|\phi(\theta(\tau))\rangle+\langle\phi(\theta(\tau))|\left[He^{- H\tau}-E_2(\tau)\right]\frac{\partial|\phi(\theta(\tau))\rangle}{\partial\theta_k}.
	\end{split}
\end{equation}
Consequently, since $\langle\phi(\theta(\tau))\mid\phi(\theta(\tau))\rangle=1$, we conclude that 
\begin{align}
    \left\|\left(\partial/\partial\tau+\left[He^{- H\tau}-E_2(\tau)\right]\right)|\phi(\theta(\tau))\rangle\right\|
    =0
\end{align}
is equivalent to the following linear equation
\begin{align}\label{eq:QITEExpressionDouble}
	\sum_{m} A_{k,m}\Dot{\theta}_m
	=C_k
\end{align}
with 
\begin{align}
	{A}_{k,m}
	=\text{Re}\left(\frac{\partial\langle\phi(\theta(\tau)|}{\partial\theta_k}\frac{\partial|\phi(\theta(\tau))\rangle}{\partial\theta_m}\right)
	\quad\text{and}\quad
	{C}_k
	=-\text{Re}\left(\frac{\partial\langle\phi(\theta(\tau))|}{\partial\theta_k}He^{- H\tau}|\phi(\theta(\tau))\rangle\right).\label{eq:qipa_C}
\end{align}
Therefore, as in the case of the standard exponential oracle, the McLachlan's principle reduces to a linear system of equations.

\subsubsection{Quantum circuit evaluations of \texorpdfstring{${A}$}{~A} and \texorpdfstring{${C}$}{~C}}
\begin{figure}[t]
	\centering
	\includegraphics[scale=1.0]{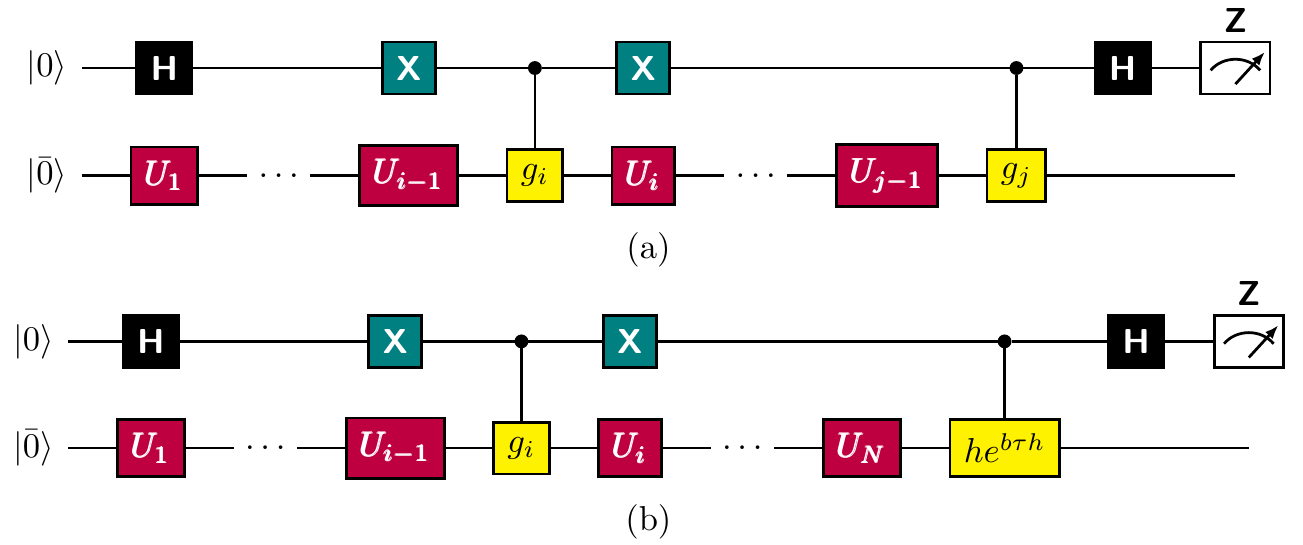}
	\caption{(a) Quantum circuit to evaluate $\textrm{Re}(\bra{\bar{0}}\bar{U}^\dagger_{k,i}\bar{U}_{l,j}\ket{\bar{0}})$ as a probability of finding the ancillary qubit in $0$. (b) Quantum circuit to evaluate $\textrm{Re}(\bra{\bar{0}}\bar{U}_{k,i}^\dagger he^{b\tau h} U\ket{\bar{0}} )$ as a probability of finding the ancillary qubit in $0$.}\label{fig:Hadamard_test_circuit_qipa}
\end{figure}
${A}$ and ${C}$ can be obtained from the circuits shown in \figref{fig:Hadamard_test_circuit_qipa}. As seen from the above \eqref{eq:qipa_C}, the main difference between QIPA and QITE, in terms of comparing the use of double-exponential and exponential cooling functions, is the presence of $e^{- H\tau}$ in $C_k$. Here, we approximate the exponential by its Taylor series expansion to the second order, as follows:
\begin{align}
	{C}_k
	=\textrm{Re}(\bra{\bar{0}}\bar{U}_{k}^\dagger HU\ket{\bar{0}})-\tau\textrm{Re}(\bra{\bar{0}}\bar{U}_{k}^\dagger H^2U\ket{\bar{0}})+\frac{\tau^2}{2}\textrm{Re}(\bra{\bar{0}}\bar{U}_{k}^\dagger H^3 U\ket{\bar{0}})+ \mathcal{O}(H^4).
\end{align}

\newpage
\section{Ansatz circuits}\label{Appen:ansatzes}
\begin{figure}[ht]
	\centering
	\includegraphics[scale=0.8]{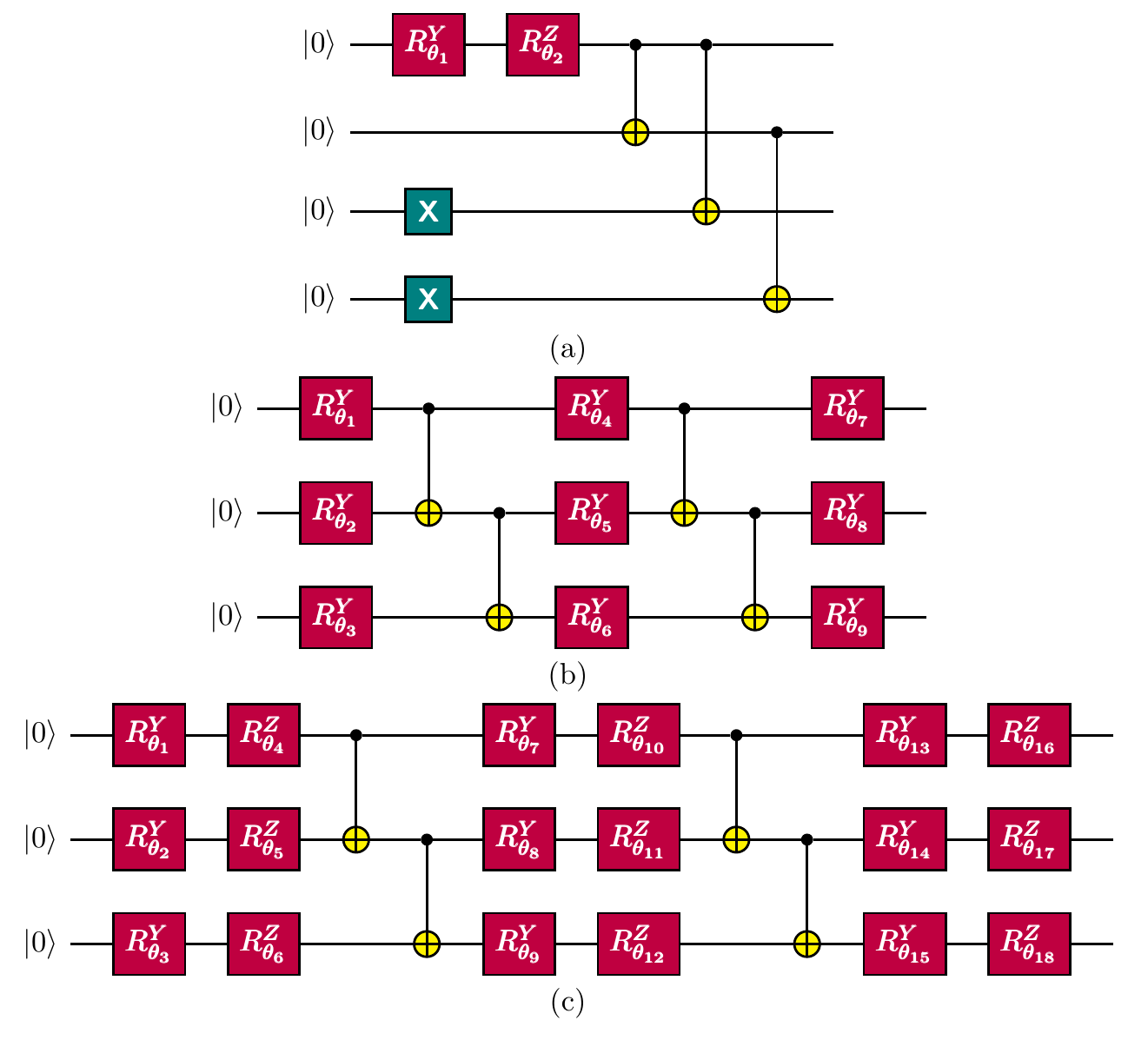}
	\caption{Ansatz circuit used for (a) the $\text{H}_2$ dissociation curve (b-c) the factorization of number $15$ and the transmon qubit ground-state search, respectively.}\label{fig:ansatzes_used}
\end{figure}

\newpage
\section{Example quantum circuit to evaluate a matrix element A}
\begin{figure}[h]
	\centering
	\includegraphics[scale=0.85]{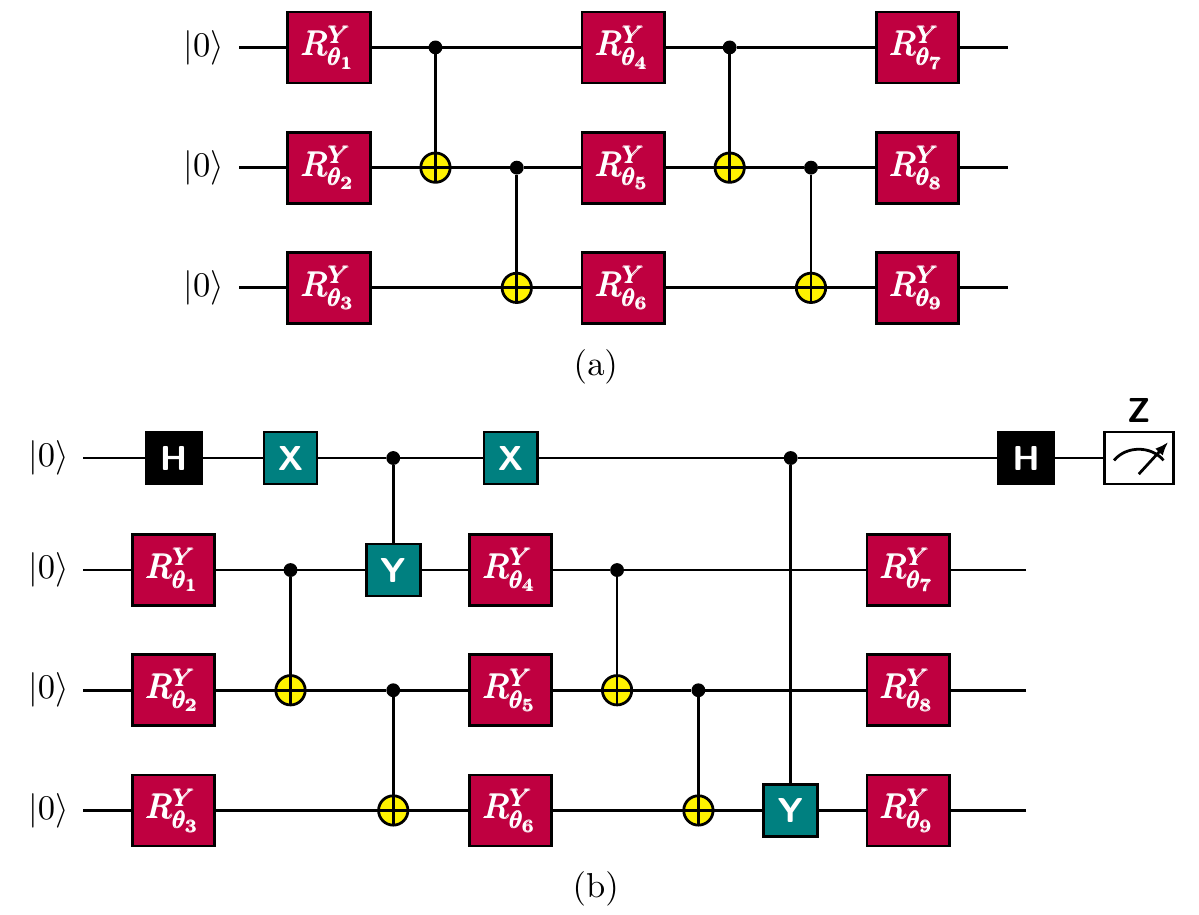}
	\caption{(a) Ansatz circuit used for the factorization of number $15$. (b) Quantum circuit to evaluate the matrix element $A_{4,9}=\textrm{Re}(\frac{\partial\bra{\phi(\tau)}}{\partial\theta_4}\frac{\partial\ket{\phi(\tau)}}{\partial\theta_9})$ as a probability of finding the ancillary qubit in $0$.}\label{fig:Hadamard_test_circuit_example}
\end{figure}

\newpage
\section{Qubit encodings}\label{Appen:qubits_encodings}
Here, we list the Pauli strings corresponding to various quantum operators present in a flux tunable transmon Hamiltonian, in the standard binary and Gray representation, respectively.
\begin{table}[htbp]
	\begin{center}
		\begin{tabular}{|c|c|c|}
			\hline
			$d=16$ & Std. Binary & Gray\\
			\hline
 			$N$ 
			&
			\begin{tabular}[c]{@{}c@{}}
				$-0.5$ $I$\\
				$-4.0$ $Z_{3}$\\
				$-2.0$ $Z_{2}$\\
				$-1.0$ $Z_{1}$\\
				$-0.5$ $Z_{0}$\\
			\end{tabular}
			&
			\begin{tabular}[c]{@{}c@{}}
				$-0.5$ $I$\\
				$-4.0$ $Z_{3}$\\
				$-2.0$ $Z_{2}Z_{3}$\\
				$-1.0$ $Z_{1}Z_{2}Z_{3}$\\
				$-0.5$ $Z_{0}Z_{1}Z_{2}Z_{3}$\\
			\end{tabular}
			\\
			\hline
			$\cos\varphi$ 
			&
			\begin{tabular}[c]{@{}c@{}}
				$+0.5$ $X_{0}$\\
				$+0.25$ $X_{0}X_{1}$\\
				$+0.25$ $Y_{0}Y_{1}$\\
				$+0.125$ $X_{0}X_{1}X_{2}$\\
				$+0.125$ $X_{0}Y_{1}Y_{2}$\\
				$+0.125$ $Y_{0}X_{1}Y_{2}$\\
				$-0.125$ $Y_{0}Y_{1}X_{2}$\\
				$+0.0625$ $X_{0}X_{1}X_{2}X_{3}$\\
				$+0.0625$ $X_{0}X_{1}Y_{2}Y_{3}$\\
				$+0.0625$ $X_{0}Y_{1}X_{2}Y_{3}$\\
				$-0.0625$ $X_{0}Y_{1}Y_{2}X_{3}$\\
				$+0.0625$ $Y_{0}X_{1}X_{2}Y_{3}$\\
				$-0.0625$ $Y_{0}X_{1}Y_{2}X_{3}$\\
				$-0.0625$ $Y_{0}Y_{1}X_{2}X_{3}$\\
				$-0.0625$ $Y_{0}Y_{1}Y_{2}Y_{3}$\\
			\end{tabular}
			&
			\begin{tabular}[c]{@{}c@{}}
				$+0.5$ $X_{0}$\\
				$+0.25$ $X_{1}$\\
				$-0.25$ $Z_{0}X_{1}$\\
				$+0.125$ $X_{2}$\\
				$-0.125$ $Z_{1}X_{2}$\\
				$+0.125$ $Z_{0}X_{2}$\\
				$-0.125$ $Z_{0}Z_{1}X_{2}$\\
				$+0.0625$ $X_{3}$\\
				$-0.0625$ $Z_{2}X_{3}$\\
				$+0.0625$ $Z_{1}X_{3}$\\
				$-0.0625$ $Z_{1}Z_{2}X_{3}$\\
				$+0.0625$ $Z_{0}X_{3}$\\
				$-0.0625$ $Z_{0}Z_{2}X_{3}$\\
				$+0.0625$ $Z_{0}Z_{1}X_{3}$\\
				$-0.0625$ $Z_{0}Z_{1}Z_{2}X_{3}$\\
			\end{tabular}
			\\
 			\hline
			$\sin\varphi$ 
			&
			\begin{tabular}[c]{@{}c@{}}
				$-0.5$ $Y_{0}$\\
				$-0.25$ $X_{0}Y_{1}$\\
				$+0.25$ $Y_{0}X_{1}$\\
				$-0.125$ $X_{0}X_{1}Y_{2}$\\
				$+0.125$ $X_{0}Y_{1}X_{2}$\\
				$+0.125$ $Y_{0}X_{1}X_{2}$\\
				$+0.125$ $Y_{0}Y_{1}Y_{2}$\\
				$-0.0625$ $X_{0}X_{1}X_{2}Y_{3}$\\
				$+0.0625$ $X_{0}X_{1}Y_{2}X_{3}$\\
				$+0.0625$ $X_{0}Y_{1}X_{2}X_{3}$\\
				$+0.0625$ $X_{0}Y_{1}Y_{2}Y_{3}$\\
				$+0.0625$ $Y_{0}X_{1}X_{2}X_{3}$\\
				$+0.0625$ $Y_{0}X_{1}Y_{2}Y_{3}$\\
				$+0.0625$ $Y_{0}Y_{1}X_{2}Y_{3}$\\
				$-0.0625$ $Y_{0}Y_{1}Y_{2}X_{3}$\\
			\end{tabular}
			&
			\begin{tabular}[c]{@{}c@{}}
				$-0.5$ $Y_{0}Z_{1}Z_{2}Z_{3}$\\
				$-0.25$ $Y_{1}Z_{2}Z_{3}$\\
				$+0.25$ $Z_{0}Y_{1}Z_{2}Z_{3}$\\
				$-0.125$ $Y_{2}Z_{3}$\\
				$+0.125$ $Z_{1}Y_{2}Z_{3}$\\
				$-0.125$ $Z_{0}Y_{2}Z_{3}$\\
				$+0.125$ $Z_{0}Z_{1}Y_{2}Z_{3}$\\
				$-0.0625$ $Y_{3}$\\
				$+0.0625$ $Z_{2}Y_{3}$\\
				$-0.0625$ $Z_{1}Y_{3}$\\
				$+0.0625$ $Z_{1}Z_{2}Y_{3}$\\
				$-0.0625$ $Z_{0}Y_{3}$\\
				$+0.0625$ $Z_{0}Z_{2}Y_{3}$\\
				$-0.0625$ $Z_{0}Z_{1}Y_{3}$\\
				$+0.0625$ $Z_{0}Z_{1}Z_{2}Y_{3}$\\
			\end{tabular}
			\\
			\hline
		\end{tabular}
	\end{center}
	\caption{Qubit encodings (standard binary and Gray code) of elementary operators used in this study, with a truncation of $d=16$. In our numerical experiments, we utilize the Gray code \cite{sawaya2020resource}.}\label{tab:enc16}
\end{table}

\end{document}